\pdfoutput=1
\documentclass[useAMS,usenatbib]{mn2e}

\usepackage[]{graphicx}

\title[FUV emission in Abell 2597 and Abell 2204]{Far Ultraviolet Emission in the A2597 and A2204 Brightest Cluster Galaxies.}
\author{J. B. R. Oonk, N. A. Hatch, W. Jaffe, M. N. Bremer, R. J. van weeren}
\begin{document}

\date{Accepted ???? Received ????; in original form ????}

\pagerange{\pageref{firstpage}--\pageref{lastpage}} \pubyear{2002}

\maketitle

\label{firstpage}

\begin{abstract}
We use the Hubble Space Telescope ACS/SBC and Very Large Telescope FORS cameras to observe the Brightest Cluster Galaxies in Abell 2597 and Abell 2204 in the far-ultraviolet (FUV) F150LP and optical U, B, V, R, I Bessel filters. 

The FUV and U band emission is enhanced in bright, filamentary structures surrounding the BCG nuclei. These filaments can be traced out to 20~kpc from the nuclei in the FUV. Excess FUV and U band light is determined by removing emission due to the underlying old stellar population and mapped with 1 arcsec spatial resolution over the central 20~kpc regions of both galaxies. 

We find the FUV and U excess emission to be spatially coincident and a stellar interpretation requires the existence of a significant amount of 10000-50000~K stars. Correcting for nebular continuum emission and dust intrinsic to the BCG further increases the FUV to U band emission ratio and implies that stars alone may not suffice to explain the observations. However, lack of detailed information on the gas and dust distribution and extinction law in these systems prevents us from ruling out a purely stellar origin. 

Non-stellar processes, such as the central AGN, Scattering, Synchrotron and Bremsstrahlung emission are investigated and found to not be able to explain the FUV and U band measurements in A2597. Contributions from non-thermal processes not treated here should be investigated.

Comparing the FUV emission to the optical H$\alpha$ line emitting nebula shows good agreement on kpc-scales in both A2597 and A2204. In concordance with an earlier investigation by \citet{Od04} we find that O-stars can account for the ionising photons necessary to explain the observed H$\alpha$ line emission.
\end{abstract}


\section{Introduction}
Regions at the centre of rich clusters where the hot T$\sim$10$^{8}$~K, thermal X-ray emitting gas is dense enough to cool radiatively within a Hubble time are called cool-cores. Cooling rates of the order of 10$^{2-3}$~$M_{\odot}$~yr$^{-1}$ in the central few hundred kpc of the cluster have been estimated from X-ray imaging for this hot gas \citep[e.g.,][]{Pe98}. However, X-ray spectroscopy implies that at most 10 per-cent of the X-ray emitting gas cools below one third of the virial temperature of a system, see \citep{Pe06} for a review. The solution most often invoked in the literature is that some form of heating balances the radiative cooling of the X-ray gas. 

At the heart of these cool-core clusters lie their Brightest Cluster Galaxies (BCG). These cD type galaxies are the most massive galaxies in the universe and peculiar in many ways. They contain substantial, cool (T$\sim$10$^{1-4}$~K) gas and dust components within about 30~kpc from their nucleus \citep*[e.g.,][]{Ja97,Ed01,Ir01,Ja01,Ed02,Wi02,Sa03,Ha05,Ja05,Wi06,Jo07,Od08,Wi09,Oo10}. Whether this cool gas is the product of the cooling process in these clusters or has an alternative origin, such as minor mergers, is unclear. However, statistically speaking these cooler gas nebulae exist only in and around BCGs situated in cool-core clusters \citep[e.g.,][]{He89}.  

These gas components at T$\sim$10$^{1-4}$~K in BCGs emit far more energy than can be explained by the simple cooling of the intracluster gas through these temperatures and some form of heating is required \citep[e.g.,][]{Fa81,He89,Ja05,Oo10}. Detailed investigations show that the primary source of ionisation and heating of the cool gas must be local to the gas \citep{Jo88,Ja05,Oo10}. This requirement combined with the observation that BCGs in cool-core clusters have significantly bluer colors in the ultraviolet (UV) to optical regime than their peers in non-cool-core clusters and in the field supports a young stellar origin for the ionizing photons \citep[][ and references therein]{Do10}.

Previous studies agree that the UV to optical emission is consistent with young stars \citep*{Cr93,Pi09,Hi10}. However, these studies suffer from poor spatial resolution, as the UV part of their datasets is based on observations with the International Ultraviolet Explorer (IUE) and GALEX. 

A number of recent papers, using high resolution FUV imaging with the Hubble Space Telescope (HST), have shown that BCGs in cool-core clusters contain clumpy, asymmetric FUV emission on scales up to $\sim$30~kpc from the nucleus \citep{Od10,Oo10a}. This is consistent with the scales on which cool gas has been detected in BCGs \citep[e.g. ][]{Ed01,Sa03,Ja05,Wi09,Oo10}. The clumpy, extended morphology of the FUV emission is consistent with a local stellar origin and comparison with radio imaging disfavours a direct relation with the central active galactic nucleus (AGN) or its outflows \citep{Od10}.

Based on FUV morphology \citet{Od10} and \citet{Od04} argue that the FUV emission is due to young stars and that the clumpiness may be related to the absence of gas rotation on large scales \citep{Wi09,Oo10}. In this picture the FUV luminosities indicate a minimum star formation rate (SFR) of the order of 1-10~M$_{\odot}$~yr$^{-1}$ in BCGs. Deriving the true SFRs is highly dependent on very uncertain extinction corrections and as we will discuss below this uncertainty limits our interpretation of the observations. 

\citet{Od10} and \citet{Od04} show that FUV continuum and the Ly$\alpha$ emission cover the same area, with the FUV continuum emission being clumpier than the Ly$\alpha$ emission. They find evidence for significant dust extinction from the high H$\alpha$/H$\beta$ ratios observed towards their BCGs. They also state that they find variations in the Ly$\alpha$/H$\alpha$ ratio, which may be attributed to a non-uniform dust distribution.

\subsection{This project}
Here we present deep far-ultraviolet (FUV) and optical imaging for the BCGs PGC 071390 in Abell 2597 (hereafter A2597) and ABELL 2204\_13:0742 (LARCS catalog by \citet{Pi06} in Abell 2204 (hereafter A2204). We perform the first spatially resolved investigation into the FUV to optical emission ratio in the central 20~kpc regions of these BCGs. 

The two BCGs studied here are part of our previous sample of BCGs in cool core clusters \citep{Ja97,Ja01,Ja05,Oo10}. The objects were selected based on their high cooling rates, strong H$\alpha$, H$_{2}$ emission and low ionisation radiation in order to minimise the role that their AGN have on the global radiation field. A2597 and A2204 have been the subject of numerous investigations in the past and have been observed at wavelengths from radio to X-rays \citep[e.g.][]{He89,Vo97,Ko99,Do00,Od04,Ja05,Wi06,Wi09,Oo10}.

This paper has two main goals: (1) To establish the morphology of the FUV emission from two BCGs and compare it to the emission structures at other wavelengths (2) To establish the nature of the stars responsible for the FUV emission (if indeed it arises from stars).  For this second purpose we compare quantitatively the FUV emission with optical emission in other bands, particularly U-band.  For this comparison there are two important intermediate steps: correction of the longer wavelength (i.e. U-band) fluxes for emission from the old stellar population of the galaxy, and correction of the U- and FUV- band fluxes for dust extinction. This last procedure in particular requires considerable discussion here.

For the above purposes we present in Section 2 descriptions of the methods and reductions of our observations with the HST ACS-SBC and the VLT FORS, and of archival data: radio data from the VLA and X-ray data from Chandra. In Section 3 we present the direct qualitative and quantitative results of the reductions in the form of images, graphs, and tables.  In Section 4 we make a preliminary analysis of the FUV to optical colors by interpreting, somewhat naively, the FUV/U ratio on the basis of models of starburst spectra. In Section 5 we present a more sophisticated analysis of the spectra including the removal of U-band emission of old stars and different dust scenarios. In Section 6 we discuss the star formation implied by the FUV emission and its relation to the ionised gas in these systems.

Because the FUV/U ratio in the purely stellar scenario is uncomfortably large, in Section 7 we consider the possibility that some of this emission is from conventional non-stellar sources. Finally in Section 8 we compare our results with those of other ultraviolet observations, and discuss in detail the possibility that the high value of the extinction corrected FUV/U ratio is the result of unconventional extinction behaviour of the dust in the UV.  Section 9 contains our conclusions.

Throughout this paper we will assume the following cosmology; $\textit{H}_{\mathrm{0}}$=72~$\mathrm{km}$~$\mathrm{s}^{-1}$~$\mathrm{Mpc}^{-1}$, $\Omega_{\mathrm{m}}$=0.3 and $\Omega_{\mathrm{\Lambda}}$=0.7. For Abell 2597 at z=0.0821 \citep*{Vo97} this gives a luminosity distance 363~$\mathrm{Mpc}$ and angular size scale 1.5~$\mathrm{kpc}$~$\mathrm{arcsec}^{-1}$. For Abell 2204 at z=0.1517 \citep{Pi06} this gives a luminosity distance 702~$\mathrm{Mpc}$ and angular size scale 2.6~$\mathrm{kpc}$~$\mathrm{arcsec}^{-1}$.

\section[]{Observations and Reduction}\label{sec_obs_red}
We have observed the central BCG in the two cool-core clusters A2597 and A2204 with ACS/SBC on the HST and with FORS on the VLT. The observations are summarised in Tables \ref{t_obslog_a2597} and \ref{t_obslog_a2204}. Similar, but less deep, FUV and optical data for A2597 has previously been presented by \citet{Vo97,Ko99,Do00,Od04}. Similar data for A2204 has not been published before. We complement this data set with archival radio and X-ray observations from the Very Large Array (VLA) and Chandra. The data reduction is described below.

\subsection{HST ACS-SBC imaging}
Hubble Space telescope (HST) FUV images were obtained with the Solar Blind Channel (SBC) of the Advanced Camera for Surveys (ACS) in the F150LP (1612~$\rm \AA$) filter (effective wavelength 1612~$\rm \AA$). The filter was selected to sample the FUV continuum emission and not include the Ly$\alpha$ line. Five pointings were performed for the A2204 field, with a total an exposure time of 3.76\,hours. Five pointings were also performed for the A2597 field. Two of the five A2597 pointings were found to be contaminated by a non-uniform dark current glow. These two pointings were excluded from further analysis. The total exposure time for the three remaining pointings on A2597 is 2.26\,hours. No dark current glows were found in any of the A2204 pointings.

Flat-fielded and dark-subtracted single exposure frames were obtained from the {\it HST} archive. Dither offsets were calculated for the exposures of each target using image cross-correlation and then the exposures were drizzled and combined using the {\sc stsdas} package {\sc multidrizzle}. Optical distortions were automatically corrected by {\sc multidrizzle} during the image stacking, but no cosmic ray correction was applied since the MAMA detectors are not affected by cosmic rays.

The combined images are convolved to an output spatial resolution of a Gaussian with 1 arcsec full width at half maximum (FWHM), in order to match the optical images. These images are then re-gridded and aligned to the optical images by using the northern star and the two companion galaxies for A2204. The FUV images of A2597 do not contain any sources that can be used to align it to the optical images. We aligned these by comparing the nuclear structures in the ACS/SBC image to the FORS U-band image. We convert the ACS/SBC units elec s$^{-1}$ to AB magnitude by applying the $\textit{ABmag}$ zeropoint (ACS instrument handbook) and we convert this to flux in units of erg~s$^{-1}$~cm$^{-2}$~Hz$^{-1}$~arcsec$^{-2}$ by applying the zeropoint for the AB system.

\subsection{Optical data}

\subsubsection{VLT FORS Imaging}
A2597 and A2204 were observed with the Focal Reducer/low dispersion Spectrograph (FORS) in imaging mode on the Very Large Telescope (VLT) in 2001 and 2002. Images were taken in the U, B, V, R and I Bessel filters under photometric conditions with a seeing better than 1 arcsec.  The reduction was performed using IRAF and a number of dedicated IDL routines. The frames are dark and flat corrected. Hot pixels, cosmics and artefacts are identified and interpolated over. A linear plane is fitted to the sky background and subtracted from the data.

All images are convolved to a common output spatial resolution of a Gaussian with 1 arcsec FWHM. The point spread function (PSF) was measured using about 20 stars in each frame, 10 of these stars are common to all observed frames. The zeropoints are determined by correcting the Landolt standard star magnitudes for the color difference between the Johnson (Landolt) and Bessel (FORS) filter systems. Multiple standard stars are available in each Landolt field and the uncertainty given for the zeropoint is taken as the maximum deviation in the zeropoints obtained from all standard stars within a given field. Only frames obtained in photometric conditions are kept in the subsequent analysis. The results are given in Tables \ref{t_obslog_a2597} and \ref{t_obslog_a2204}.

Using bright stars near the BCGs that are visible in all bands we align the FORS images to the 2MASS World Coordinate System (WCS). The FORS Bessel magnitudes are converted to AB magnitudes. The conversion factor \textit{(FORS\_AB)$_{conv}$} is calculated using the FORS Bessel filter curves and the HYPERZ program \citep*[][; H. Hildebrandt priv. comm.]{Bo00}. The results are listed in Table \ref{t_ab_conv}. We then convert AB magnitude to flux in units of erg~s$^{-1}$~cm$^{-2}$~Hz$^{-1}$~arcsec$^{-2}$ by applying the zeropoint for the AB system.

\subsubsection{VLT FORS Spectroscopy}\label{s_fors_spec}
A2597 and A2204 were observed with FORS in a 2~arcsec wide, long-slit spectroscopy mode on the VLT in 1999 and 2002. The A2597 observations were taken along minor axis of the BCG and the A2204 observations were taken along an axis running the north-south. Both set of observations intercepted the nucleus of the BCG. The data will presented in a forthcoming paper (Oonk et al. in prep.). For the purposes of this paper we will only use spatially integrated spectra. In the case of A2597 we summed the spectrum over a 16$\times$2~arcsec$^{2}$ region and for A2204 we summed the spectrum over a 12.5$\times$2~arcsec$^{2}$ region. Within these regions the lines ratios do not vary strongly as a function of position. 

Here we use these spectra to investigate (i) the contamination of the U, B, V, R and I Bessel filters by line emission and (ii) the extinction using the Balmer decrements. The spectra were reduced using a set of dedicated IDL scripts. Emission line fluxes were measured using Gaussian fitting. Broadband fluxes were obtained by convolving the spectra with FORS Bessel filter curves. The contribution of the line emission to the broadband flux is given in Table \ref{t_contam} and the measured Balmer decrements are given in Table \ref{t_dust}.

\subsection{Radio data}
A2597: Archival VLA 5~GHz observations of A2597 (project code: AC742) were reduced with the NRAO Astronomical Image Processing System (AIPS). The A-configuration observations were taken in single channel continuum mode with two 50~MHz wide IFs centred around 4850~MHz. The total on source time was 405 min. The data was flux calibrated using the primary calibrator 0137+331. We used the \citet*{Pe99} extension to the \citet{Ba77} scale to set the absolute flux scale. Amplitude and phase variations were tracked using the secondary calibrator 2246-121 and applied to the data. The data was imaged using robust weighting set to 0.5, giving a beam size of 0.64 arcsec by 0.49 arcsec. The one sigma map noise is 20~$\mu\mathrm{Jy}$~$\mathrm{beam}^{-1}$. Radio maps of A2597 at 0.33, 1.4, and 8.4~$\mathrm{GHz}$ were previously published by \citep{Bi08,Cl05,Sa95}.
\\ \\
A2204: Archival VLA 5 and 8~GHz observations of A2204 (project code: AT211) were reduced with the NRAO Astronomical Image Processing System (AIPS). The 5 GHz observations were taken in A-configuration in single channel continuum mode with two 50~MHz wide IFs centred around 4850~MHz. The 8~GHz observations were taken in B- and C-configuration in single channel continuum mode with two 50~MHz wide IFs centred around 8115~MHz. The total on source time was 31 min and 113 min for the 5 and 8~GHz observations respectively. The data was flux calibrated using the primary calibrator 1331+305. Amplitude and phase variations were tracked using the secondary calibrator 1651+014 and applied to the data. The data was imaged using robust weighting set to 0.0. This gives a 0.44 arcsec by 0.41 arcsec beam size and one sigma noise of 60~$\mu\mathrm{Jy}$~$\mathrm{beam}^{-1}$ at 5~GHz observations and a 0.97 arcsec by 0.86 arcsec beam size and one sigma noise of 20~$\mu\mathrm{Jy}$~$\mathrm{beam}^{-1}$ at 8~GHz. Observations of A2204 at 1.4, 4.8 and 8~$\mathrm{GHz}$ were previously published by \citep{Sa09}.

\subsection{X-ray data}
We have retrieved all publicly available X-ray data from the \textit{CHANDRA} archive. For A2597 we combined three separate observations having a total exposure time of 153.7~$\mathrm{ks}$ (project codes 7329; 6934; 922). For A2204 we combined three separate observations having a total exposure time of 98.1~$\mathrm{ks}$ (project codes 7940; 6104; 499). \textit{CHANDRA} data for A2597 and A2204 has previously been published by \citep{Mc01,Cl05,Ja05,Sa09}.

\section[]{Results}

\subsection{FUV: A2597 and A2204.}
\begin{figure*}
    \includegraphics[width=0.60\textwidth, angle=90]{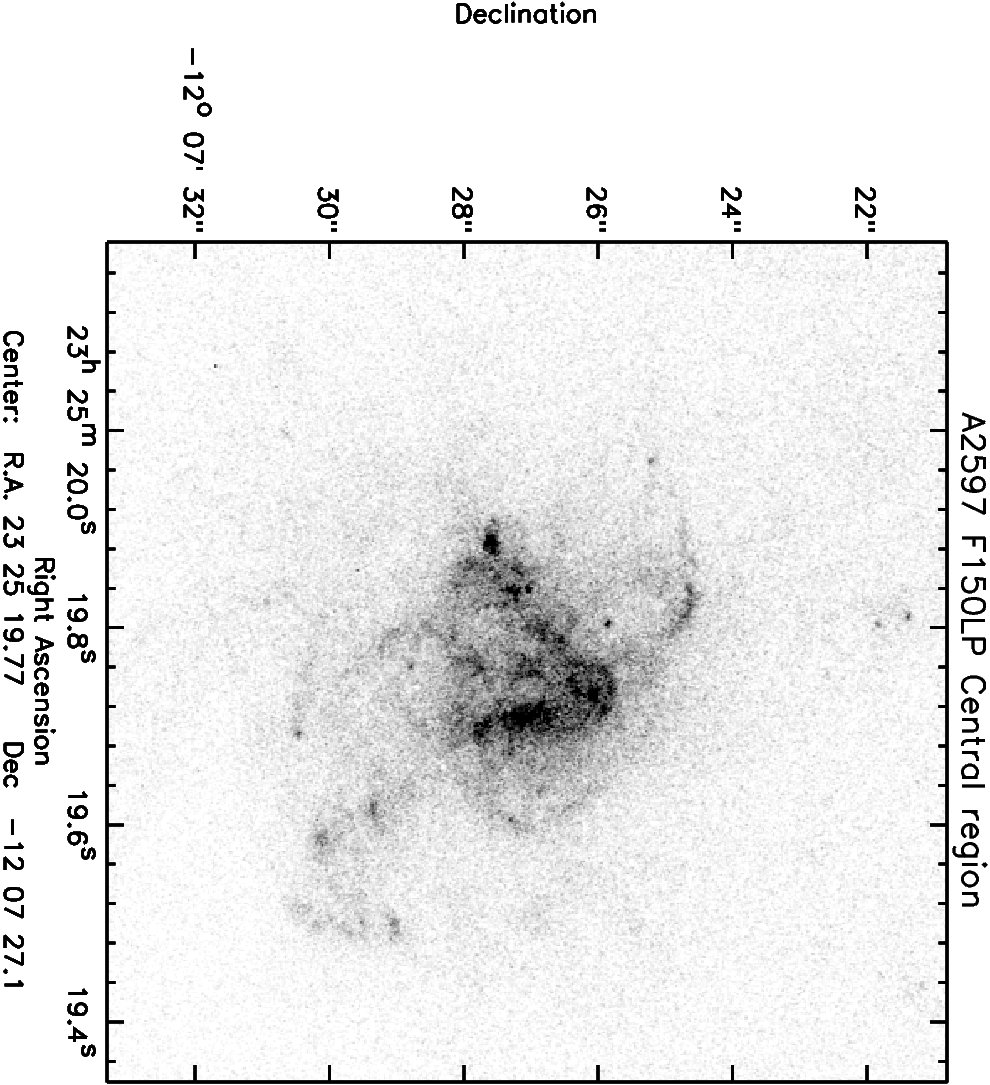}
    \includegraphics[width=0.64\textwidth, angle=90]{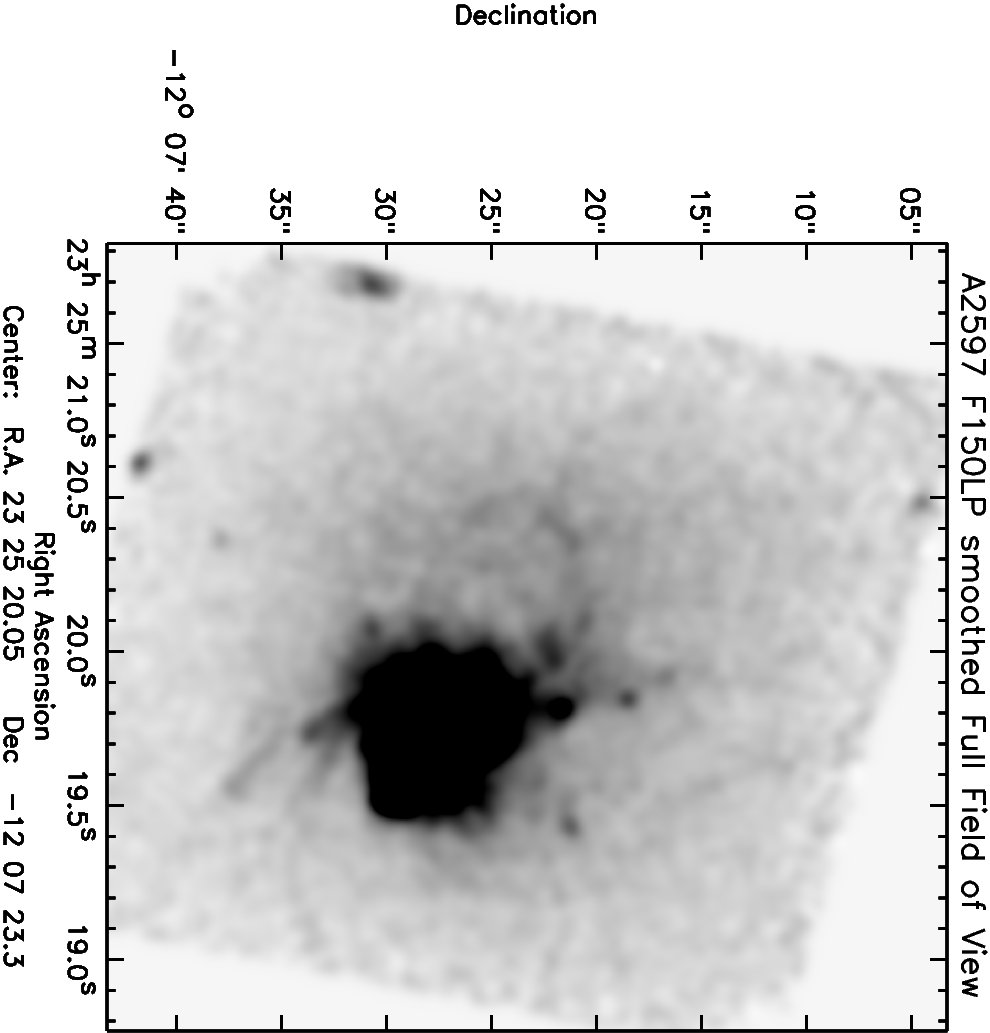}
  \vspace{0.5cm}
  \caption{ABELL 2597 FUV F150LP emission. (\textit{Top}) Central bright FUV continuum emission at the intrinsic resolution of HST. (\textit{Bottom}) FUV continuum emission convolved to an output spatial resolution of 1 arcsec FWHM and scaled to enhance the low surface brightness structures. The BCG nucleus is at ($\alpha;\delta$)=(23 25 19.719 ; -12 07 26.83) (J2000).}\label{f_a2597}
\end{figure*}

The combined F150LP images of A2597 and A2204 show FUV continuum emission out to 20 kpc from their respective BCG nuclei, see Figs. \ref{f_a2597}, \ref{f_a2597_ovl}, \ref{f_a2204} and \ref{f_a2204_ovl}. This emission is observed to originate in knots that are part of narrow, kpc-scale filaments. These knots and filaments are embedded in lower-level, diffuse FUV emission centered on the BCG nucleus. The most prominent knots identified by us are given in Fig. \ref{f_fuv_knots} and Tables \ref{t_a2597_knots} and \ref{t_a2204_knots} in Appendix \ref{fuv_knots}. 

In A2597 the filaments appear to be winding around the BCG nucleus in a spiral-like manner, perhaps indicative of ongoing gas in- and outflows. There are three main filamentary structures extending towards the north-east (NE), the south-east (SE) and the south-west (SW) from the nucleus. With the exception of the SE filament the FUV emission extends mostly along the projected minor axis of the A2597 BCG. The optical nucleus lies along the SW filament.

The central knots observed in the our ACS-SBC FUV continuum map of A2597 match those seen in the lower signal to noise HST-STIS FUV continuum image by \citet{Od04}. The central FUV structures observed here also match the observed optical and near-infrared line emission observed with HST by \citet{Ko99} and \citet{Do00}. The extension of the SW filament towards ($\alpha,\delta$)=(23:25:19.5,-12:07:30) is marginally visible in the Ly$\alpha$+FUV continuum image by \citet{Od04}, but it is not visible in the \citet{Ko99} and \citet{Do00} images. 

Convolving the FUV emission to match the resolution of the deep, ground-based H$\alpha$ map by \citet{Ja05} we find that there is good agreement between H$\alpha$ and FUV continuum emission in the central 40$\times$40~kpc$^{2}$ in A2597, see Fig. \ref{f_a2597_ovl}.

There have been no earlier investigations of A2204 in the FUV at the resolution of HST. There are many small clumps and filaments extending mostly radially away from the nucleus. There are two main filamentary structures. The first filament runs south-east to north-west along the major axis of the BCG. The second filament runs from the south to the north. Both filaments pass through the nucleus. 

Convolving the FUV emission to the match the resolution of the H$\alpha$ map by \citet{Ja05} shows that also in this object there is good agreement between H$\alpha$ and FUV continuum emission in the central 40$\times$40~kpc$^{2}$, see Fig. \ref{f_a2204_ovl}.

\subsection{Optical: A2597 and A2204}
The combined optical images of A2597 show that the BCG has a blue core, see Fig. \ref{f_uvr}. In Section \ref{s_totfu} we show that this core consists of three blue filamentary structures extending outwards from the nucleus towards the north-east, south-east and south-west. These structures closely follow the morphology of the FUV emission. The optical images are not as deep as the FUV image and only allow us to trace these filaments out to about 10~kpc from the nucleus.

The combined optical images of A2204 show that this BCG also has a blue core, see Fig. \ref{f_uvr}. In Section \ref{s_totfu} we show that there are two blue filamentary structures intersecting the nucleus along an axis running north-south and an axis running south-east to north-west. This is in good agreement with the FUV emission. The quality of the optical imaging again only allows us to trace these filaments out to about 10~kpc from nucleus.

The optical images furthermore show that the A2204 cluster is a strongly lensing system. Many blue, lensed galaxies are observed and these become even more pronounced if the B-band data is added (not shown here). The lensed galaxy (z=2.74, \citet{Wi06}) 8 arcsec north-east shows up prominently in the U-band image but it is not visible in the FUV image.

\subsection{Radio and X-ray emission}\label{s_res_rx}
The radio source in the A2597 BCG is known as PKS~2322-12. We show a very deep 5~GHz radio map of A2597 in Fig. \ref{f_a2597_ovl}. The radio structures are in good agreement with the previously published deep, but lower resolution, 1.3~GHz map by \citet{Cl05}. We now clearly see that besides the well known northern and southern radio lobes a much weaker double lobe-like structure appears at a roughly orthogonal position angle. Whether this structure is due to an older outflow or perhaps a backflow from the current jets can not be clarified from this data alone. \citet{Cl05} show that more extended radio emission is present in A2597 at lower frequencies. 

Previous investigations of A2597 claim a correlation between the radio emission, the UV-optical excess emission and the central emission line gas \citep{Mc93,Ko99,Do00,Od04}. In particular they show that the U-band excess light follows the radio lobes in the central 10~kpc, and that some of the FUV and emission line filaments trace parts of the radio lobes.

In \citet{Oo10} we found a strong dynamical interaction between the emission line gas and the current radio jet in the central few kpc of A2597. Outside of this central region there was no sign of a causal relation between the two. The FUV observations presented here are consistent with this picture. We find that the peak of the FUV emission agrees with the radio core in A2597 and that the central FUV filaments overlap with the radio structures. This is most evident for the brightest parts of the northern and southern filaments that appear to curve along the inner edges of the radio lobes. A new feature observed here is that the south-eastern FUV filament is oriented along the low surface brightness double lobe-like radio structure.

The radio source in A2204 is known as VLSS~J1632.7+0534. Fig. \ref{f_a2204_ovl} shows that the 5~GHz radio emission extends to the north and south of the nucleus. This double lobe structure is confirmed by the very deep 8~GHz observations shown in in Fig. \ref{f_a2204_8ghz}. In the 8~GHz map we find two new radio features at the 3~sigma level. The first feature is about 4~arcsec north of the nucleus and, in projection, coincides with the bright northern FUV knot. It is discussed further in Section \ref{s_sfr}. The second feature is a narrow arc about 3~arcsec south-west of the nucleus. This arc does not coincide with any known structures at other wavelengths. Both features are low signal-to-noise and need to be confirmed by future observations.

The relation between the FUV and radio emission in A2204 is similar to A2597. In A2204 we find that the peak of the FUV emission agrees with the radio core and that the north-south FUV filament is aligned with the radio lobes. \citet{Wi09} show that the emission line gas in the central few kpc of A2204 has a high velocity dispersion. It is likely that the gas here is stirred up by the AGN, in the same manner as in A2597, but this needs to be confirmed by observations a higher spatial resolution.

The observations presented here for A2597 and A2204 are consistent with the idea of a strong interaction between the radio source and its immediate surroundings in the central few kpc of the BCG. However, outside of this region the FUV emission, and the emission line gas, does not show an obvious causal relation with the current radio outflows and as such remain to be explained.

The general relation between FUV and radio emission, if any exists, is currently not clear for cool-core BCGs as a class of objects. \citet{Od10} present FUV and radio observations for a sample of 7 BCGs. All of their sources show FUV and radio emission, but only half of them show enhanced FUV continuum emission at the location of the radio source. In our observations, as well as in those by \citet{Od10}, it is found that in almost all BCGs the current radio outflows are active on scales less than 10~kpc whereas the FUV emission is present on scales of about 30~kpc. More and deeper observations, in particular in the radio, are necessary to investigate the relation between the FUV emission and the radio outflows in these objects further.

The radio images of A2597 and A2204 presented here show that previously unknown or unresolved radio structures become visible when imaging the BCGs to ever-greater depths. At present it is difficult to determine the details of these structures as matched multi-frequency observations at sufficient depth and spatial resolution are lacking. At 5~GHz the radio emission traces young relativistic electrons with a synchrotron lifetime $\sim$10$^{7}$~yr for a magnetic field strength of 30~$\mu$G.

The X-ray emission in both clusters is centered on the BCG. The peak of the X-ray emission agrees with the peak emission observed in the radio, FUV and optical images. The X-ray emission in these clusters is discussed in detail in \citet{Sa09} and \citet{Cl05,Mc01}.

\subsection{Surface brightness profiles}
In Fig. \ref{f_pslit} we compare the FUV, H$\alpha$, V-band, X-ray emission and 5~GHz radio continuum with each other along 2~arcsec wide pseudo-slits centred on the nuclei of the BCGs. The V-band emission traces the older stellar population in both BCGs. The H$\alpha$ emission is taken from \citet{Ja05} and shown superimposed on the smoothed FUV emission in Figs. \ref{f_a2597_ovl} and \ref{f_a2204_ovl}. The surface brightness profiles obtained from these pseudo-slits have been normalised at the position of the optical nucleus and have been computed after convolving all images to a common spatial output resolution of 1~arcsec FWHM.

In A2597 the slits are placed along the projected major (PA=-40~degrees) and minor (PA=+50~degrees) optical axis of the BCG. Along the major axis of the BCG the FUV and H$\alpha$ and radio emission are more centrally concentrated than the stellar V-band and thermal X-ray emission. Northwards along the minor axis, excess FUV, H$\alpha$ and radio emission is observed relative to the V-band stellar emission. No excess emission relative to the V-band stellar light is observed south along the minor axis. Overall, the 5~GHz radio emission is more and the X-ray emission is less centrally concentrated than the FUV, H$\alpha$ and V-band light. The FUV and H$\alpha$ emission trace each other better than the stellar V-band light. The sharp decrease in radio emission outside of the nucleus is in part due to our focus on the 5~GHz radio emission alone. \citet{Cl05} show that the radio source is significantly larger at lower radio frequencies. To properly account for the non-thermal, relativistic, electron component in BCGs matched multi-frequency radio imaging needs to be performed.

In A2204 the slits are placed along an axis running east-west and an axis running north-south. There are no regions with both strong FUV and H$\alpha$ excesses observed relative to the stellar light along these slits. The H$\alpha$ emission appears to be in excess relative to the FUV and stellar V-band light. This is unlikely to be physical. A possible explanation follows from the filters used by \citet{Ja05} to measure H$\alpha$. These filters are so narrow that part of the broad H$\alpha$ line wings at the nucleus are not included. This will then lead to an artificially broadened H$\alpha$ profile. For A2204 we also find that the radio emission is more and the X-ray emission is less centrally concentrated than the FUV, H$\alpha$ and stellar light. The northern radio lobe is largely missed by our north-south slit, this explains the steep drop in the radio surface brightness profile here. The radio source in A2204 is observed to be more extended at lower frequencies \citep{Sa09}.

\section{Total FUV/U emission}\label{s_totfu}
BCGs in cool-core clusters are well known to have large extended ionised gas nebulae and bluer continuum colors than non-cool-core cD galaxies \citep[e.g. ][]{Mc92,Bil08}. 

\citet*{Jo88} find that to explain the observed distribution of H$\alpha$ line emission in NGC~1275, the BCG in the Perseus galaxy cluster, that one needs a source of heating and excitation which is local to the gas. The same is true for other BCGs \citep[e.g.][]{He89}. A natural explanation for these observations would be that hot, young stars are forming in BCGs \citep{Cr93,Pi09,Hi10}.

In the previous section we showed that the ionised gas emission in A2597 and A2204 on large scales matches the FUV emission. In this section we will show that the central 20~kpc regions of the BCGs in these clusters are very blue by investigating  the FUV$_{\nu}$/U$_{\nu}$ flux ratio and by, somewhat naively, comparing it to single stellar population (SSP) models from \citet*{Br03}. We use the FUV$_{\nu}$/U$_{\nu}$ ratio as these are the only bands that are free of contamination by line emission. In Section \ref{s_stellar} we will investigate in more detail what kind of stars are required in order to reproduce the observed FUV to optical colors. In Section \ref{s_starformation} we will study the relation between these stars and the ionised gas in these systems.

Prior to computing the FUV$_{\nu}$/U$_{\nu}$ ratio we remove the background emission from the FUV images using the background regions shown in Fig. \ref{f_uvr}. For consistency we use the same regions to define and remove a zero-level from the optical images. In the optical images these background regions still lie within regions of stellar emission, but the optical colors are not affected by selecting the background in this way.

\subsection{Bruzual \& Charlot 2003 SSP models}
In Fig. \ref{f_totfu} and \ref{f_sbrst} we model the total observed FUV to U band emission with the emission ratio expected from a single stellar population (SSP). The aim of this SSP investigation is not to determine what the exact stellar population is, but simply to show that the emission in the filamentary structures in the cores of the BCGs is blue and indicative of a young stellar population.

The SSPs used in this investigation are a set of solar metallicity template spectra published by \citet{Tr03} that are based on the models by \citet*{Br03}. The ages of these populations vary from 5~Myr to 11~Gyr. We have calculated the FUV$_{\nu}$/U$_{\nu}$ ratio for these models as function of SSP age using different amounts of extinction for A2597 and A2204. The results are shown in Fig. \ref{f_sbrst}.   

If we take into account extinction due to the MW foreground only we find that the central 10~kpc emission in A2597 and A2204 can be explained with SSPs models that have SSP ages less than 100~Myr, see Fig. \ref{f_totfu}. In this analysis most of the blue knots in A2597 and A2204 have SSP ages less than 30~Myr. The most extreme knots, i.e. the eastern knot in the south-eastern filament in A2597 and northern knot in the north-south filament in A2204, have SSP ages less than 5~Myr. 

We note here that the SSPs are double-valued in terms of the FUV$_{\nu}$/U$_{\nu}$ ratio for ages older than 640 Myr in A2597 and 290 Myr in A2204. This behaviour of the FUV$_{\nu}$/U$_{\nu}$ ratio is due to the UV-upturn for stellar populations with ages above a 1 Gyr. For all FUV$_{\nu}$/U$_{\nu}$ ratios in the range where the SSP models are double valued we have set the SSP age in Fig. \ref{f_totfu} to the youngest value allowed in this range. 

Allowing for additional extinction intrinsic to the BCG further lowers the allowed SSP ages for the regions observed. In Fig. \ref{f_totfu} we present the FUV$_{\nu}$/U$_{\nu}$ ratio for SSPs reddened by different amounts of dust. If all of the extinction attributed to the BCG were to lie in front of the filaments then all of the filamentary emission in the central 10~kpc of both A2597 and A2204 would require an SSP age less than 5~Myr. We will discuss the presence of the dust and to correct for it in more detail in Section \ref{s_dust}. 

We conclude here that the observations are marginally consistent with the existence of a very young stellar population. Modelling the filamentary emission with a SSP is not very realistic as there is a considerable underlying older stellar population in the BCGs. Any contribution to the FUV and U band light from the underlying old stellar population will lead to a lower FUV$_{\nu}$/U$_{\nu}$ ratio and hence to an overestimate of the age of the young stellar component. Trying to distinguish multiple stellar populations using our current set of broadband data, with only two line emission free bands, is a highly degenerate process for these complex systems and we will not pursue it here. Instead we will attempt to further constrain the nature of the blue filaments by removing the old stellar population and investigating only the FUV$_{\nu}$/U$_{\nu}$ ratio of the excess emission in terms of simple stellar and non-stellar models.

\section{Excess FUV/U: Stellar Origin ?}\label{s_stellar}
In this section we will investigate whether young stars can account for the observed colors in A2597 and A2204. We do this by obtaining the excess FUV$_{\nu,exc}$/U$_{\nu,exc}$ continuum emission ratio and compare it with main sequence stellar models by \citet*{Ku93} (hereafter K93). The K93 stellar models will allow us to assign stellar temperatures to the stars necessary to explain the observed emission. In this work we focus on stars, however we also present results for blackbody (BB) models. These BB models can be used to describe optically thick emission processes and are given for comparison purposes.

The FUV to U-band ratio is a good discriminator of stellar temperature, because (i) the ratio increases strongly with increasing temperature and (ii) it is single valued in terms of temperature. We show this in the context of BB models in Fig. \ref{f_bb_z0} and this is equally true for the K93 stellar models. We have verified that the stellar temperatures derived via our broad band method are consistent with the stellar temperatures derived from FUV spectroscopy for a number of main sequence in the nearby HII region NGC~604 in M33, see Appendix \ref{app_n604}. 

In order to obtain the excess ratio we will first investigate line contamination of our images and then remove emission due to the old stellar population from our FUV and U-band images. Having obtained the excess ratio we calculate the stellar temperatures necessary to explain the observations. We then end this section by discussing the implications of nebular continuum emission and dust intrinsic to the BCG. 

\subsection{Contamination by line emission}\label{s_contaml}
BCGs contain a significant amount of gas at a large range of temperatures. Emission from this gas is usually dominated by line emission and not by the underlying nebular continuum. This emission may contribute the observed FUV$_{\nu}$/U$_{\nu}$ ratio. In this section we will discuss the line emission and in Section \ref{s_nebcont} we will discuss the nebular continuum. 

Optical to millimetre spectroscopy shows prominent line emission from gas at temperatures T$\sim$10$^{1-4}$~K \citep[e.g.][]{Vo97,Ed01,Ja05}. We have taken optical spectra of A2597 and A2204 with FORS on the VLT (Oonk et al. in prep.). These spectra show that the [OII] 3727~$\rm \AA$ line is redshifted out of the U-band for both A2597 and A2204, see Fig. \ref{f_fors_spec_filt}. No other known strong emission lines are present that could contaminate the U-band images. Contamination of the continuum by line emission does affect the V, R and I-band images. The fraction of line to continuum emission, as measured from our FORS spectra, is summarised in Table \ref{t_contam}.

The FUV F150LP images of A2597 and A2204 are not affected by Ly$\alpha$ emission. However, the images could be affected by other emission lines such as CIV at 1549~$\rm \AA$. After Ly$\alpha$ this CIV line is usually the most prominent emission line in the FUV regime. This line is typically observed in regions where gas is heated to temperatures T$\sim$10$^{5}$~K. Regions near powerful AGN or heated by strong shocks are examples in which CIV at 1549~$\rm \AA$ is seen in emission. Below we elaborate on why we believe that there is no contamination by line emission in the FUV band. 

\subsubsection{CIV emission in A2597 and A2204}
CIV 1549~$\rm \AA$ line emission was recently suggested by \citet{Sp09} to explain the FUV emission of the south-eastern filaments in M87. If strong line emission is present in our FUV images this means that our method will overestimate the stellar temperatures in A2597 and A2204. 

In the case of A2597 an off-nuclear FUV spectrum is presented in \citet{Od04}. This spectrum shows no evidence for line emission in the wavelength range sampled by our F150LP images. We can calculate what the strength of the CIV 1549~$\rm \AA$ line would have to be in the \citet{Od04} spectrum if the FUV emission observed by us in this region is solely due to this line. From our F150LP image we find F$_{\lambda}\sim$10$^{-17}$~erg~s$^{-1}$~cm$^{-2}$~$\rm \AA^{-1}$~arcsec$^{-2}$. This is in good agreement with \citet{Od04}. The width of the F150LP filter is 113~$\rm \AA$ and thus the integrated F150LP flux is F$\sim$10$^{-15}$ erg~s$^{-1}$~cm$^{-2}$~arcsec$^{-2}$.

If we assume that the line has the same FWHM as the Ly$\alpha$ line measured by \citet{Od04} then the peak flux of this line would be F$_{peak}\sim$10$^{-16}$~erg~s$^{-1}$~cm$^{-2}$~$\rm \AA^{-1}$~arcsec$^{-2}$. This would be comparable to the Ly$\alpha$ line and should easily have been visible in the FUV spectrum presented by \citet{Od04}. These authors calculate the F(Ly$\alpha$)/F(CIV)$\leq$0.07 in their spectrum.

We thus conclude that the CIV 1549~$\rm \AA$ line or any other line in the wavelength range sampled by the F150LP filter can not be responsible for a significant part of the FUV flux observed by us in A2597. For A2204 there is no measured FUV spectra with sufficient quality to investigate the presence of line emission. 

In order to further assess the presence of line emission in our FUV band we have re-processed the M87 FUV images from \citet{Sp09}. We find that CIV emission is not a necessary requirement. Normal main sequence stars with stellar temperatures T$\approx$10000~K are able to explain the observations, see Appendix \ref{app_m87_film}. We thus conclude, by extension of the A2597 and M87 results, that it is unlikely that a significant part of the FUV emission observed in A2204 would be due to line emission. 

\subsection{Removing the old stellar population}\label{old_star_corr}
BCGs are complex stellar systems. The U band and to a smaller degree the FUV band emission do not just come from young stars. Older stars, such as K-giants, contribute to the observed emission. In order to determine whether a young stellar population is consistent with the observed FUV$_{\nu}$/U$_{\nu}$ ratio we need to first remove the older population. 

We do this by identifying regions in our images that are dominated by old stars. These regions are indicated by the blue squares in Fig. \ref{f_uvr} or equivalently by the black squares in Figs. \ref{f_a2597_vrat} and \ref{f_a2204_vrat}. The area in which we can find reliable old-star regions is limited by the field of view of the FUV observations. 

In the case of A2204 there is a companion elliptical galaxy to the south-west of the BCG. This elliptical is part of the galaxy cluster and shows low-level diffuse FUV emission expected from an old stellar population. We will refer to this old elliptical as A2204-SW. The other elliptical companion north of the A2204 BCG contains a bright FUV point-source associated with its nucleus. This is not consistent with an old stellar population and hence we do not use it.

In the case of A2597 there is no elliptical companion galaxy within the field of view of the FUV observations. Instead we use two regions within the outer regions of the BCG itself to identify old stellar regions. We will refer to the average of these two regions as A2597-OFF.

Having identified the old stellar regions in our observations, we can determine their FUV/V and U/V ratios and use the V image to remove the emission coming from the underlying old stellar population in our FUV and U images. We use the V band image, instead of the R or I band image, because the line contamination in this band is small and well determined from our spectroscopic observations. 

However, first we perform a cross-check of the old stellar regions defined by us. We do this by comparing the U/V ratios determined here with the U/V ratio for a number of nearby elliptical galaxies using the larger field of view of the optical images. The nearby ellipticals selected are indicated by the red squares in Fig. \ref{f_uvr}. We will refer to this sample of nearby ellipticals as A2597-COMP and A2204-COMP and we find that their average U/V ratio agrees well with that of A2597-OFF and A2204-SW, see Tables \ref{t_fuv_ratio} and \ref{t_u_ratio}.
 
\begin{figure*}
    \includegraphics[width=0.45\textwidth, angle=90]{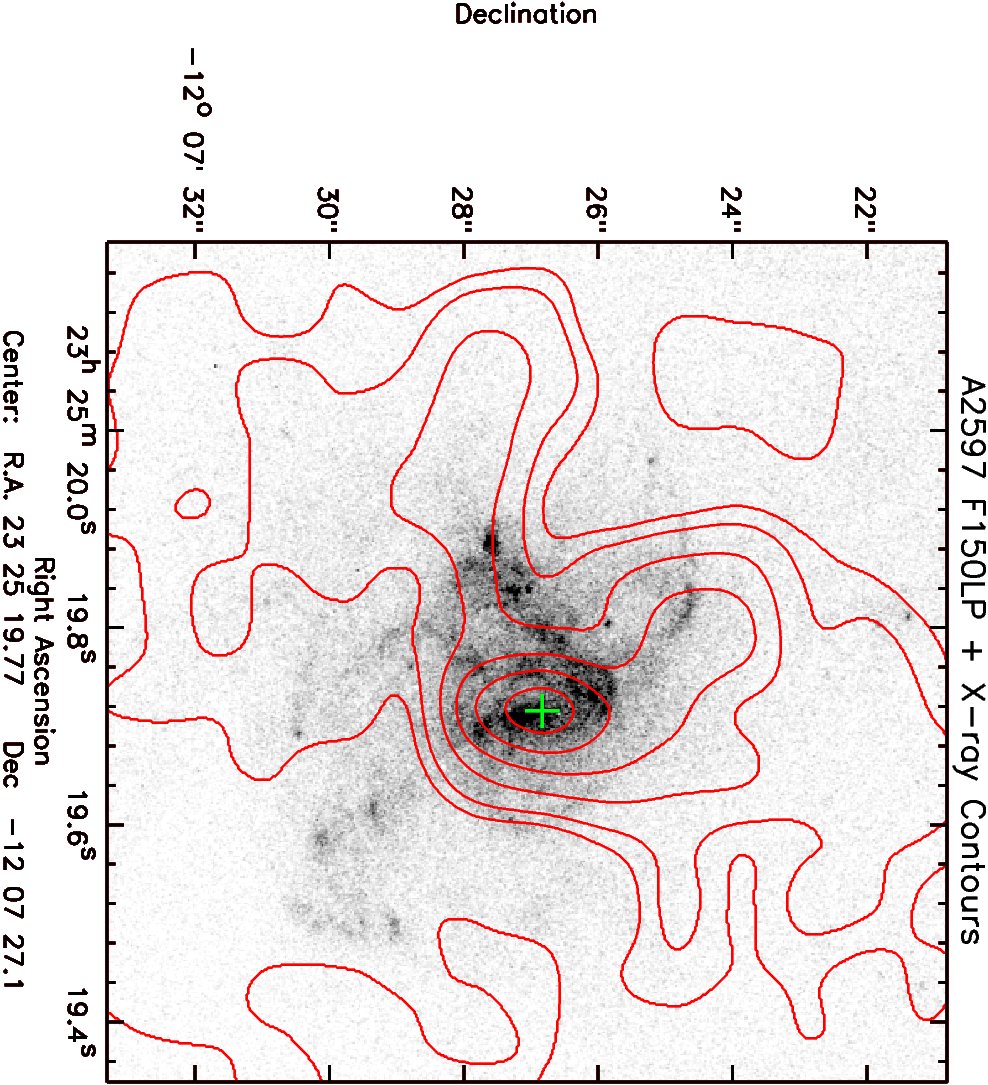}
    \includegraphics[width=0.45\textwidth, angle=90]{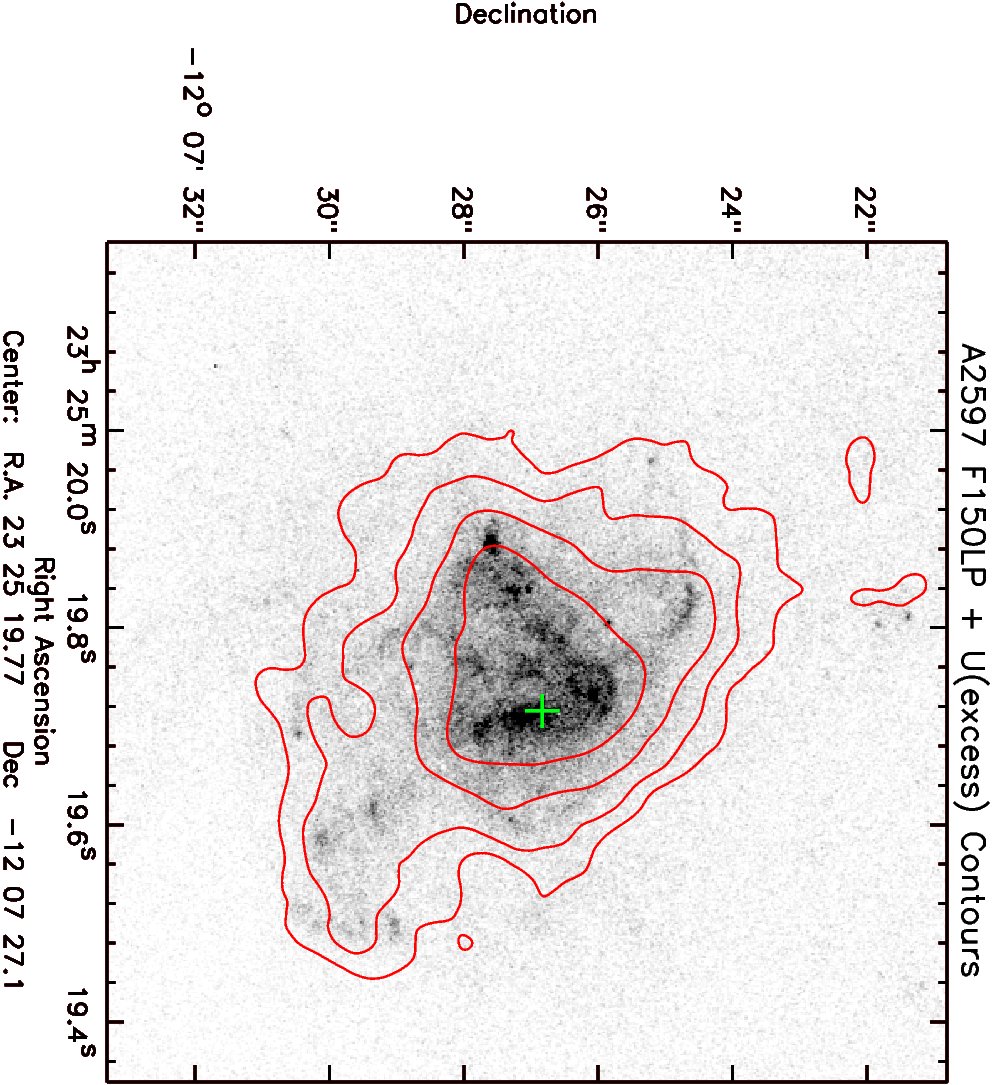}
    \includegraphics[width=0.45\textwidth, angle=90]{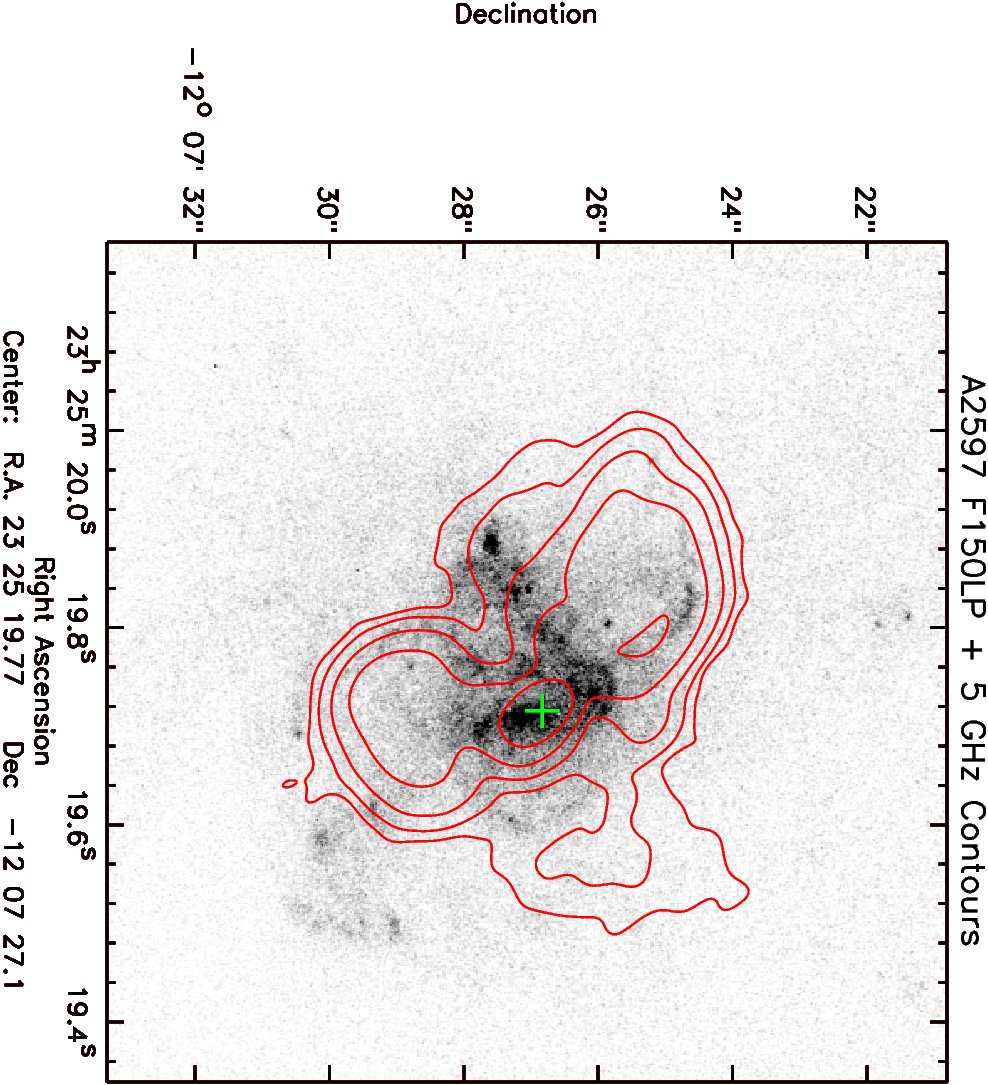}
    \includegraphics[width=0.45\textwidth, angle=90]{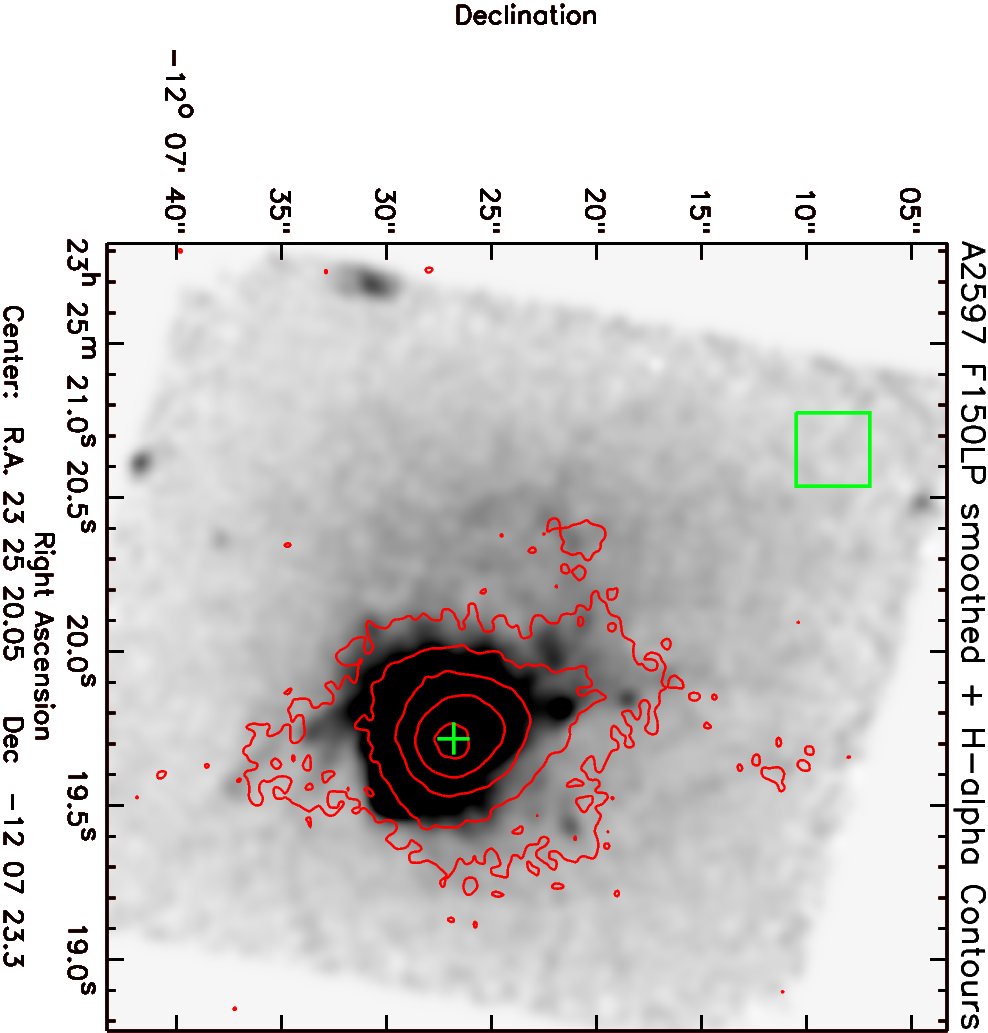}
  \vspace{0.5cm}
  \caption{ABELL 2597 FUV F150LP overlays. (\textit{Top-Left}) CHANDRA X-ray contours on top of the FUV emission. (\textit{Top-Right}) FORS U~band excess contours on top of FUV emission. (\textit{Bottom-Left}) VLA 5~GHz contours at [1,4,16,64,256]$\times$10$^{-4}$~Jy/beam on top of the FUV emission. (\textit{Bottom-Right}) H$\alpha$ contours from \citet{Ja05} on top of the convolved FUV emission. The BCG nucleus is indicated by the green cross. The region used for background subtraction is indicated by the green rectangle.}\label{f_a2597_ovl}
\end{figure*}

\begin{figure*}
    \includegraphics[width=0.60\textwidth, angle=90]{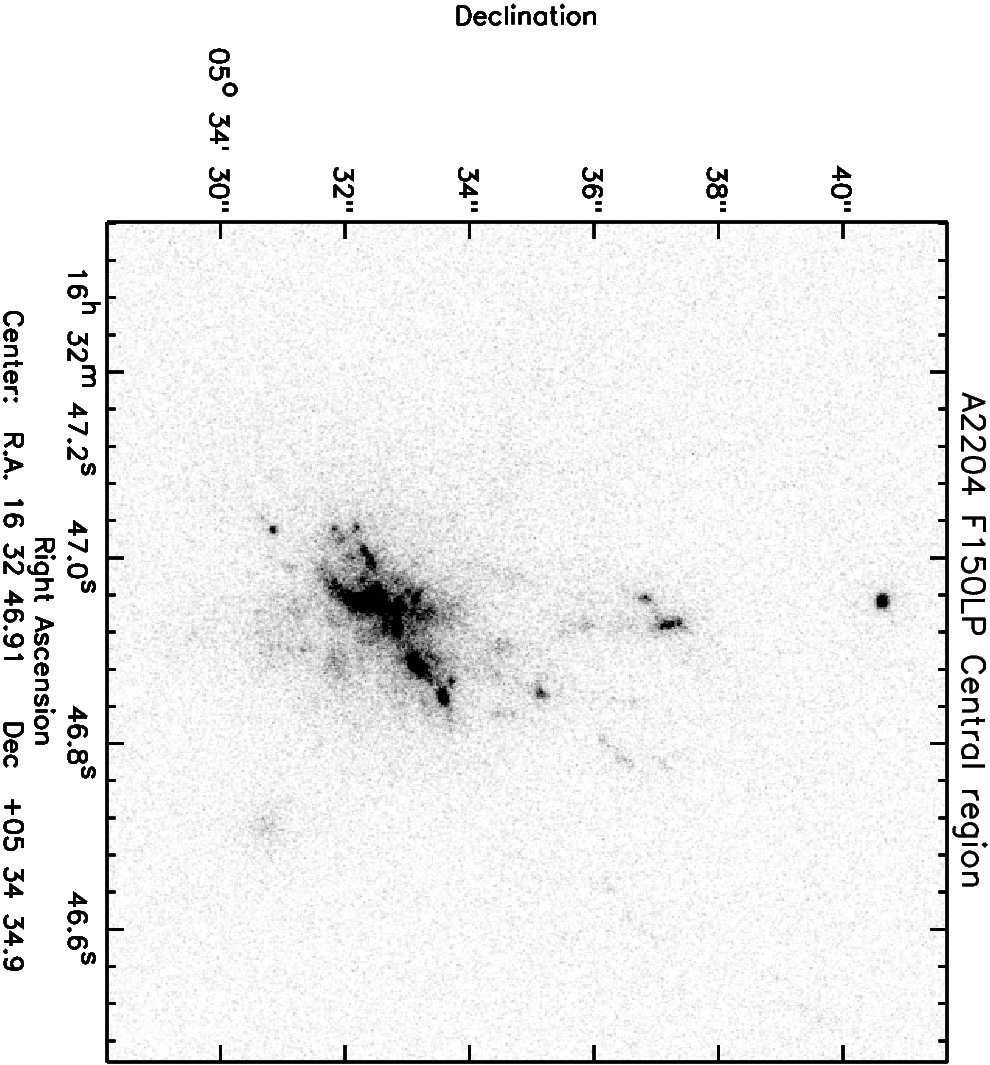}
    \includegraphics[width=0.60\textwidth, angle=90]{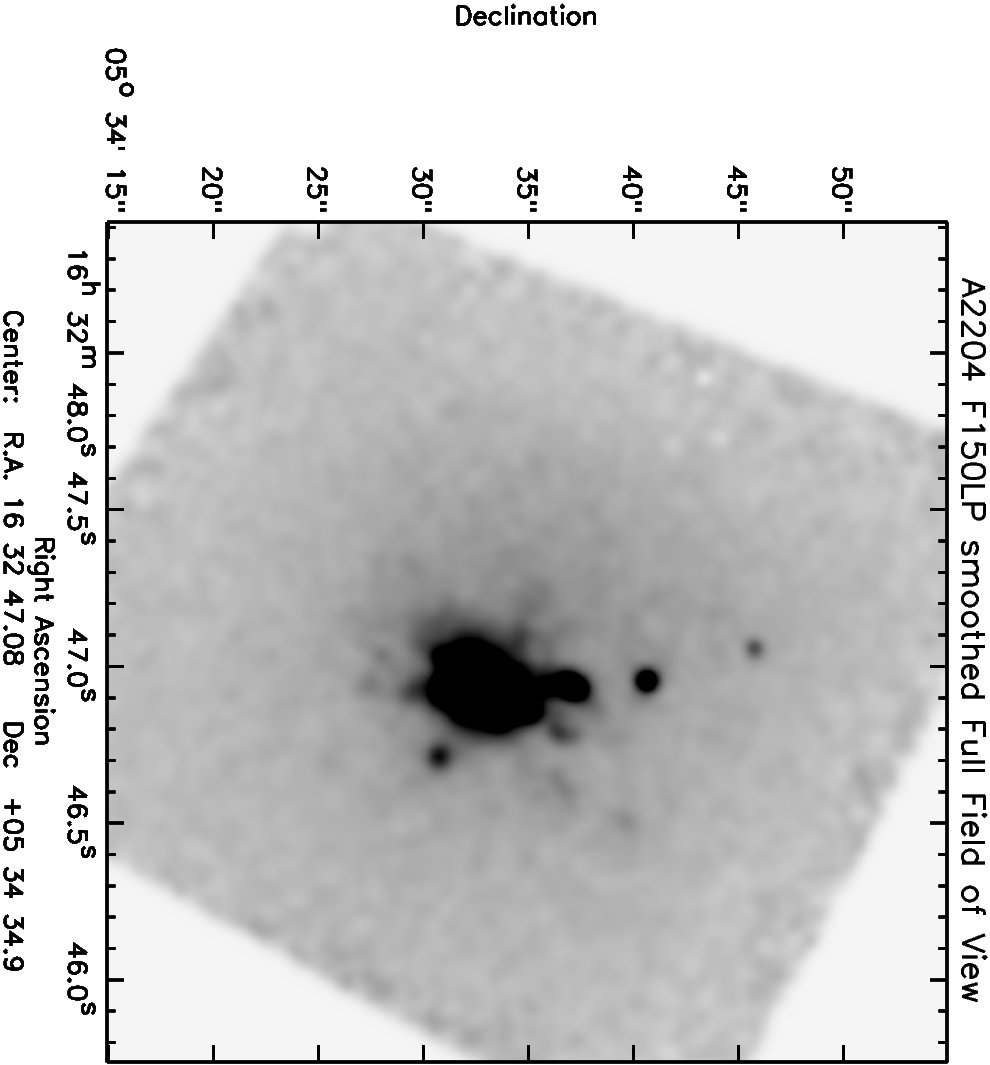}
  \vspace{0.5cm}
  \caption{ABELL 2204 FUV F150LP emission. (\textit{Top}) Central bright FUV continuum emission at the intrinsic resolution of HST. (\textit{Bottom}) FUV continuum  emission convolved to an output spatial resolution of 1 arcsec FWHM and scaled to enhance the low surface brightness structures. The BCG nucleus is at ($\alpha;\delta$)=(16 32 46.938 ; +05 34 32.81) (J2000).}\label{f_a2204}
\end{figure*}

\begin{figure*}
    \includegraphics[width=0.45\textwidth, angle=90]{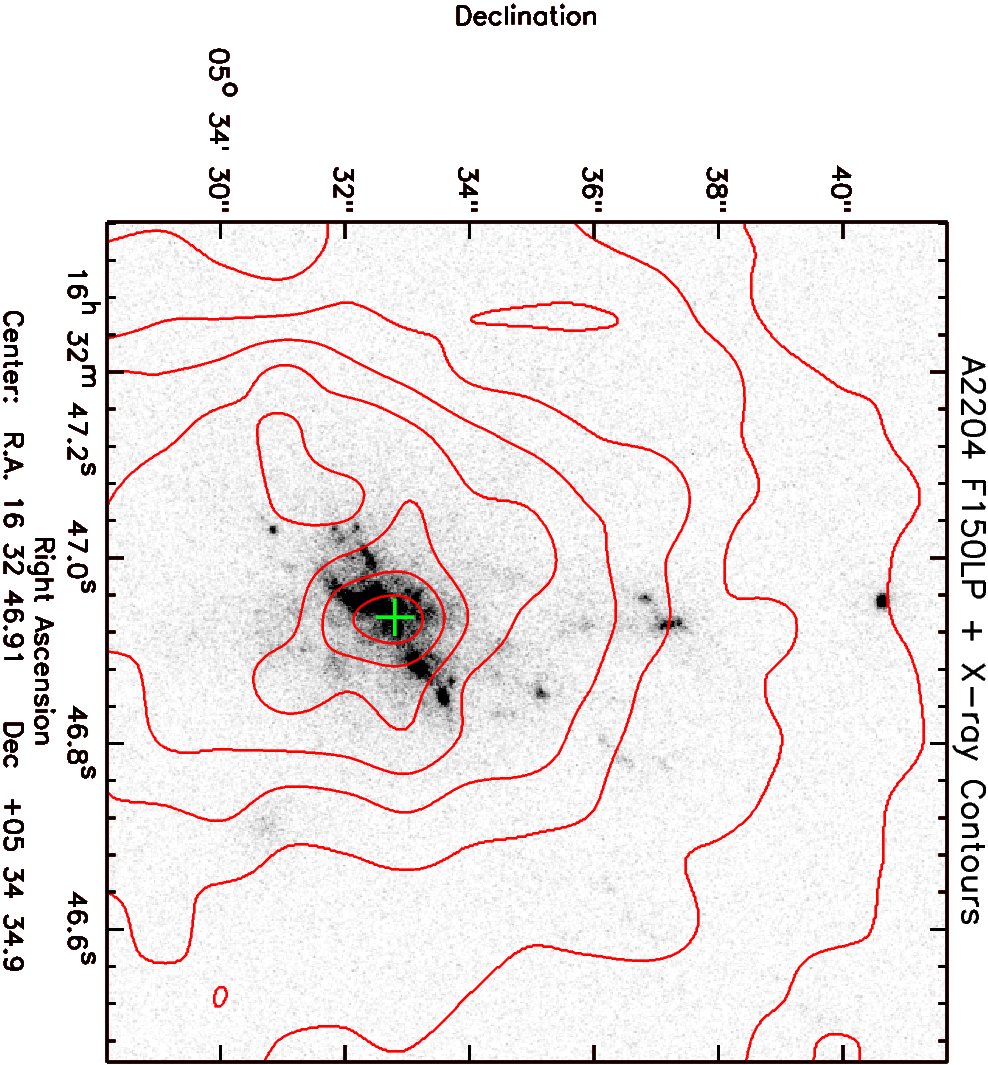}
    \includegraphics[width=0.45\textwidth, angle=90]{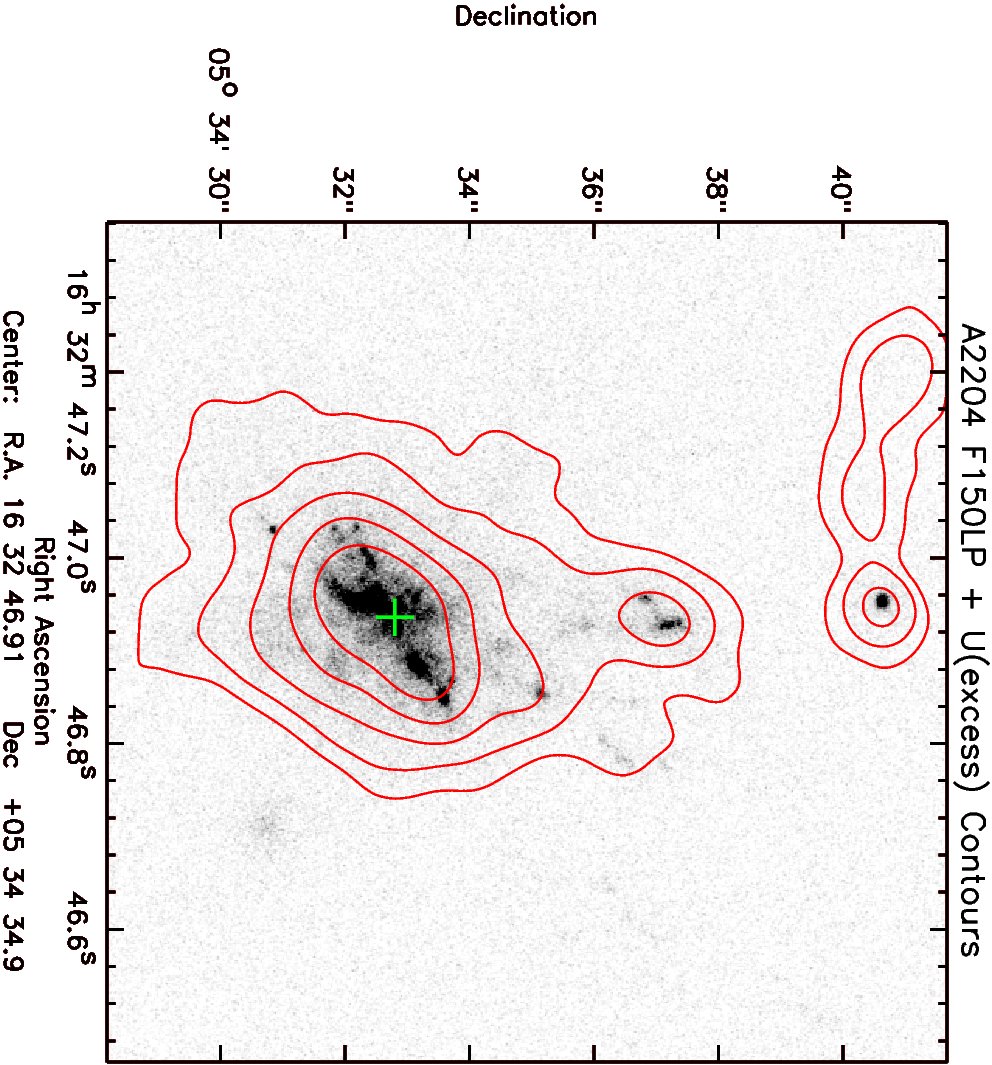}
    \includegraphics[width=0.45\textwidth, angle=90]{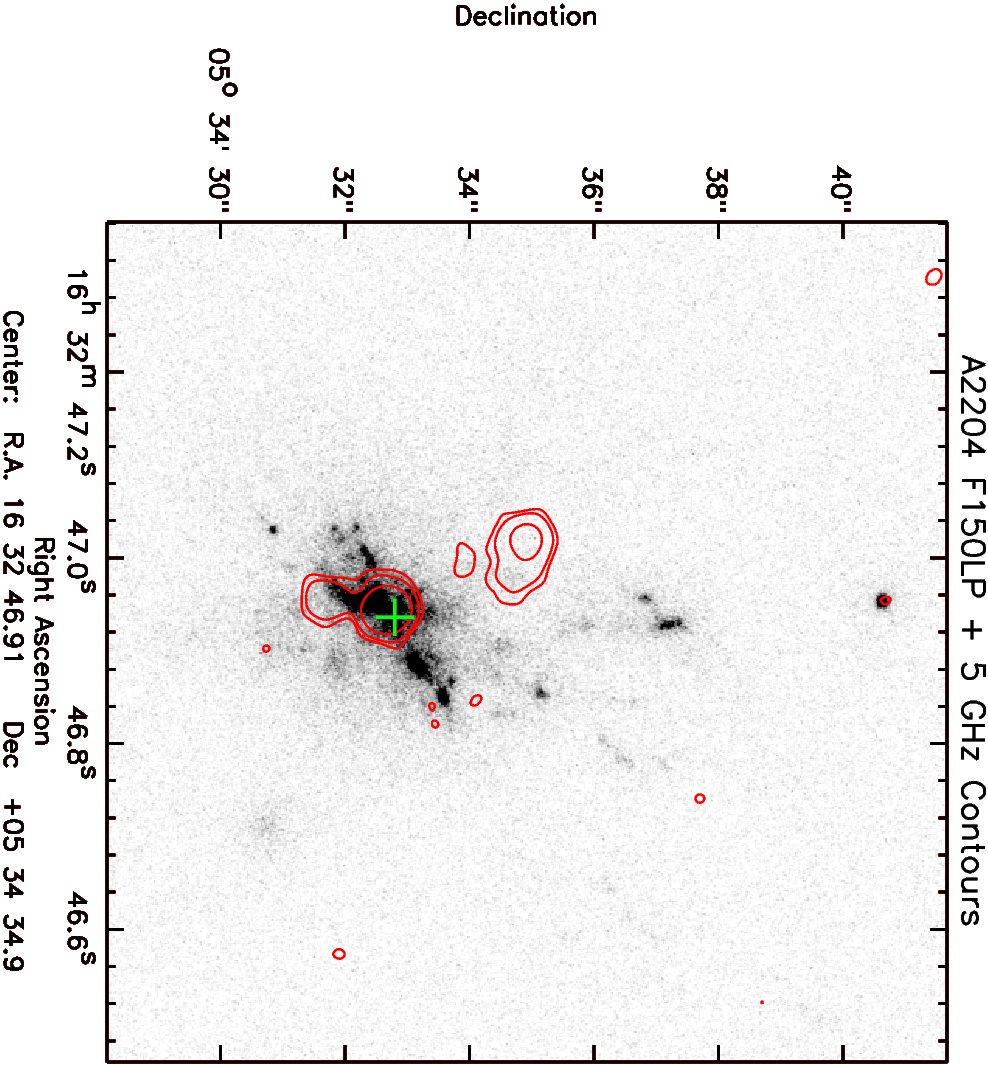}
    \includegraphics[width=0.45\textwidth, angle=90]{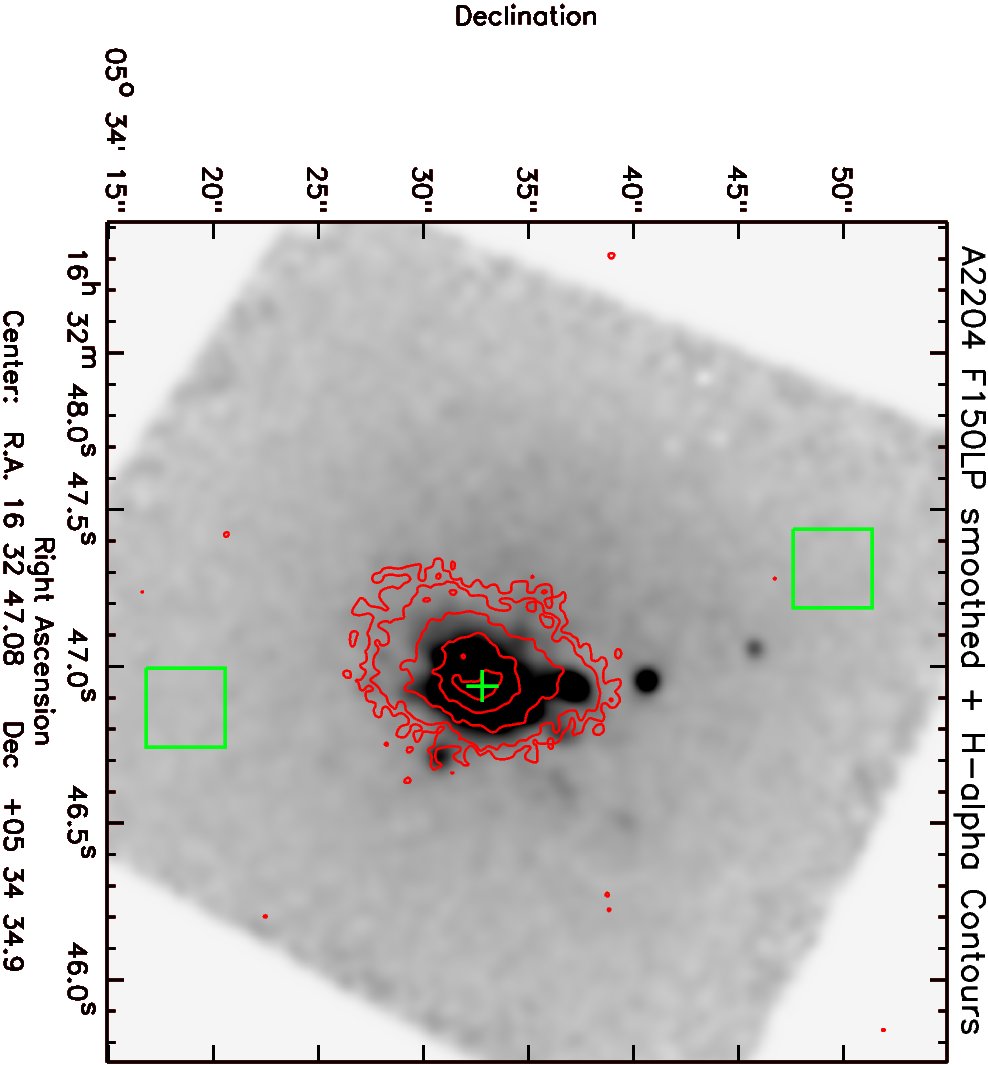}
  \vspace{0.5cm}
  \caption{ABELL 2204 FUV F150LP overlays. (\textit{Top-Left}) CHANDRA X-ray contours on top of the FUV emission. (\textit{Top-Right}) FORS U~band excess contours on top of FUV emission. (\textit{Bottom-Left}) VLA 5~GHz contours at [2,4,16,256]$\times$10$^{-4}$~Jy/beam on top of the FUV emission. (\textit{Bottom-Right}) H$\alpha$ contours from \citet{Ja05} on top of the convolved FUV emission. The BCG nucleus is indicated by the green cross. The regions used for background subtraction are indicated by the green rectangles. The lensed galaxy north-east of the BCG \citep[e.g.][]{Wi06} is visible in the U-band but not in the FUV band.}\label{f_a2204_ovl}
\end{figure*}

\begin{figure*}
    \includegraphics[width=0.60\textwidth, angle=90]{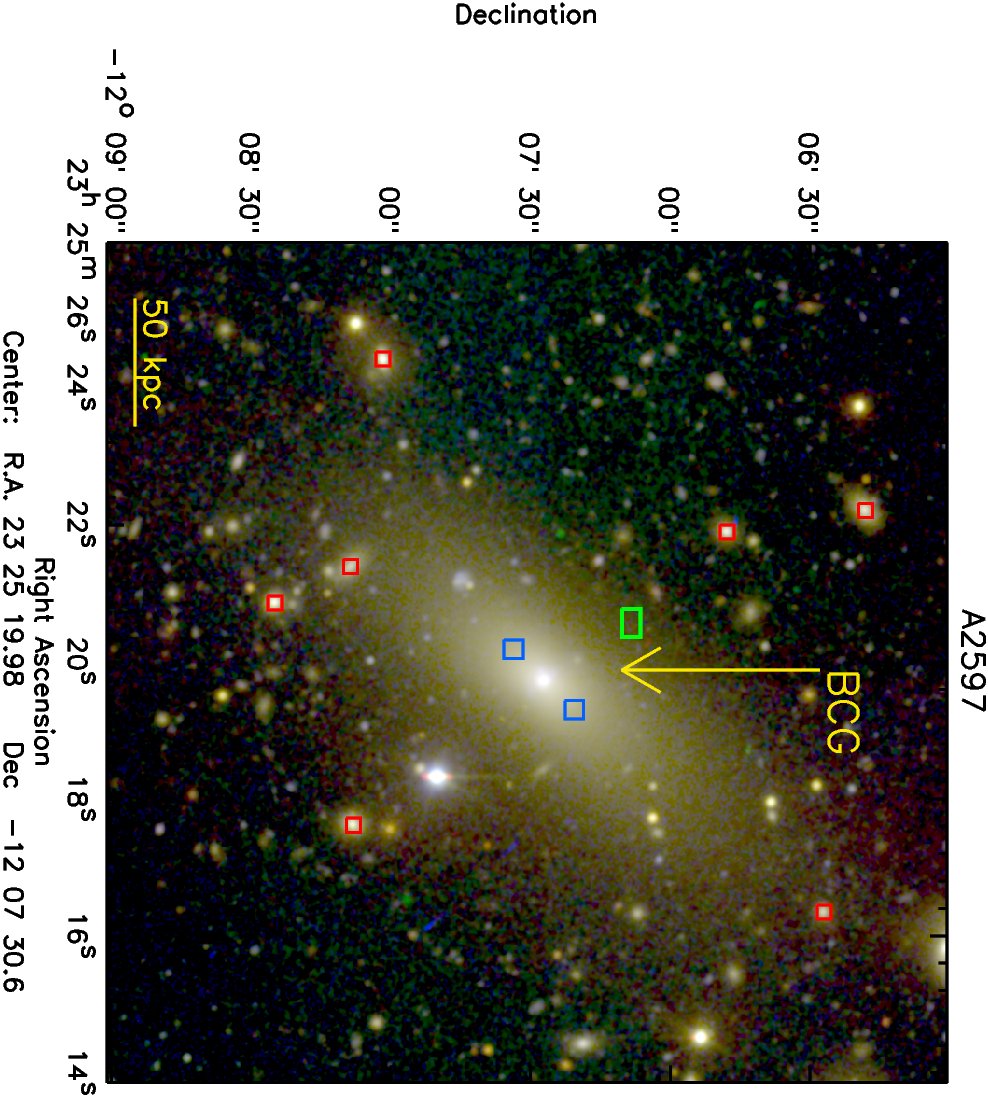}
    \includegraphics[width=0.60\textwidth, angle=90]{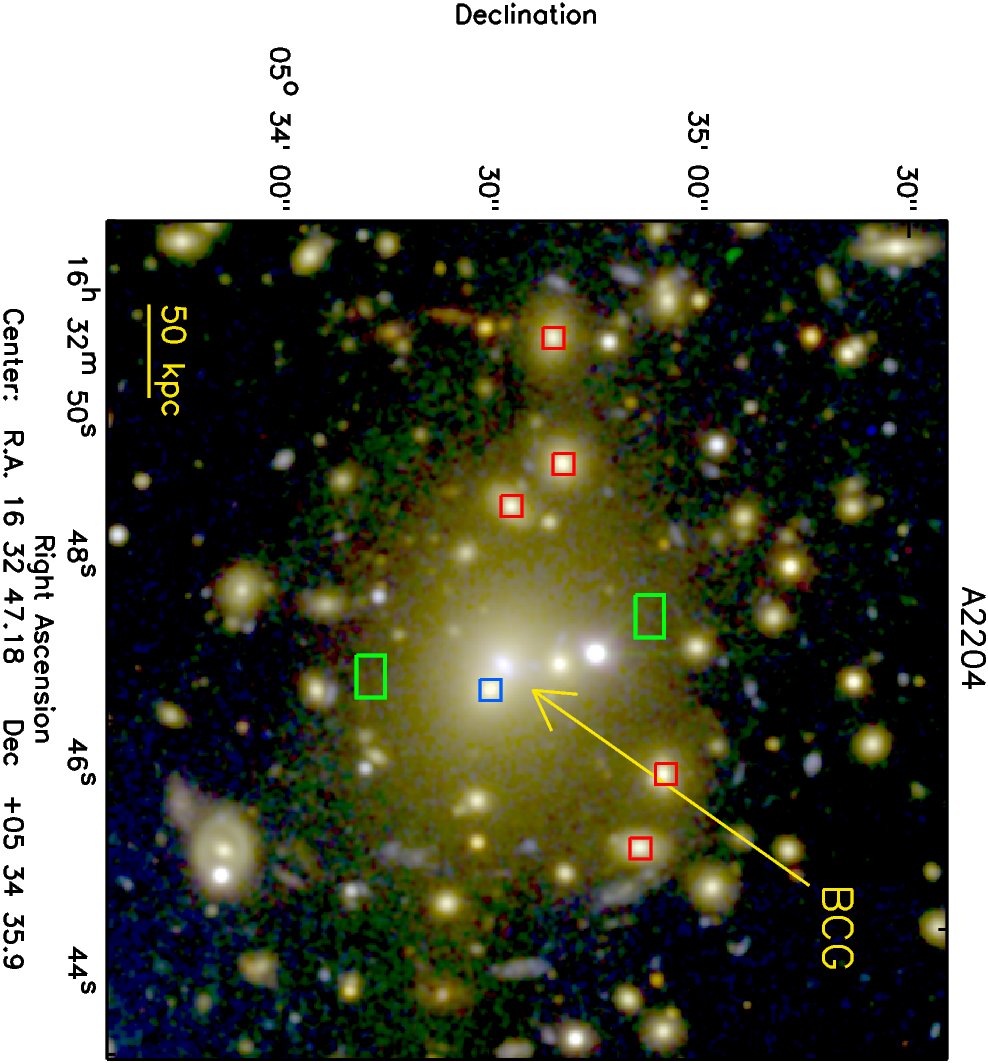}
  \vspace{0.5cm}
  \caption{FORS UVR 3-color images. ABELL 2597 (\textit{Top}) and ABELL 2204 (\textit{Bottom}). The blue squares indicate the old star regions where we estimated the FUV/V and U/V ratios. The red squares indicate the nearby cluster ellipticals that were used as comparison sample for the U/V ratios determined in the blue squares. The green squares show regions used for determining the background in the FUV and optical images. The many distorted, blue galaxies in the A2204 image show that this cluster is a strongly lensing system.}\label{f_uvr}
\end{figure*}

\begin{figure*}
    \includegraphics[width=0.60\textwidth, angle=90]{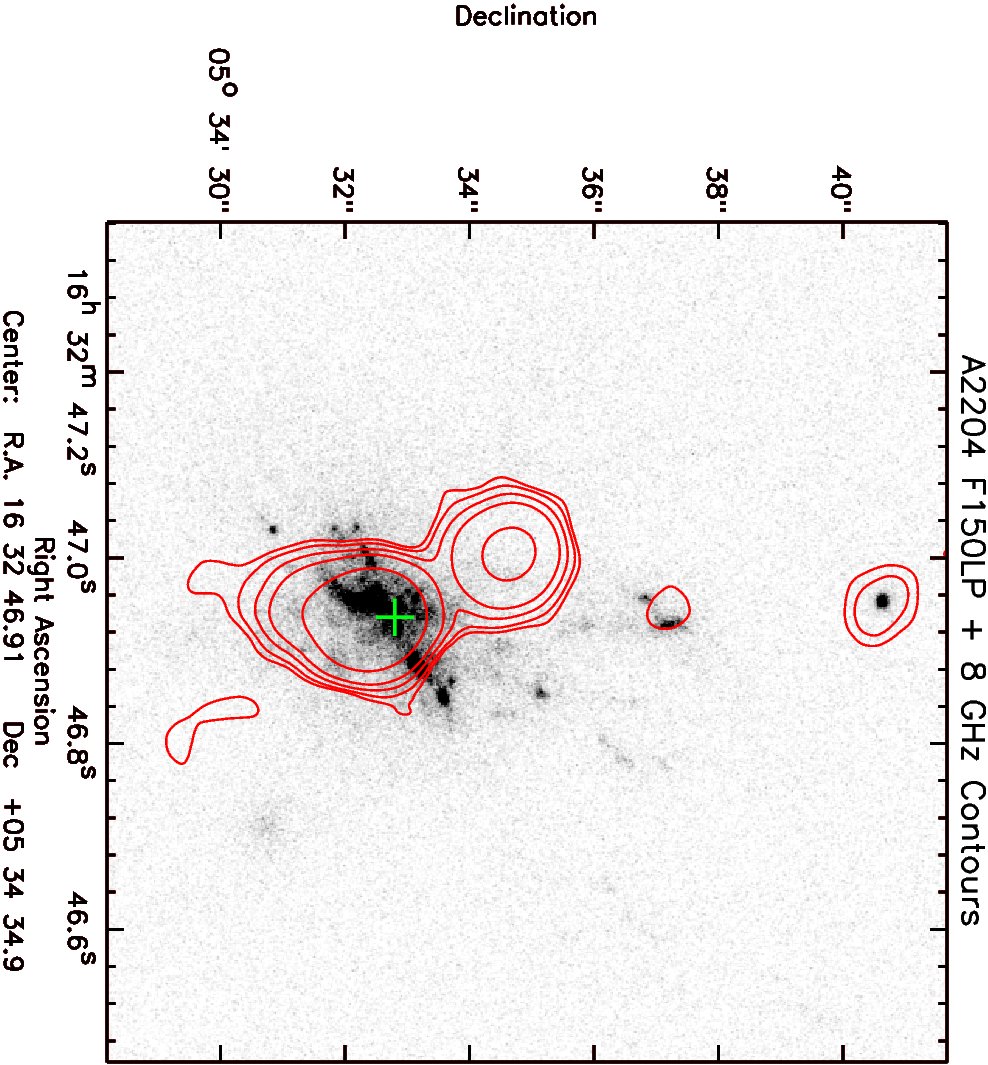}
  \vspace{0.5cm}
  \caption{A2204 VLA 8~GHz radio continuum emission. Radio contours are given at [0.6,1,2,4,16,256]$\times$10$^{-4}$~Jy/beam on top of FUV emission. The one sigma map noise is 0.2$\times$10$^{-4}$~Jy/beam. This dataset is significantly deeper than the 5~GHz data presented in Fig. \ref{f_a2204_ovl}.}\label{f_a2204_8ghz}
\end{figure*}

Next we need to consider line contamination of the FORS V band image. The V band line contamination is due to H$\beta$ and OIII emission. The deep narrow-band H$\alpha$ images by \citet{Ja05} show that the old stellar regions selected by us do not contain significant line emission, see also Fig. \ref{f_a2597_ovl}. Similarly from off-nuclear FORS spectroscopy (not shown here) we know that the line contamination of the V band decreases sharply along the major-axis of the BCG, see also Fig. \ref{f_pslit}. We thus conclude that that the U/V and the FUV/V ratios determined in the old-star regions are not contaminated by line emission.

We then assume that the line contamination within the FUV, U-band bright filamentary regions is independent of location and well described by the average value quoted in Table \ref{t_contam}. This assumption is supported by our H$\alpha$ images and FORS spectra. These show that the line contamination in the central 20~kpc region along the minor axis of the galaxy does not vary significantly on the scales relevant to this investigation, i.e. $\sim$1~arcsec, see also Fig. \ref{f_a2597_ovl}. 

We thus subtract the emission fraction due to line contamination from our V band images and use these line subtracted V-band images to compute the FUV and U excess images, see Figs. \ref{f_a2597_vrat} and \ref{f_a2204_vrat}. We note that the line contamination correction of the V-band has a small but systematic effect on the FUV$_{\nu,exc}$/U$_{\nu,exc}$ excess emission ratio in that it lowers this ratio about 5 per-cent. This is easily explained by noting that the old stellar population contributes more to the U than to the FUV emission. From here on we will denote the total light images as FUV$_{\nu,tot}$ and U$_{\nu,tot}$ and the excess light images as FUV$_{\nu,exc}$ and U$_{\nu,exc}$.

In Table \ref{t_laper} we give the FUV and U band fluxes integrated over large 18~kpc radial apertures for (i) the total light, (ii) the excess light, (iii) the nebular continuum and (iv) the H$\alpha$ flux in A2597 and A2204. We see that on large scales only about 20 percent of the total observed FUV flux can be accounted for by the old stellar population in both targets. Similarly we find that about 80 percent of the U flux in A2597 and about 60 percent of the U flux in A2204 can be accounted for by the old stellar population.

\subsection{The FUV$_{\nu,exc}$/U$_{\nu,exc}$ excess ratio}
Having subtracted the emission contributed by the old stellar population from the FUV$_{tot}$ and U$_{tot}$ images we obtained the two excess images FUV$_{exc}$ and U$_{exc}$, see Figures \ref{f_a2597_yngfu} and \ref{f_a2204_yngfu}. These images show that excess FUV and U band light trace each other very well in the central 20~kpc regions of A2597 and A2204.

We will now investigate whether the observed FUV$_{\nu,exc}$/U$_{\nu,exc}$ ratio is consistent with the emission ratio expected for young stars by comparing it to K93 stellar models. A2597 and A2204 are at redshifts z=0.0821 and z=0.1517 respectively. We redshift the K93 stellar models by these amounts and compute the FUV$_{\nu}$/U$_{\nu}$ ratio by convolving the model spectra with the F150LP and U Bessel filters. For comparison reasons we also compute the FUV$_{\nu}$/U$_{\nu}$ ratio for a set of BB models. The results for both models are given in Fig. \ref{f_bcg_bb_k93}. 

To improve the signal to noise we have re-binned the FUV$_{exc}$ and U$_{exc}$ images for A2597 and A2204 by a factor 3 so that one pixel now corresponds to 0.6$\times$0.6~arcsec$^{2}$. We find in the central region region of A2597 that the observed FUV$_{\nu,exc}$/U$_{\nu,exc}$ ratio is between 1.7 and 2.6. Comparing with the K93 stellar models and invoking extinction by MW foreground dust only, we find that a FUV$_{\nu,exc}$/U$_{\nu,exc}$ ratio between 1.7 and 2.6 corresponds to main-sequence stars with temperatures between 28000 and 50000~K. 

We have plotted the observed FUV$_{\nu,exc}$/U$_{\nu,exc}$ ratio, its corresponding K93 stellar temperature, and their uncertainties in Fig. \ref{f_a2597_yngfu}. The uncertainty in the ratio takes into account the absolute flux uncertainty and the poissonian noise for all involved images. The uncertainty in the ratio and in the temperature are non-linear. For the maps presented here we have chosen to represent the uncertainty by the average of the one sigma upper and lower deviation. The K93 models allow for stellar temperatures in range 0-50000~K. If either the lower or the upper deviation falls outside of this range, then the uncertainty given is set to the deviation that does fall within the allowed range.

There are a number of pixels that have a ratio greater than 2.6 and thus a best-fit temperature above the value allowed for normal main-sequence O-stars. These pixels have a low signal-to-noise mainly because of their low U band flux. The corresponding large error on these pixels allows us to reconcile them with a stellar origin. There is not much sub-structure in the temperature map of A2597. One can argue that there is a small decrease in temperature beyond the central 10$\times$10~kpc$^{2}$ but the low statistics and higher noise here do not allow for strong statements. 

In A2204 we find that the observed FUV$_{\nu,exc}$/U$_{\nu,exc}$ ratio is between 1 and 1.8. Comparing with the K93 stellar models, taking into account extinction by MW foreground dust only, this corresponds to main-sequence stars with temperatures between 20000 and 40000~K. The results are plotted in Fig. \ref{f_a2204_yngfu}. The temperatures derived for the FUV and U excess emission in A2204 are bit lower than for A2597. The A2204 temperature map does show some substructure in that the knots west and north of the nucleus have the highest temperatures being at 30000 to 40000~K. The remaining parts of the central region in A2204 have a temperature corresponding to 20000 to 30000~K.

Interpreting the observed FUV$_{\nu,exc}$/U$_{\nu,exc}$ ratio terms of BB models requires temperatures above 35000~K in A2597 and above 30000~K in A2204, if extinction by MW foreground dust only is taken into account.

\subsection{Nebular continuum emission}\label{s_nebcont} %%%%%
Figs. \ref{f_a2597_ovl} and \ref{f_a2204_ovl} show that the ionised gas, as traced by the H$\alpha$ recombination line, is co-spatial with the FUV emission on kpc-scales. This ionised gas has a temperature T$\sim$10$^{4}$~K and will emit continuum emission in the FUV-optical wavelength range. The amount of nebular continuum emission depends not only on the temperature and density of this gas but also on whether the gas is ionisation or density bounded. 

The maximum amount of nebular continuum is produced when the gas is ionisation bounded. We have estimated the contribution of nebular continuum to the FUV flux in this case based on the H$\alpha$ measurements by \citet{Ja05} using the \textit{NEBCONT} program \citep{Ho04}. As our input we use an electron temperature T$_{e}$=10$^{4}$~K, electron density n$_{e}$=10$^{2}$~cm$^{-3}$, 10\% He abundance and the total extinction estimated from the Balmer decrements measured in our FORS spectra. The results for 18~kpc radial apertures are presented in Table \ref{t_laper}. 

We find that 5\% and 11\% of the total FUV band flux in A2597 and A2204 respectively is due to nebular continuum emission. This percentage increases to 6\% and 13\% if we consider only the excess FUV emission. The contribution of nebular emission to the total U band flux is 6\% and 15\% for A2597 and A2204 respectively. This increases to 38\% and 39\% if we consider only the excess U emission.

Removing the nebular continuum emission increases the FUV$_{\nu,exc}$/U$_{\nu,exc}$ ratio by a factor 1.5 and 1.4 for A2597 and A2204 respectively. Increasing the observed FUV$_{\nu,exc}$/U$_{\nu,exc}$ ratio by a factor 1.5 means that most of the central 20~kpc emission in A2597 can no longer be explained by stars. In the case of A2204 removing the nebular continuum means that the FUV bright clumps north and north-west of the nucleus can no be explained by stars, but most of the remaining emission in the central 20~kpc can still be reconciled with a stellar origin.

Above we have calculated the maximum contribution coming from the ionised gas in the form of nebular continuum emission. We note that currently it is not clear if the gas is ionisation or density bounded. In the latter case the continuum emission contributed to the FUV and U-bands by the ionised gas could be lower. \citet{Od94} argue that the H$\alpha$ emission arises from the ionised skins on HI clouds and that as such the gas nebula in A2597 is ionisation bounded. This picture is consistent with low spatial resolution measurements of H$\alpha$, HI and OI~6300~\AA. Recent high spatial resolution HST measurements of NGC~1275, the nearby BCG in the Perseus galaxy cluster, show that the H$\alpha$ emission line filaments are very thin, i.e. they have pc-scale widths \citep{Fa08}. Such very thin filaments may allow for a density bounded state of the gas in NGC~1275 and by extension in A2597 and A2204. Further observations are necessary to investigate the condition of this gas.

\subsection{Dust intrinsic to the BCG}\label{s_dust}
Our FORS optical images of A2597 and A2204 do not show strong evidence for dust in these systems. However, archival optical images from HST do show that these BCGs are morphologically disturbed at their centres, see Fig. \ref{f_hst_add}. This patchy morphology is indicative of obscuration by dust in these systems. 

This is confirmed by measurements of the Balmer decrements in our FORS spectra, see Table \ref{t_dust}. These decrements deviate strongly from case~B recombination and can be be described by dust extinguishing light according to an average Milky Way (MW) type extinction law, such as described by \citet*{Ca89} (hereafter C89).

Results by \citet*{Vo97} confirm that in the optical regime the dust in A2597 can be described by an average MW type extinction law with A$_{V}\sim$1. For A2204 our Balmer decrements imply a higher extinction than would be obtained from the measurements presented in \citet{Cr99}. Contrary to \citet{Cr99} we measure the Balmer ratios within the same slit which is why we prefer our measurement for A2204. Interestingly we note that our spectra do not show strong changes in the Balmer decrements over the central $\sim$15 kpc regions in A2597 and A2204. A disturbed morphology and Balmer decrements that deviate strongly from case~B recombination are typical for cool-core BCGs as a class of objects \citep[e.g.][]{Cr99,Od10}.

The C89 extinction law aims to reproduce the global extinction properties for the diffuse interstellar medium in our galaxy. It is parametrized from 0.125 to 3.5 $\mu$m covering the range in wavelengths important to us. Although there is no reason to a priori assume that the dust properties in the BCGs is the same as in the MW, there is no evidence from either optical spectroscopy or the far-infrared spectral energy distributions \citep[][Oonk et al. in prep.]{Vo97,Ed10} that the dust in BCGs is different. 

Below we will investigate how extinction by dust affects the FUV and U band fluxes. In order to do so we will always use the C89 extinction law to describe the total extinction as being due to two components; (i) foreground MW dust and (ii) dust intrinsic to the BCG. The extinction due to the MW towards A2597 and A2204 is A$_{V,MW}$=0.1 and A$_{V,MW}$=0.3 respectively \citep{Sc98}. In order to explain the observed Balmer decrements we require the dust component intrinsic to the BCG to be A$_{V,BCG}$=1.3 for A2597 and A$_{V,BCG}$=1.6 for A2204, see Table \ref{t_dust}. These values have been calculated by invoking the C89 extinction law and by taking redshift effects into account.

Up to this point we have considered extinction due to MW foreground dust only. In Fig. \ref{f_bcg_bb_k93} we plot the FUV$_{\nu}$/U$_{\nu}$ ratio for varying amounts of extinction. Dust has a dramatic effect upon the FUV$_{\nu}$/U$_{\nu}$ ratio. The highest ratio values observed in A2597 and A2204 can be reconciled with the K93 stellar models in the case where the only dust in front of these regions is the MW foreground. The average ratio value observed in the central 10~kpc of A2597 is about 2.1 and about 1.1 in A2204. If half of the dust intrinsic to the BCG or more is front of the emitting regions then normal main sequence stars as described by the K93 models can no longer explain the observed emission ratio.

This leads to the following possible explanations; (i)~There is little to no dust in front of the high FUV$_{\nu,exc}$/U$_{\nu,exc}$ areas and their emission is due to hot main-sequence stars. (ii)~The emission is due to exotic stars, such as Wolf-Rayet stars or White Dwarfs, with temperatures above 50000~K. (iii)~A non-stellar source contributes to the observed wavelength regime or alternatively this source affects the Balmer decrement measurements. (iv)~A MW type extinction law, although consistent with the optical Balmer decrements, is not applicable to the FUV wavelength range studied here. 

Explanations (ii-iv) will be discussed in Sections \ref{s_non_stellar} and \ref{s_discuss}. Here we will shortly discuss explanation (i). A clumpy, dusty medium, such as observed in A2597 and A2204, may give rise to a selection effect in that we preferentially observe FUV emission coming from regions with low extinction along the line of sight. There are two ways in which this could arise, (a) the sightlines along these regions have intrinsically very low extinction, or (b) the Balmer decrements trace the dust over a longer column than the observed FUV emission i.e. we only observe the FUV emission coming from the front part of the line of sight.

\begin{figure*}
\hspace{1.0cm}
    \includegraphics[width=0.35\textwidth]{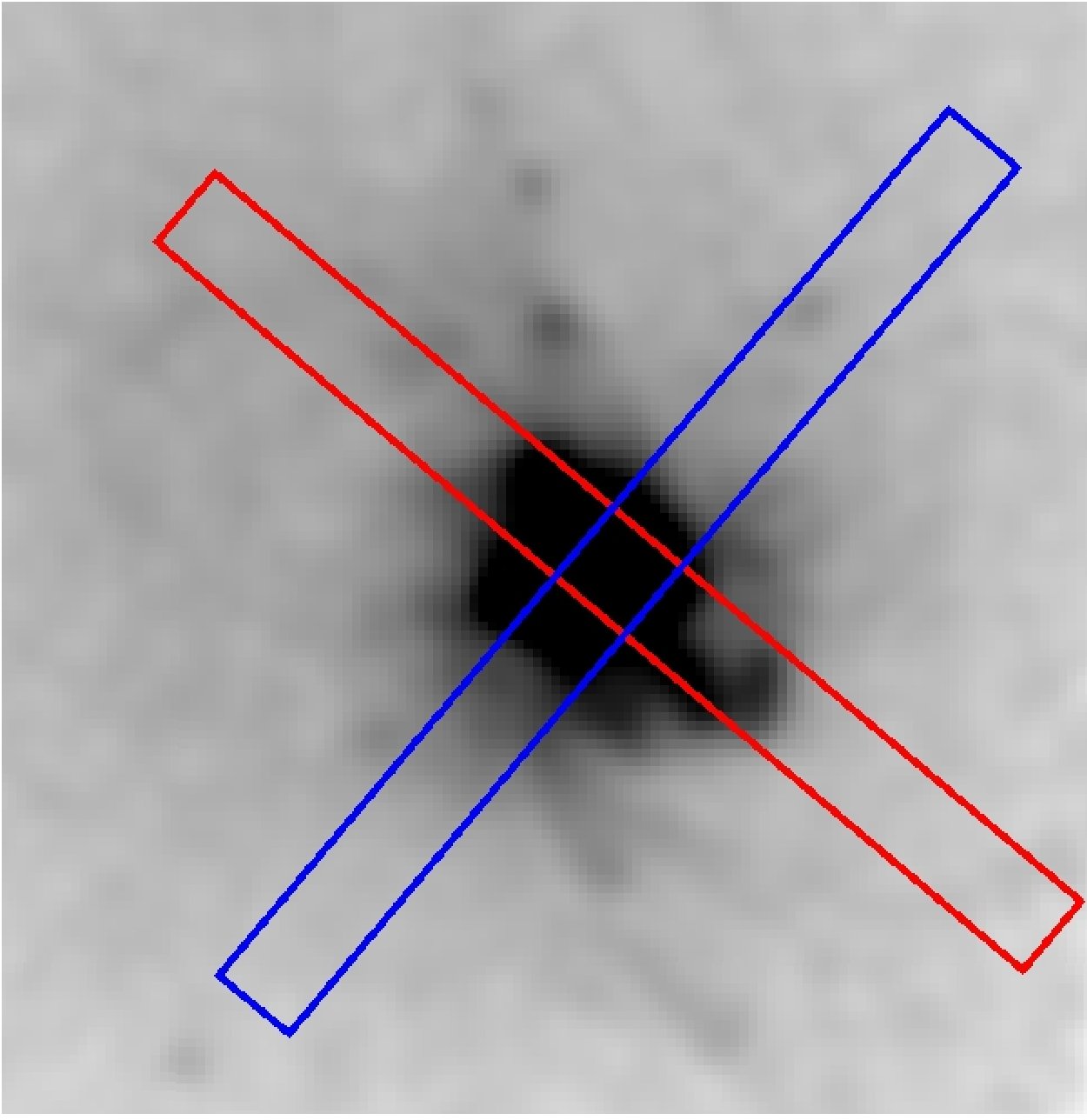}
\vspace{0.5cm}
\hspace{2.0cm}
    \includegraphics[width=0.35\textwidth]{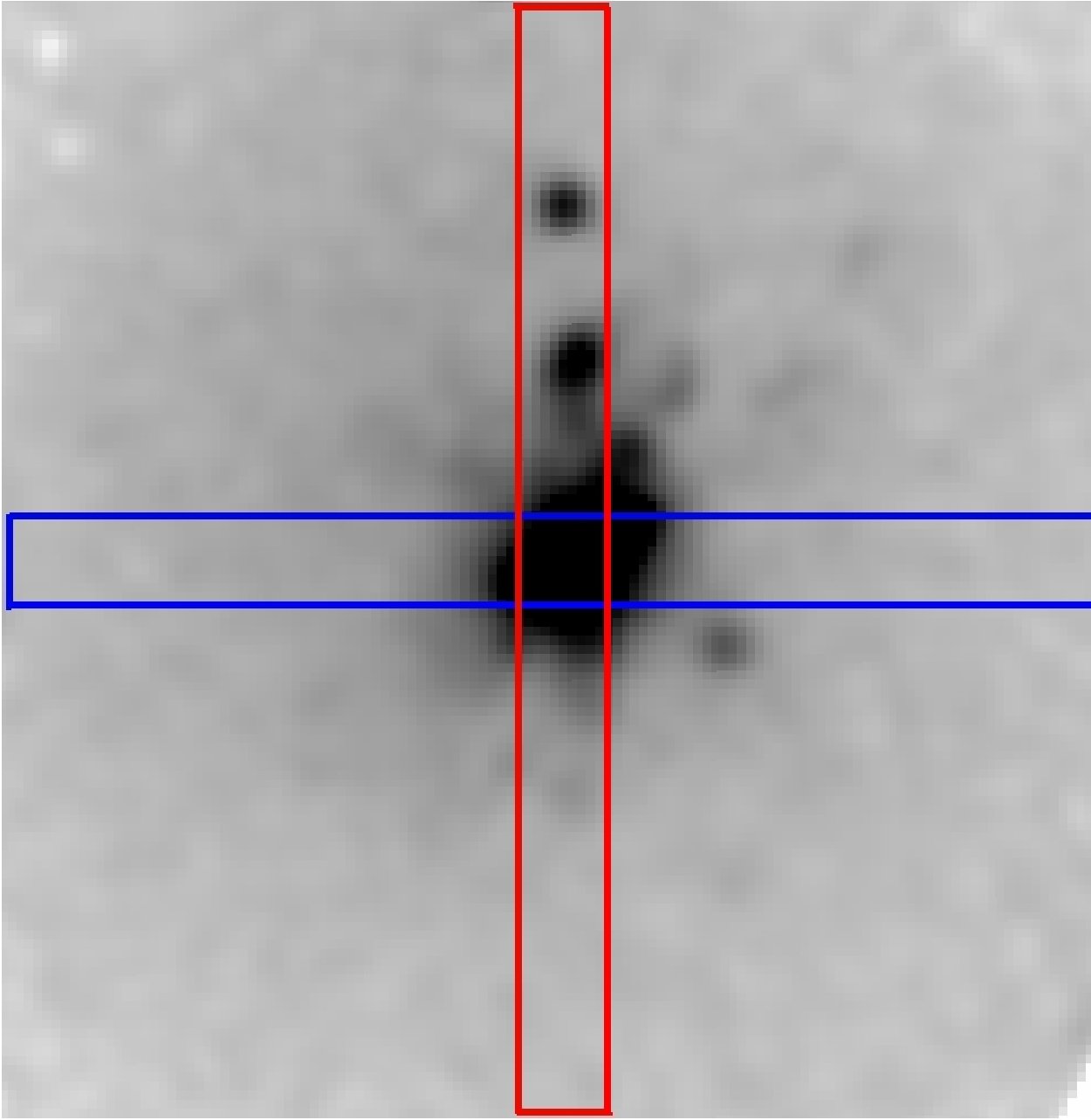}
\vspace{0.5cm}
    \includegraphics[width=0.35\textwidth, angle=90]{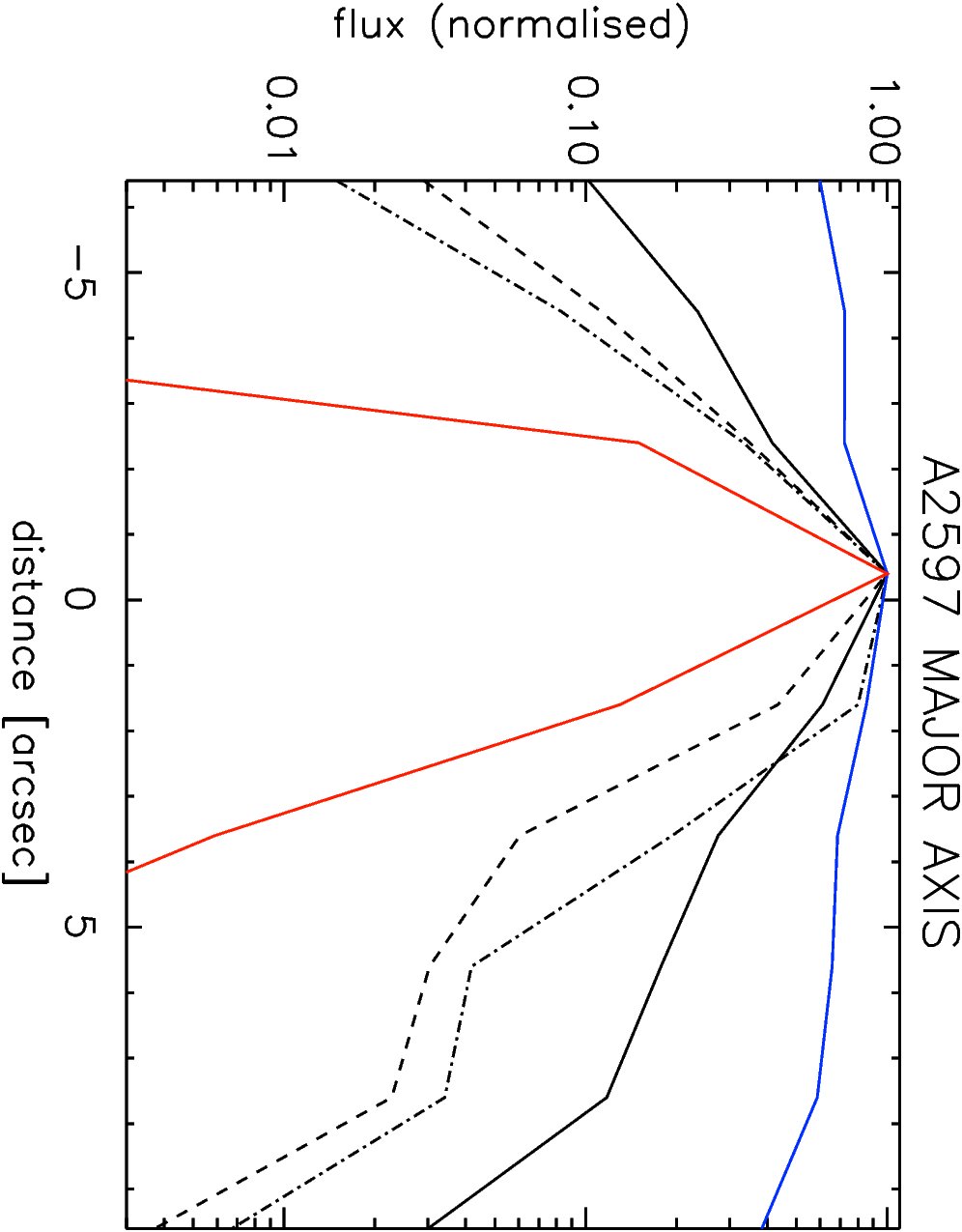}
    \includegraphics[width=0.35\textwidth, angle=90]{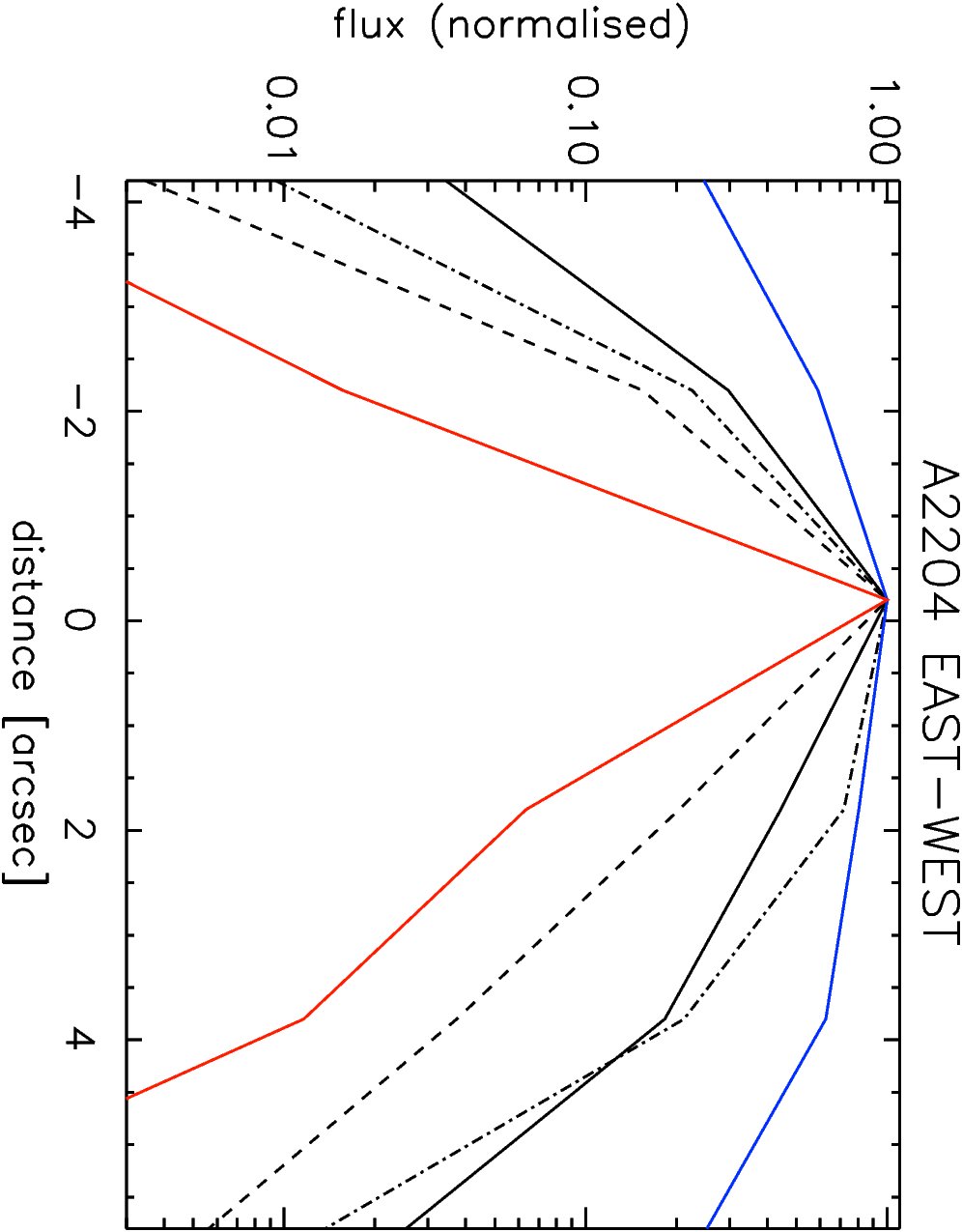}
    \includegraphics[width=0.35\textwidth, angle=90]{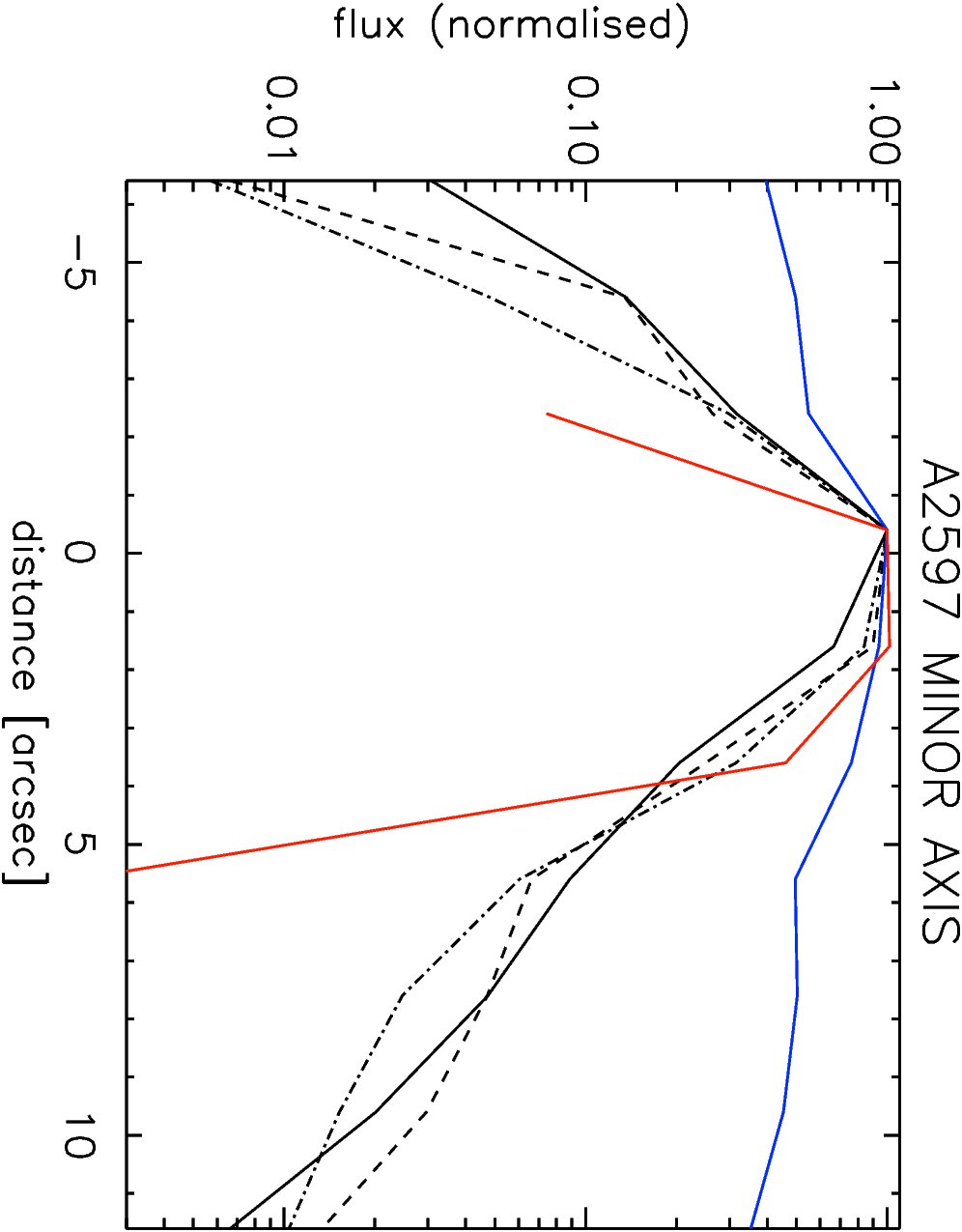}
    \includegraphics[width=0.35\textwidth, angle=90]{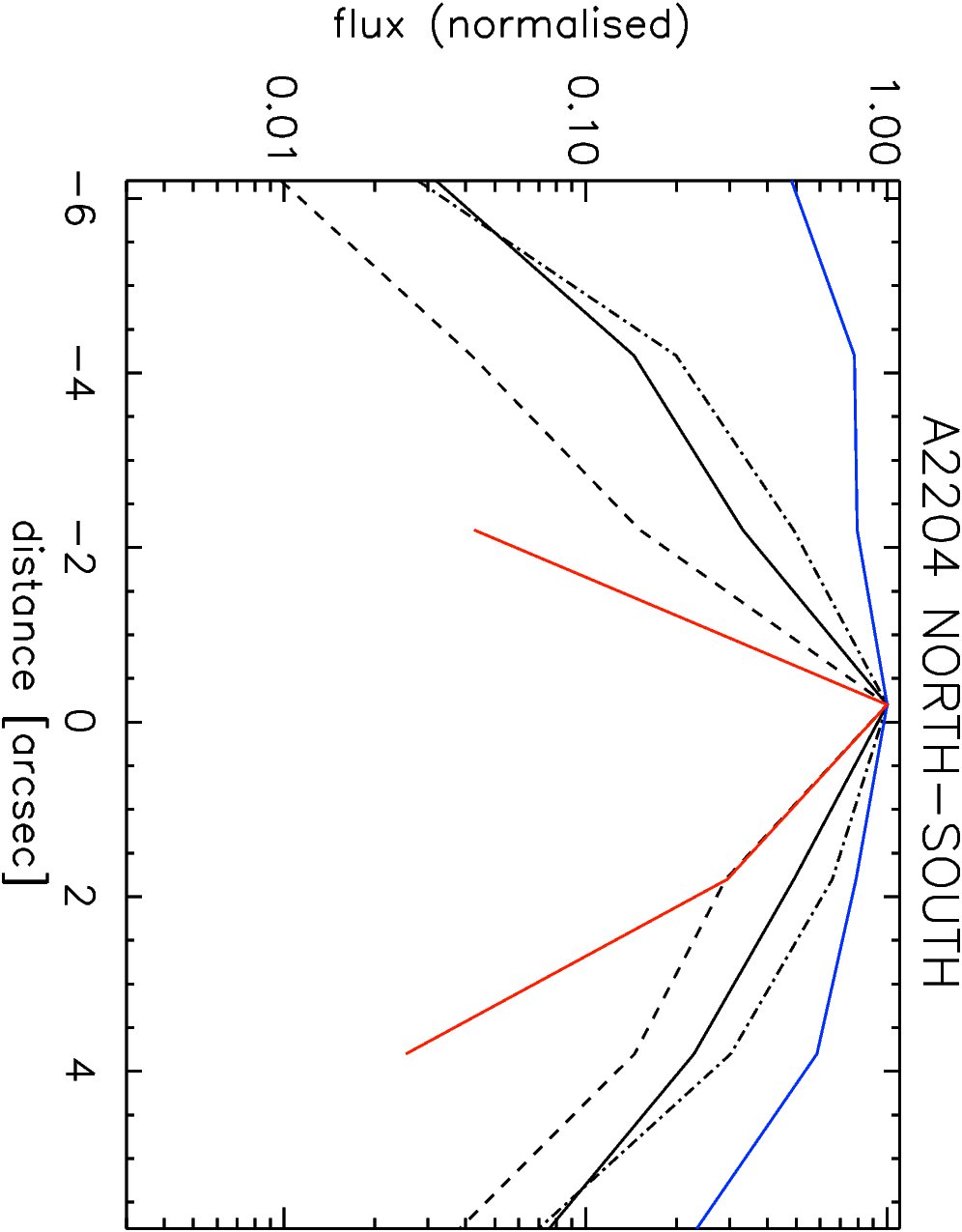}
  \vspace{0.5cm}
  \caption{Surface brightness profiles for A2597 (\textit{Left}) and A2204 (\textit{Right}). (\textit{Top-Left}) Position of pseudo slits on A2597. The image is 25 arcsec on a side and the slits drawn are 25 arcsec long and 2 arcsec wide. (\textit{Top-Right}) Position of pseudo slits on A2204. The image and slit sizes are the same as for A2597. The surface brightness profiles are determined along drawn pseudo slits in 2~arcsec steps along these slits. The solid black line is FORS V band emission. The dashed black line is the background subtracted F150LP FUV emission. The black dot-dash line is H$\alpha$ emission from \citet{Ja05}. The blue solid line is CHANDRA X-ray emission and the red solid line is VLA 5~GHz radio emission. (\textit{Middle-Left}) A2597 pseudo slit along the major axis of the BCG (blue slit). Positive distances are north-west of the BCG nucleus. (\textit{Middle-Right}) A2204 pseudo slit along an axis running east-west through the BCG  (blue slit). Positive distances are east of the BCG nucleus. (\textit{Bottom-Left}) A2597 pseudo slit along the minor axis of the BCG (red slit). Positive distances are north-east of the BCG nucleus. (\textit{Bottom-Right}) A2204 pseudo slit along an axis running north-south through the BCG (red slit). Positive distances are north of the BCG nucleus. All profiles have been extracted from images convolved to a common spatial resolution of 1~arcsec FWHM.}\label{f_pslit}
\end{figure*}

\begin{figure*}
    \includegraphics[width=0.36\textwidth, angle=90]{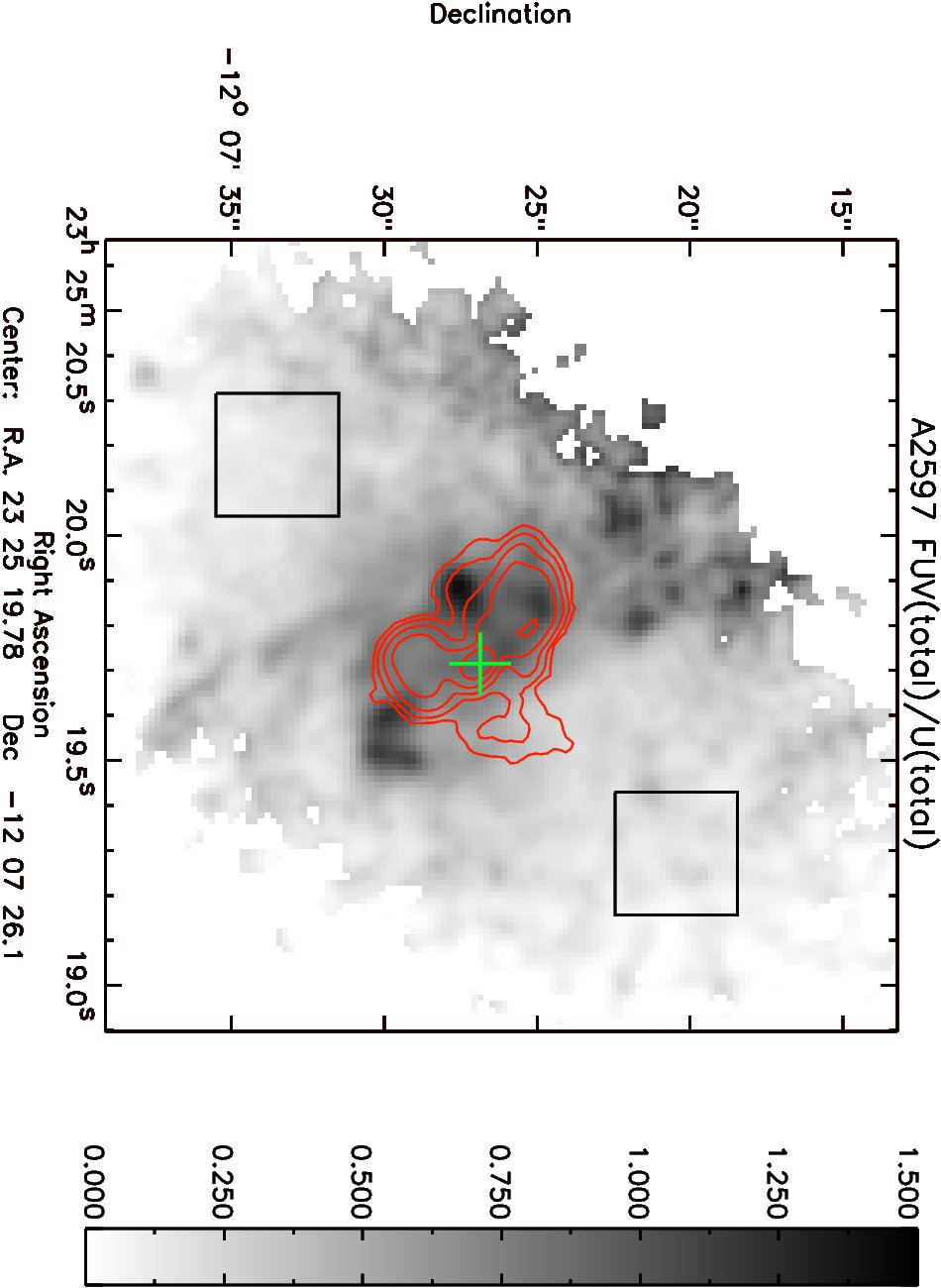}
    \includegraphics[width=0.36\textwidth, angle=90]{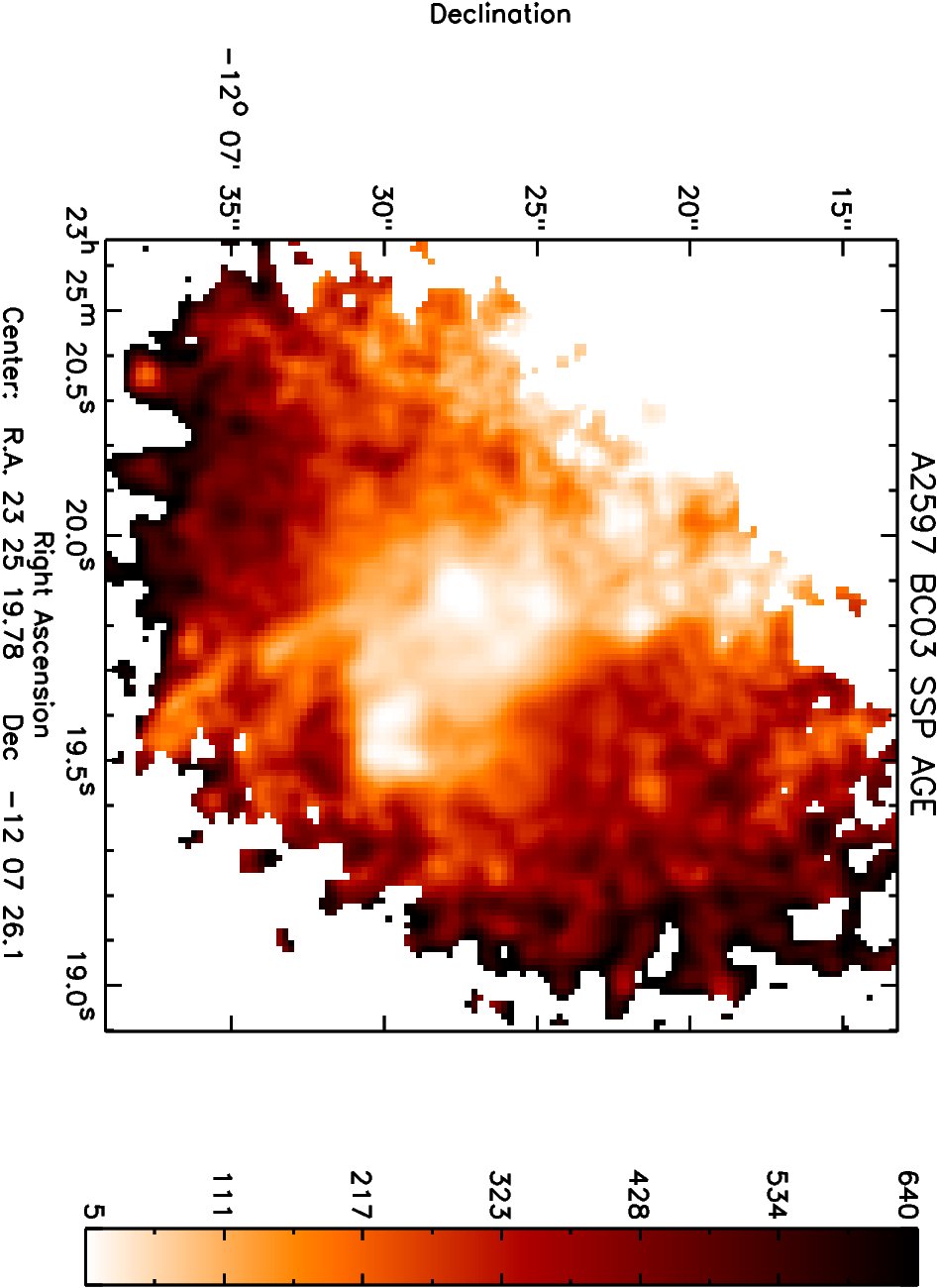}
    \includegraphics[width=0.36\textwidth, angle=90]{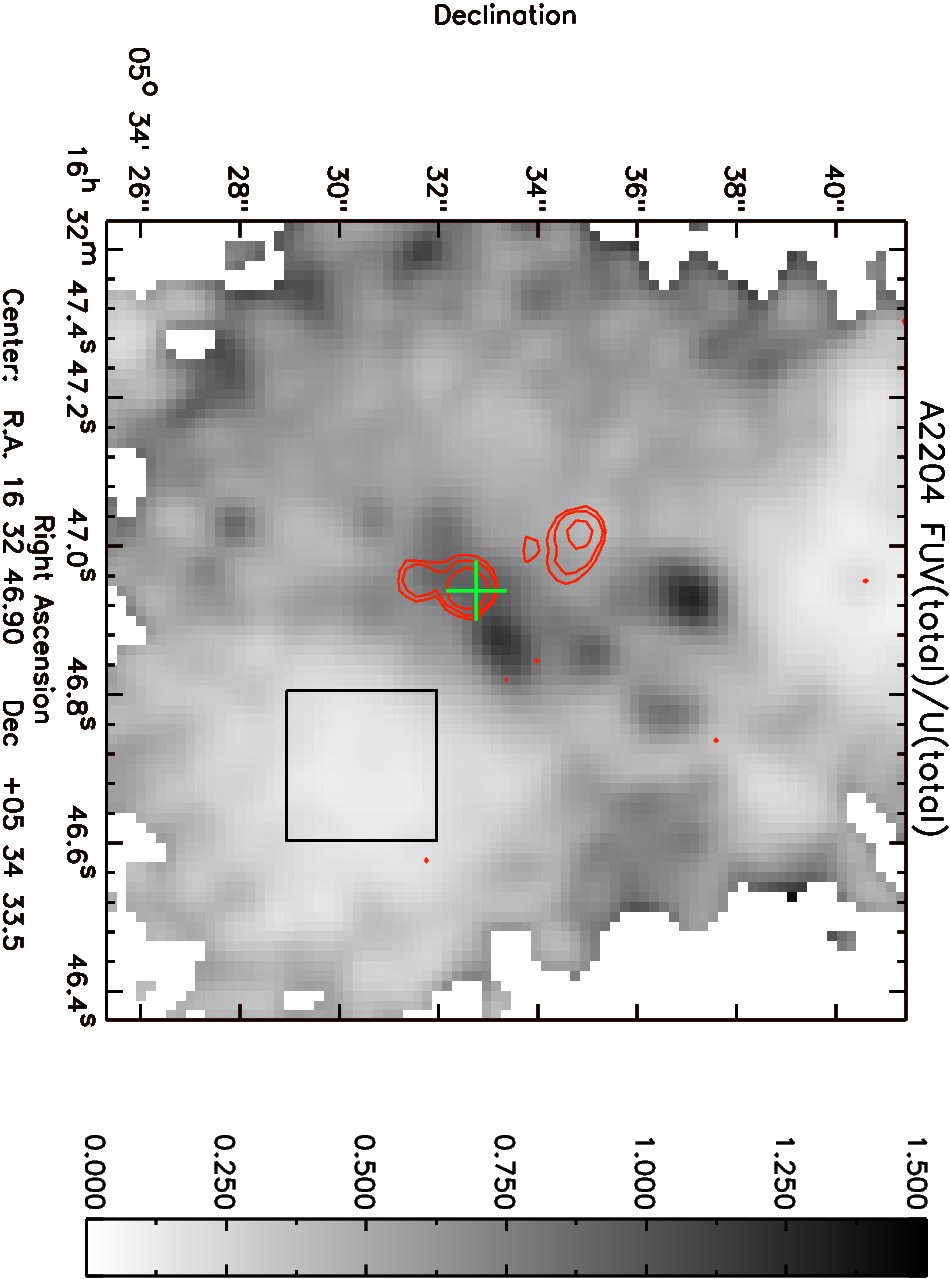}
    \includegraphics[width=0.36\textwidth, angle=90]{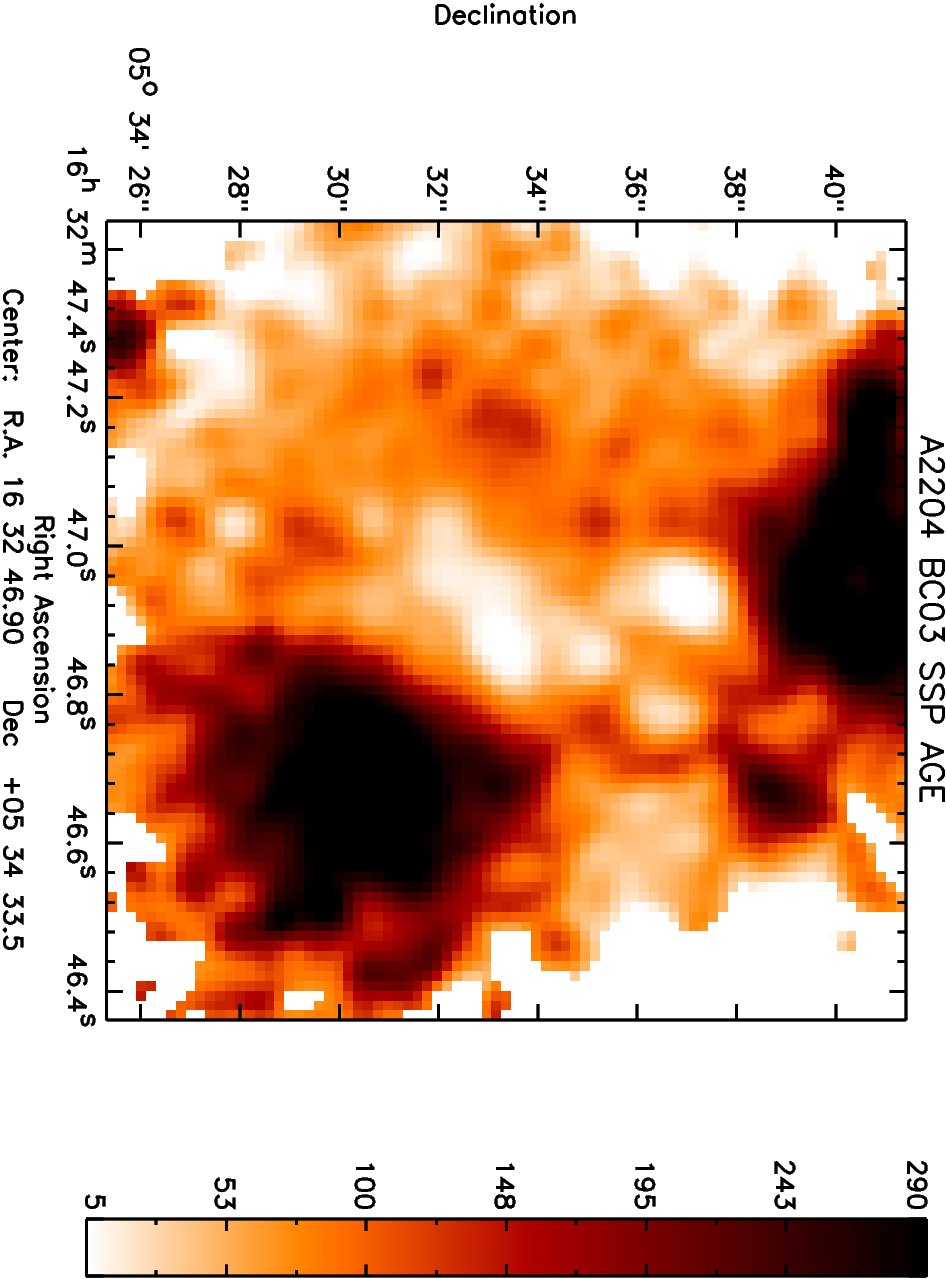}
  \vspace{0.5cm}
  \caption{The FUV$_{\nu,tot}$/U$_{\nu,tot}$  ratio and corresponding SSP age for A2597 and A2204. (\textit{Top-Left}) A2597 observed FUV$_{\nu,tot}$/U$_{\nu,tot}$. (\textit{Top-Right}) A2597 SSP age in units of Myr. (\textit{Bottom-Left}) A2204 observed FUV$_{\nu,tot}$/U$_{\nu,tot}$. (\textit{Bottom-Right}) A2204 SSP age in units of Myr. In computing the SSP age we have taken extinction due to the MW foreground only into account, see Fig. \ref{f_sbrst}. The SSP age vs FUV$_{\nu,tot}$/U$_{\nu,tot}$ ratio is double valued for ages greater than 640 and 290 Myr in A2597 and A2204 respectively. This is due to the UV upturn for a stellar population older than 1 Gyr. For any flux ratios in the range where the corresponding SSP age is double valued we have set the SSP age to the youngest age possible in this range. The green crosses indicate the BCG nuclei. The black squares indicate the regions where the U/V and FUV/V ratios for the old stellar population have been determined. Red contours show the 5~GHz radio continuum emission. The contours levels are the same as in Figs. \ref{f_a2597_ovl} and \ref{f_a2204_ovl}.}\label{f_totfu}
\end{figure*}

\begin{figure*}
    \includegraphics[width=0.45\textwidth, angle=90]{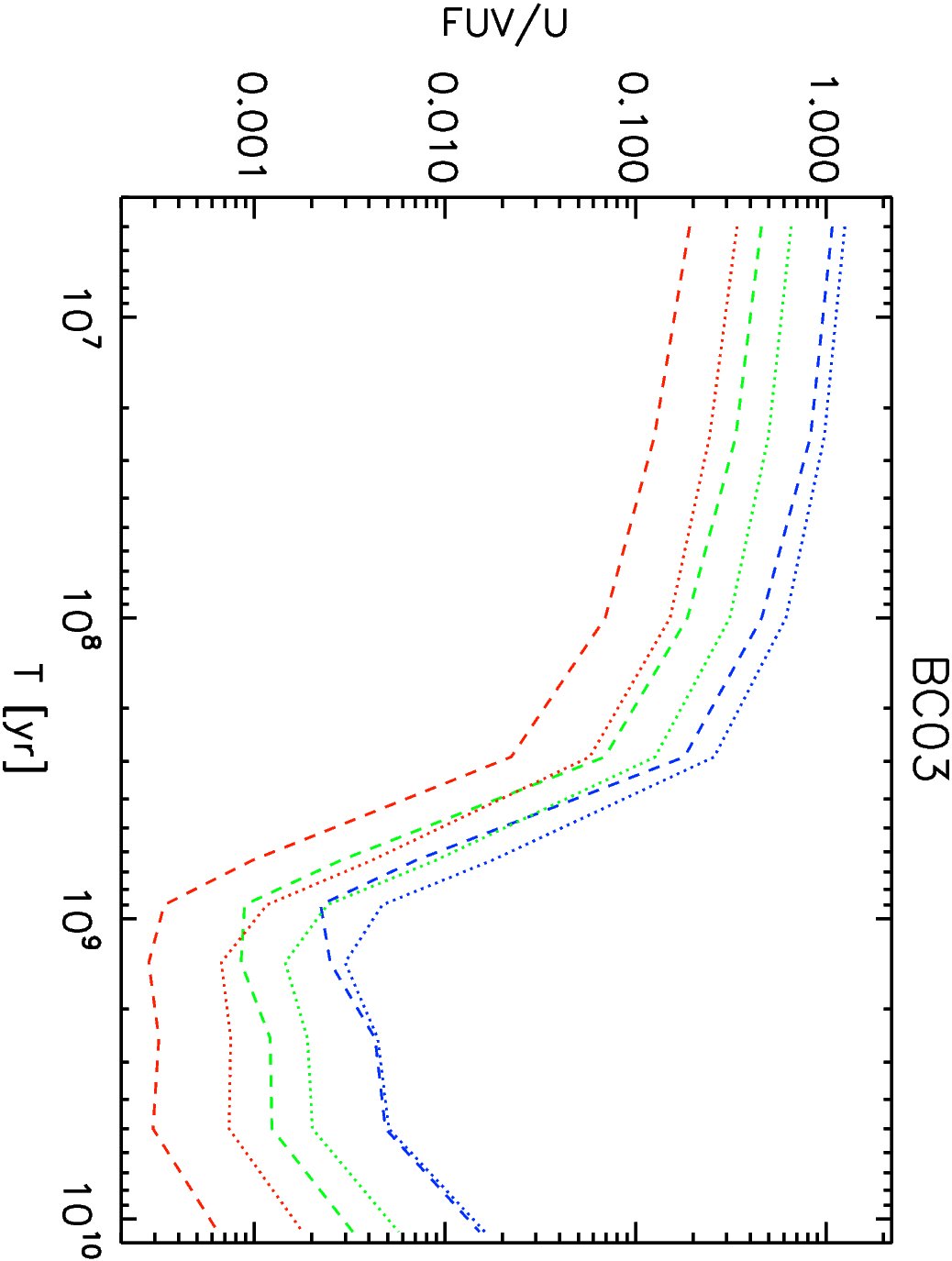}
  \vspace{0.5cm}
  \caption{The FUV$_{\nu}$/U$_{\nu}$ ratio as a function of SSP age for \citet{Br03} SSP models. Curves are shown for two redshifts and three different amounts of extinction. The dotted and dashed curves correspond to z=0.0821 (A2597) and z=0.1517 (A2204) respectively. The blue curves correspond to MW foreground extinction only. The red curves assume that all of the extinction intrinsic to the BCG is in front of the FUV and U emitting regions. The green curves assume that only half of the extinction intrinsic to the BCG is in front of the FUV and U emitting regions. Dust extinction has been computed using the C89 extinction law for both the MW and the BCG component. The model ratios are computed for the HST-ACS/SBC F150LP and VLT-FORS U\_Bessel filters.}\label{f_sbrst}
\end{figure*}

\begin{figure*}
    \includegraphics[width=0.45\textwidth, angle=90]{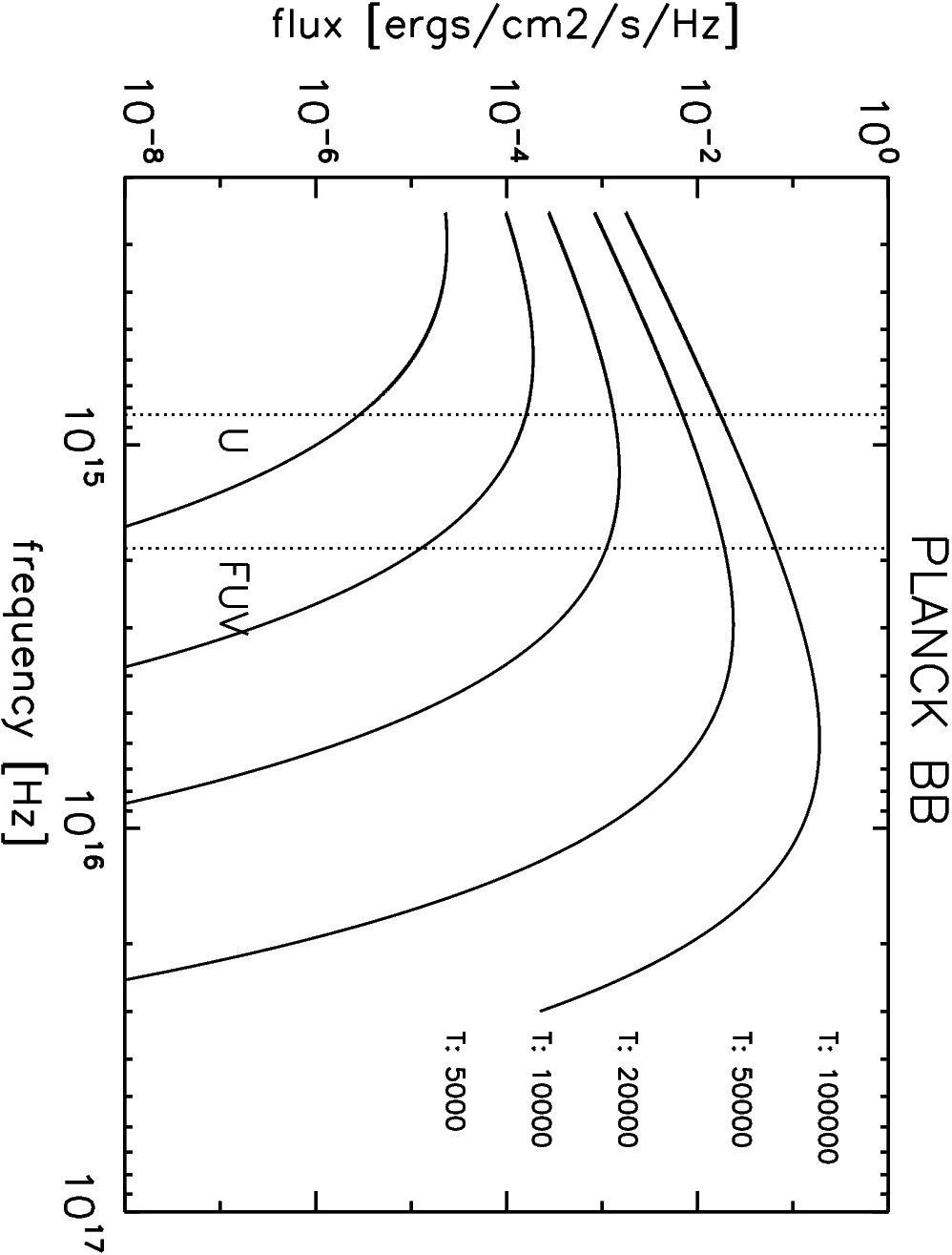}
  \vspace{0.5cm}
  \caption{Blackbody (BB) models in the FUV - Optical regime. The dashed lines show the peak transmission wavelengths of the U (U\_Bessel) and FUV (F150LP) filters. This image illustrates how the FUV flux changes relative to the U flux as we increase the temperature for a BB model. The fast change in their relative ratio as a function of temperature shows that it is a good discriminator of temperature. We note that the K93 stellar models as function of stellar temperature qualitatively show the same behaviour in the FUV to optical regime as the BB models. The FUV to U band ratio is thus a good discriminator for these models too.}\label{f_bb_z0}
\end{figure*}

\begin{figure*}
    \includegraphics[width=0.6\textwidth, angle=90]{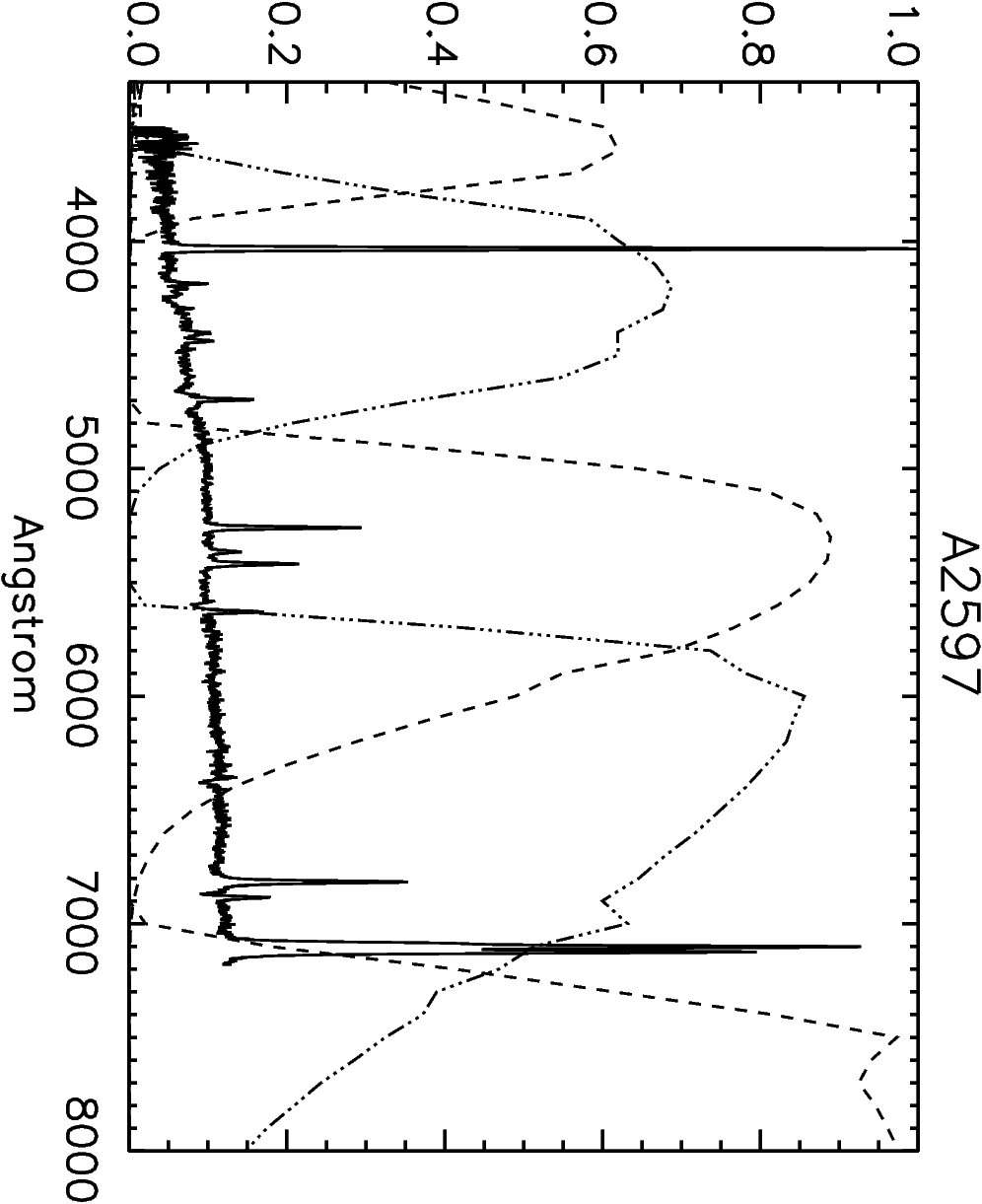}
    \includegraphics[width=0.6\textwidth, angle=90]{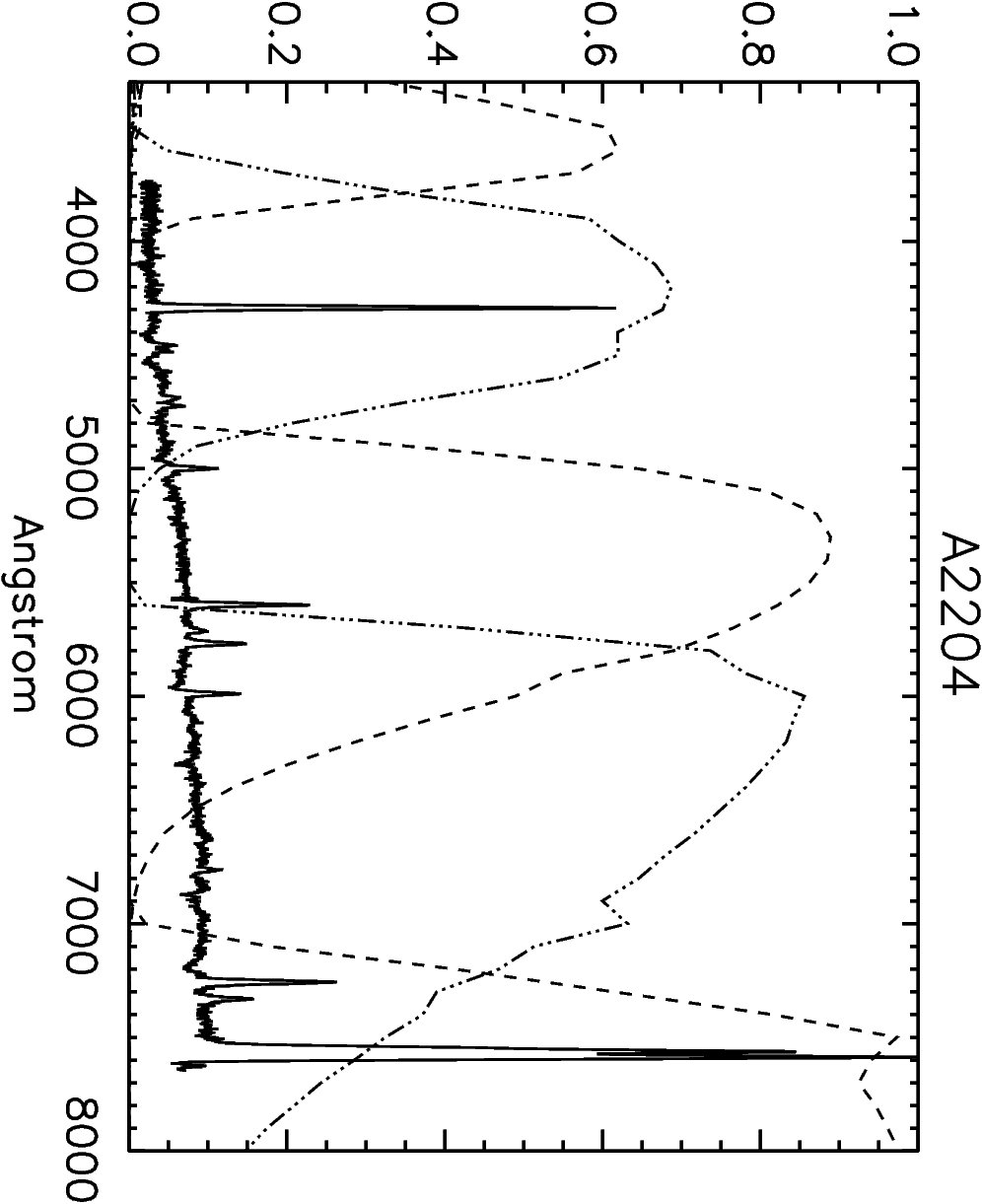}
  \vspace{0.5cm}
  \caption{Line Contamination of the FORS U, B, V, R and I Bessel filters. (\textit{Top}) Spatially integrated (16$\times$2~arcsec$^{2}$) spectrum of A2597. (\textit{Bottom}) Spatially integrated (12.5$\times$2~arcsec$^{2}$) spectrum of A2204. Both spectra are centred on their respective BCG nucleus. The dashed and dot-dash curves show the shape of FORS U, B, V, R and I Bessel filters.}\label{f_fors_spec_filt}
\end{figure*}

\begin{figure*}
    \includegraphics[width=0.36\textwidth, angle=90]{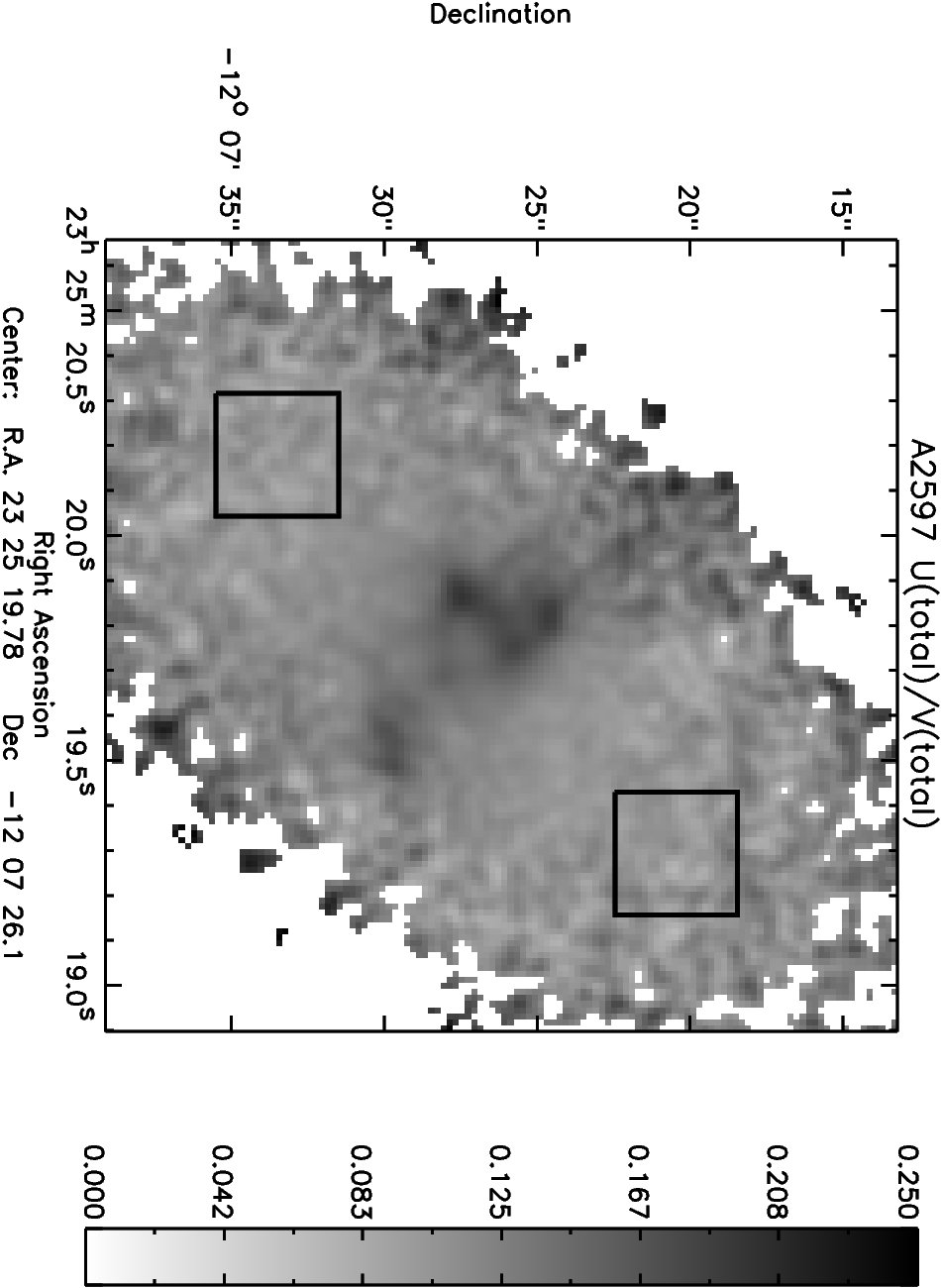}
    \includegraphics[width=0.36\textwidth, angle=90]{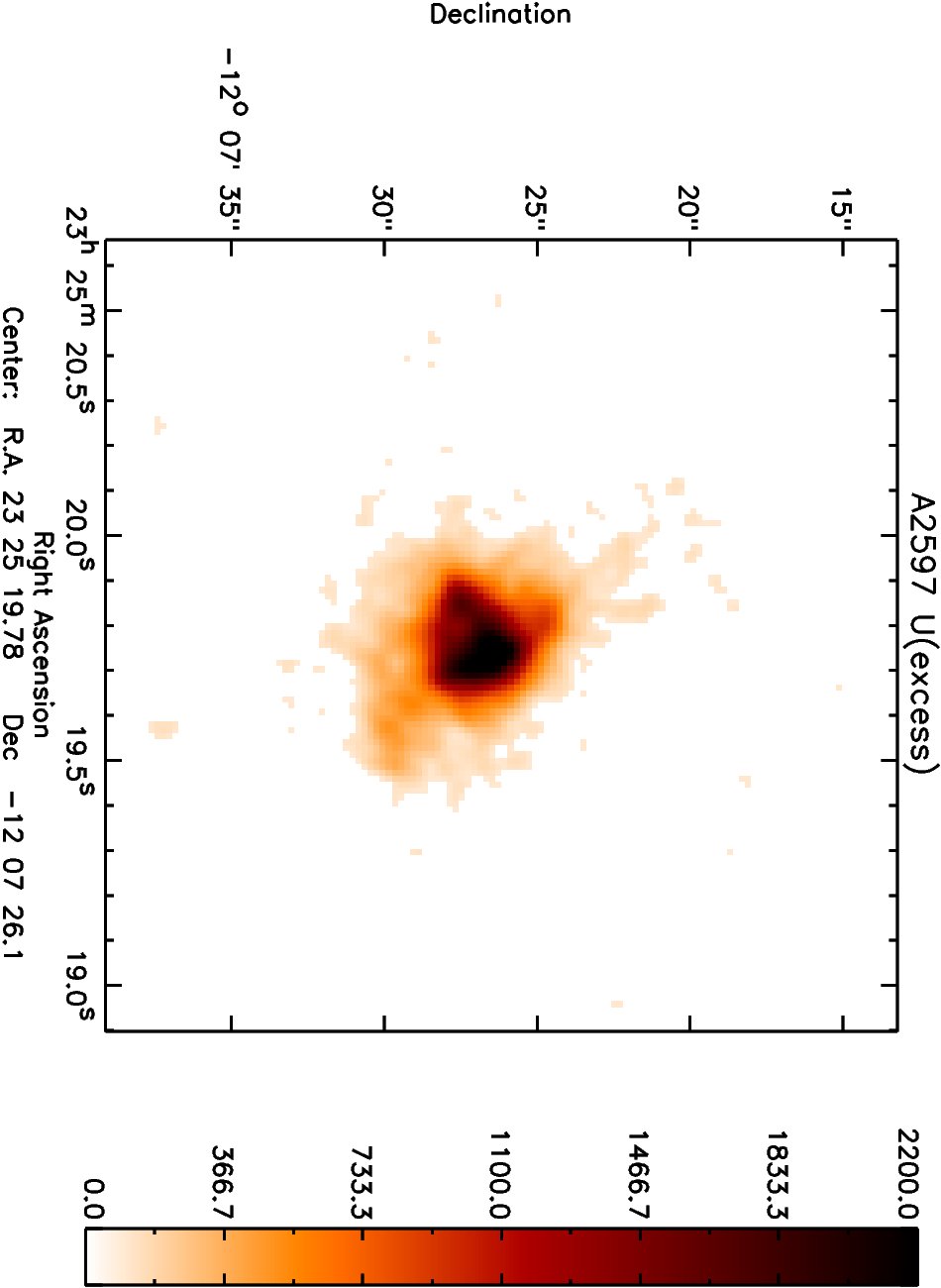}
    \includegraphics[width=0.36\textwidth, angle=90]{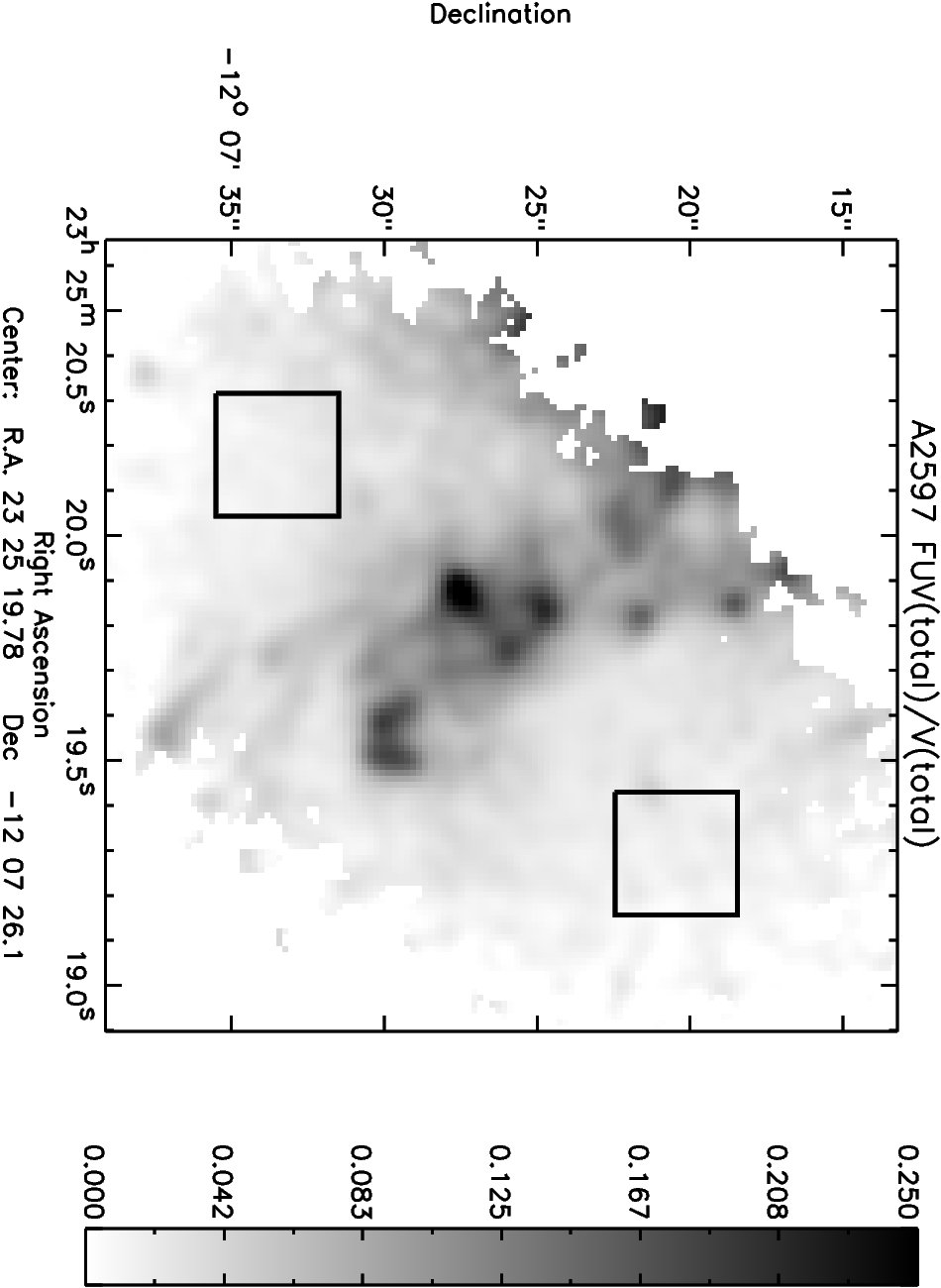}
    \includegraphics[width=0.36\textwidth, angle=90]{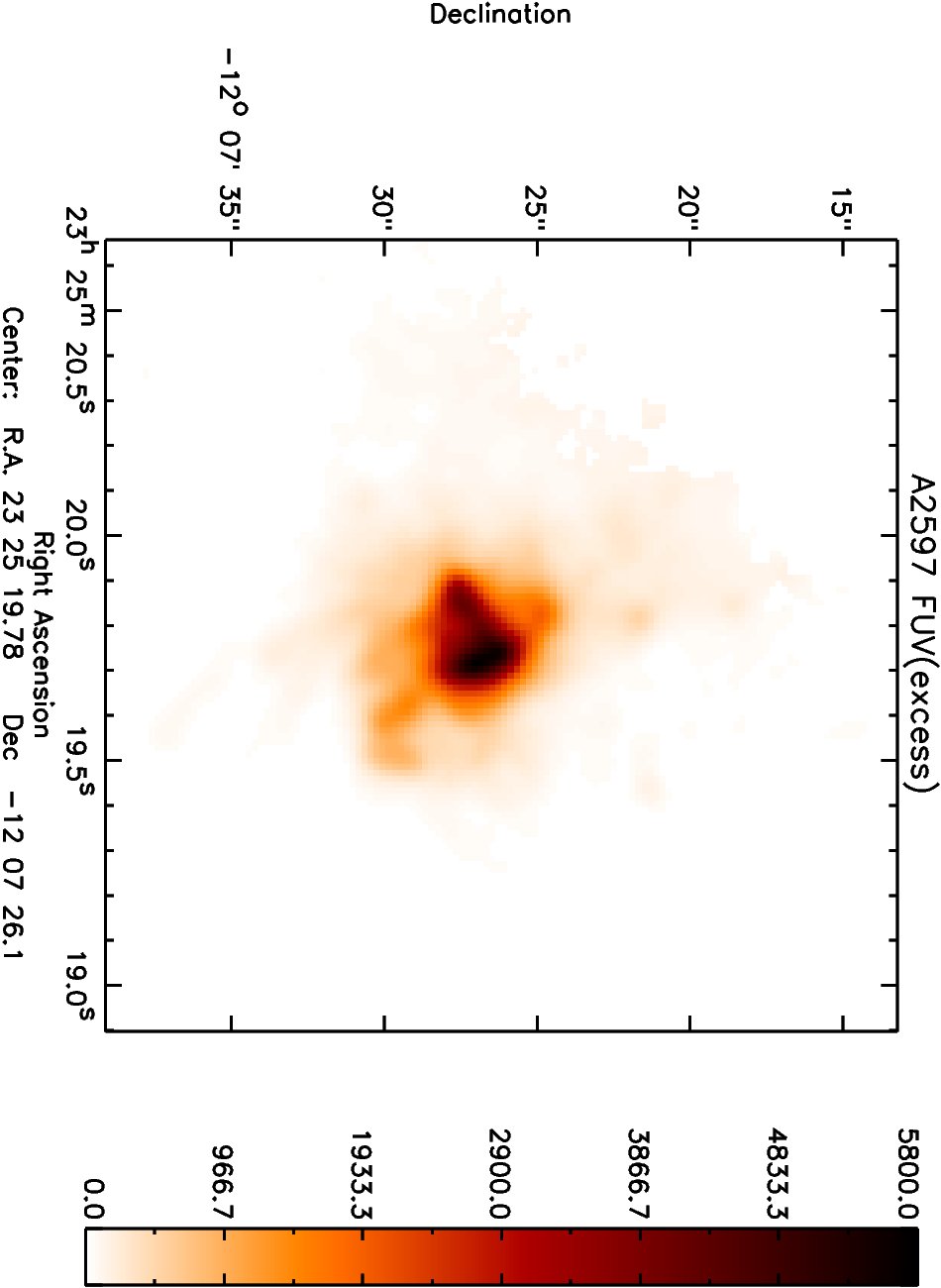}
  \vspace{0.5cm}
  \caption{Excess emission in A2597. (\textit{Top-Left}) U$_{\nu,tot}$/V$_{\nu,tot}$ ratio. (\textit{Top-Right}) Excess U$_{\nu,exc}$ band emission after removal of the old stellar population. (\textit{Bottom-Left}) FUV$_{\nu,tot}$/V$_{\nu,tot}$ ratio. (\textit{Bottom-Right}) Excess FUV$_{\nu,exc}$ emission after removal of the old stellar population. These images have been corrected for background emission and line contamination. The black squares indicate the regions where the U$_{\nu,tot}$/V$_{\nu,tot}$ and FUV$_{\nu,tot}$/V$_{\nu,tot}$ ratios for the old stellar population have been determined.}\label{f_a2597_vrat}
\end{figure*}

\begin{figure*}
    \includegraphics[width=0.36\textwidth, angle=90]{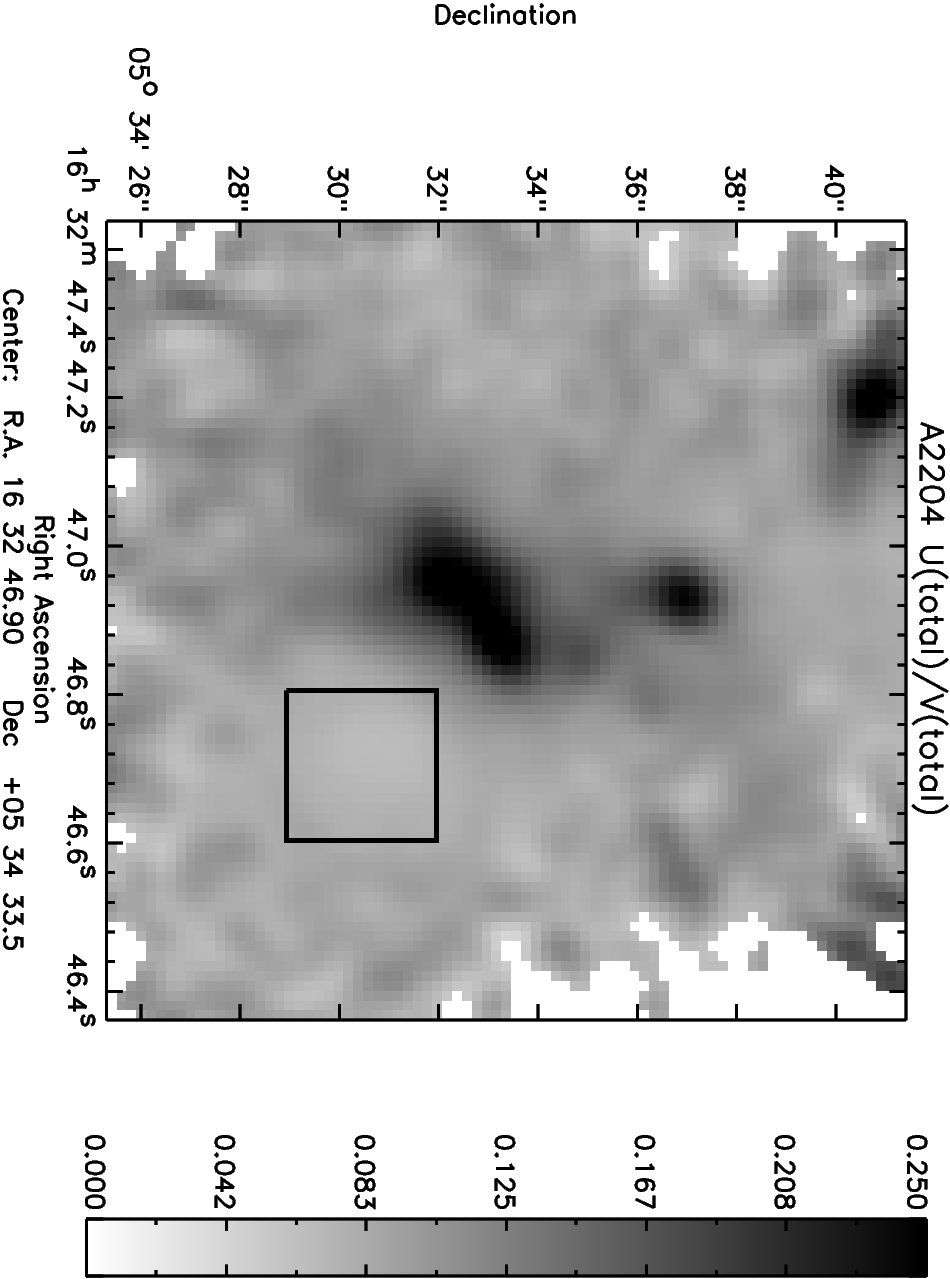}
    \includegraphics[width=0.36\textwidth, angle=90]{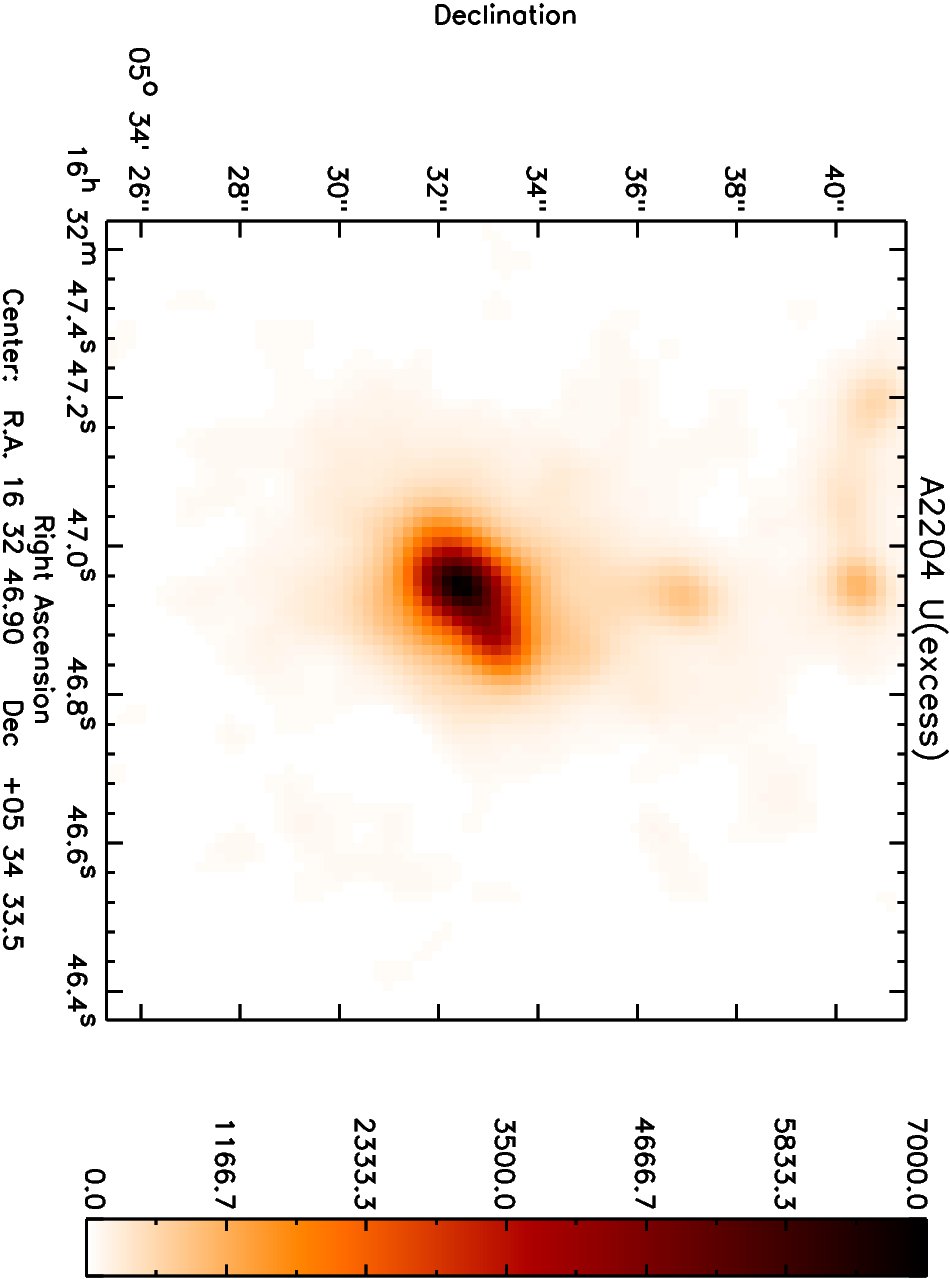}
    \includegraphics[width=0.36\textwidth, angle=90]{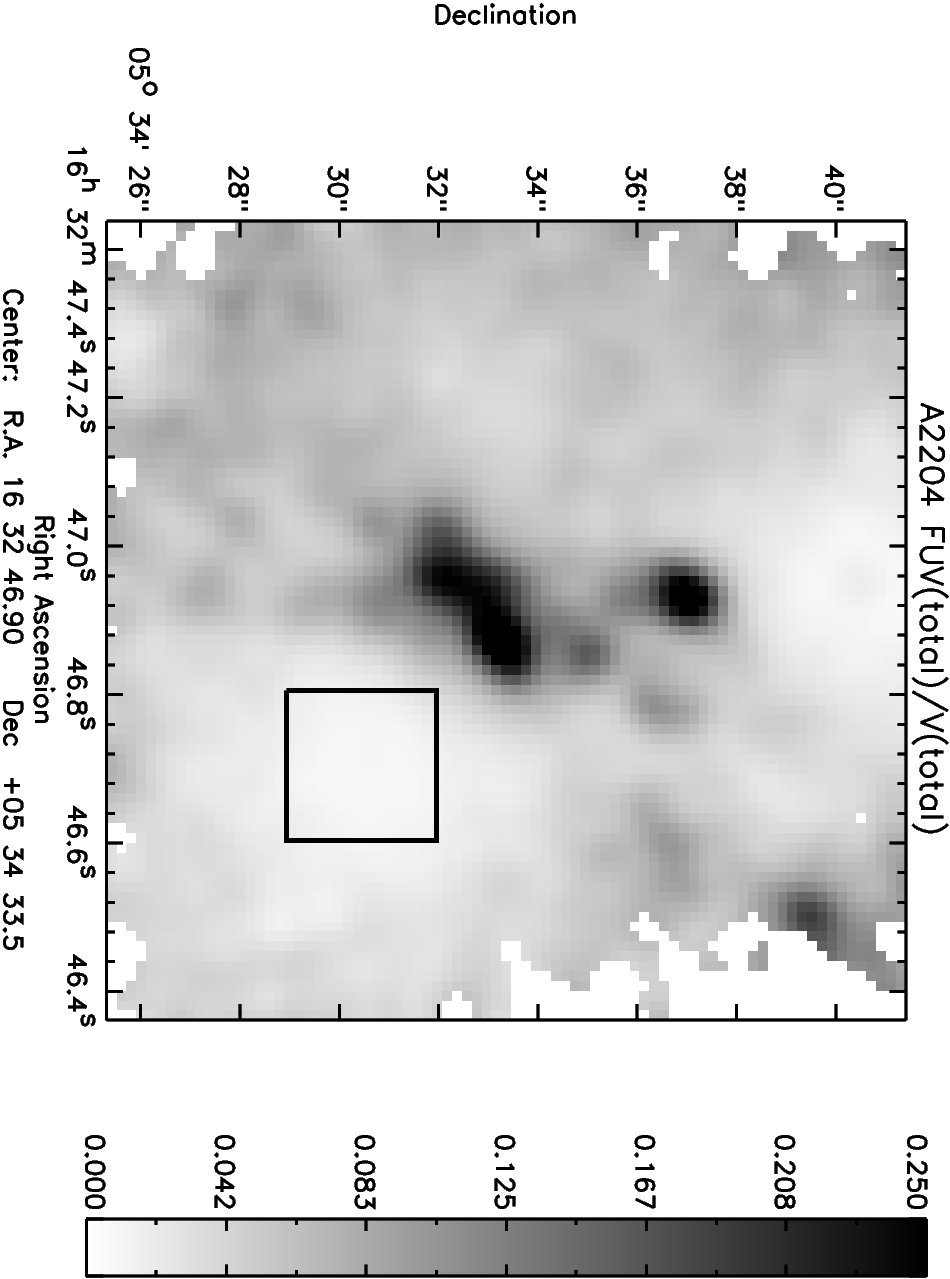}
    \includegraphics[width=0.36\textwidth, angle=90]{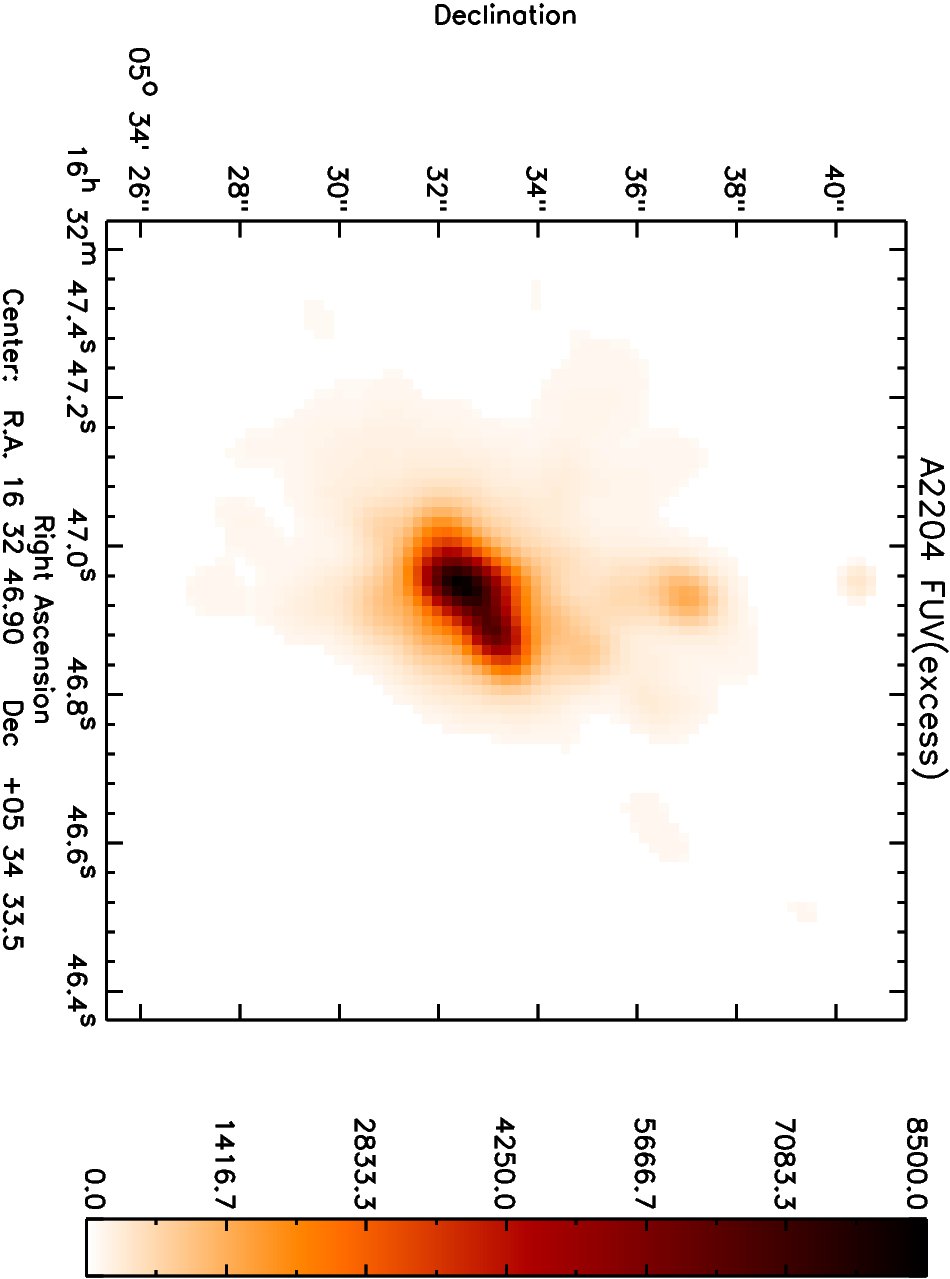}
  \vspace{0.5cm}
  \caption{Excess emission in A2204: (\textit{Top-Left}) U$_{\nu,tot}$/V$_{\nu,tot}$ ratio. (\textit{Top-Right}) Excess U$_{\nu,exc}$ emission after removal of the old stellar population. (\textit{Bottom-Left}) FUV$_{\nu,tot}$/V$_{\nu,tot}$ ratio. (\textit{Bottom-Right}) Excess FUV$_{\nu,exc}$ emission after removal of the old stellar population. The images have been corrected for background emission and line contamination. The black squares indicate the regions where the U$_{\nu,tot}$/V$_{\nu,tot}$ and FUV$_{\nu,tot}$/V$_{\nu,tot}$ ratios for the old stellar population have been determined.}\label{f_a2204_vrat}
\end{figure*}

\begin{figure*}
    \includegraphics[width=0.36\textwidth, angle=90]{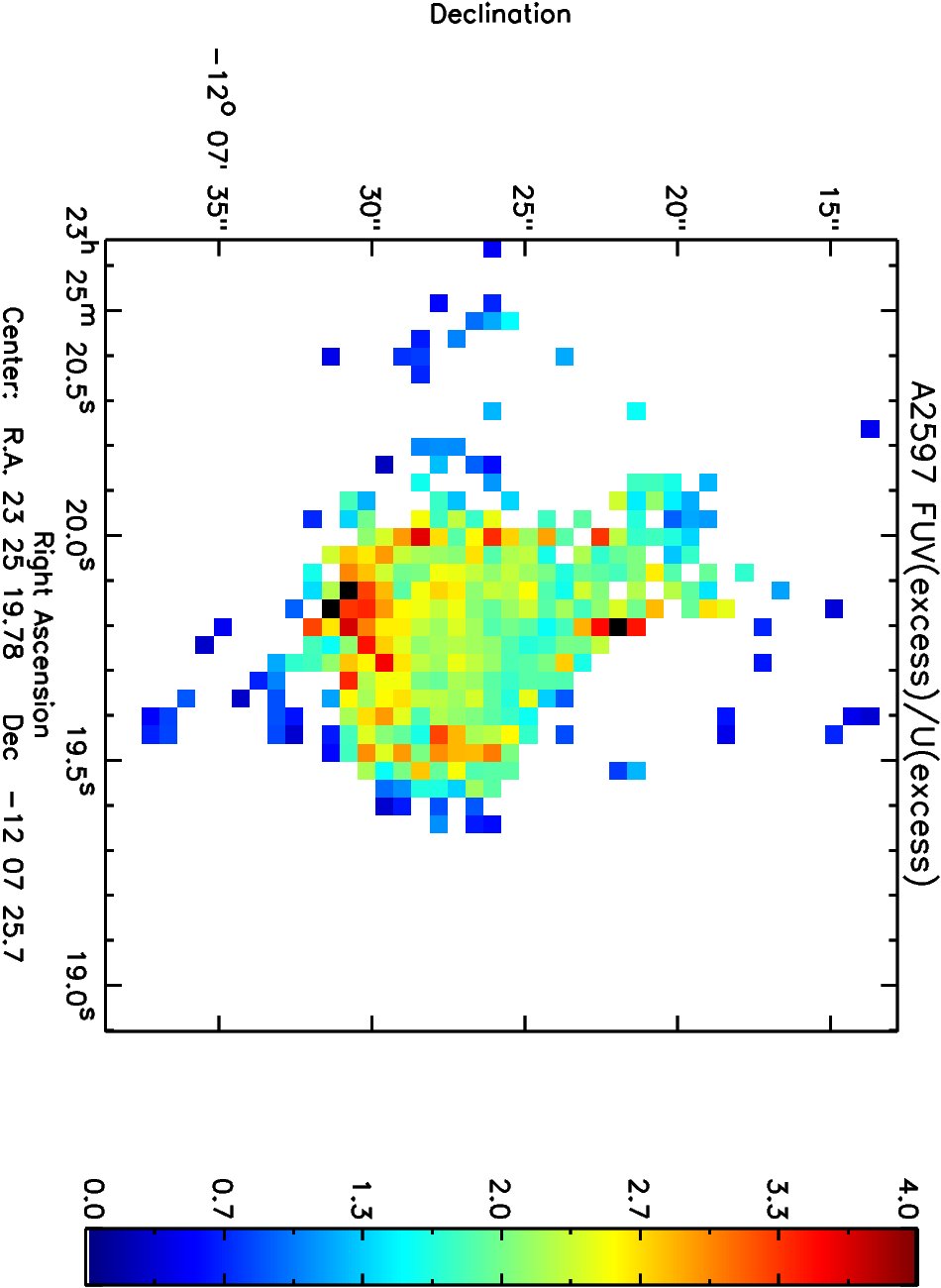}
    \includegraphics[width=0.36\textwidth, angle=90]{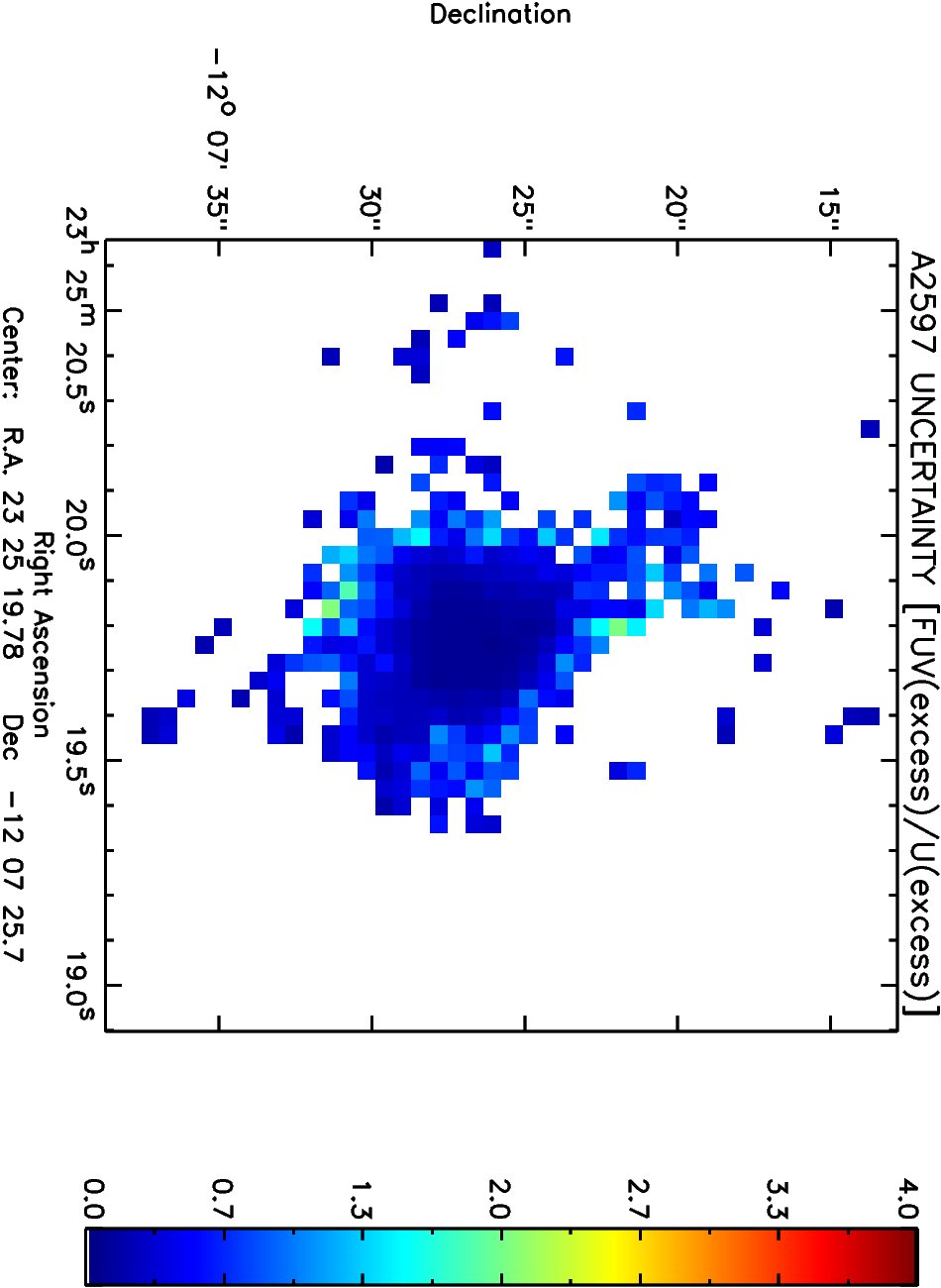}
    \includegraphics[width=0.36\textwidth, angle=90]{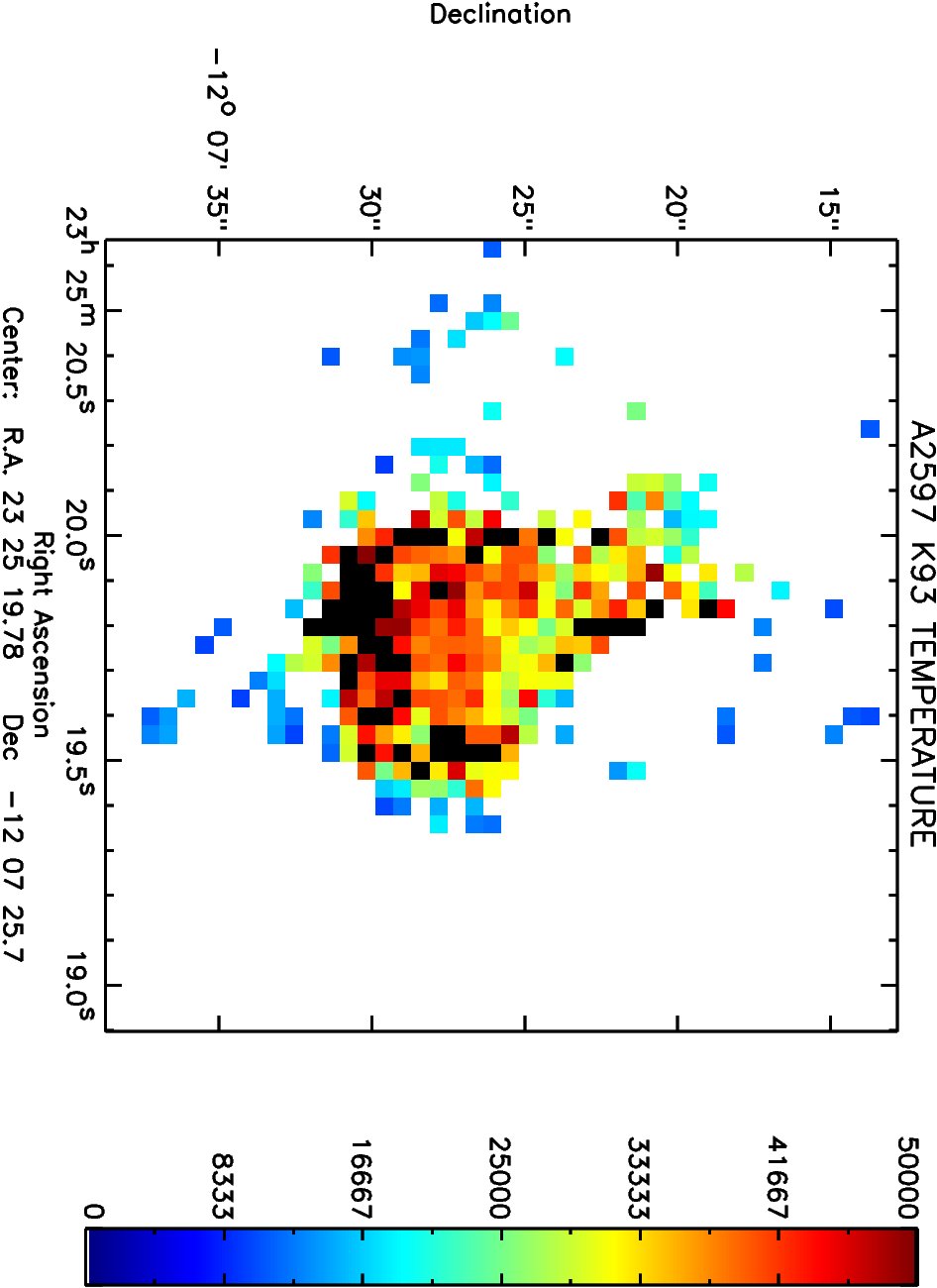}
    \includegraphics[width=0.36\textwidth, angle=90]{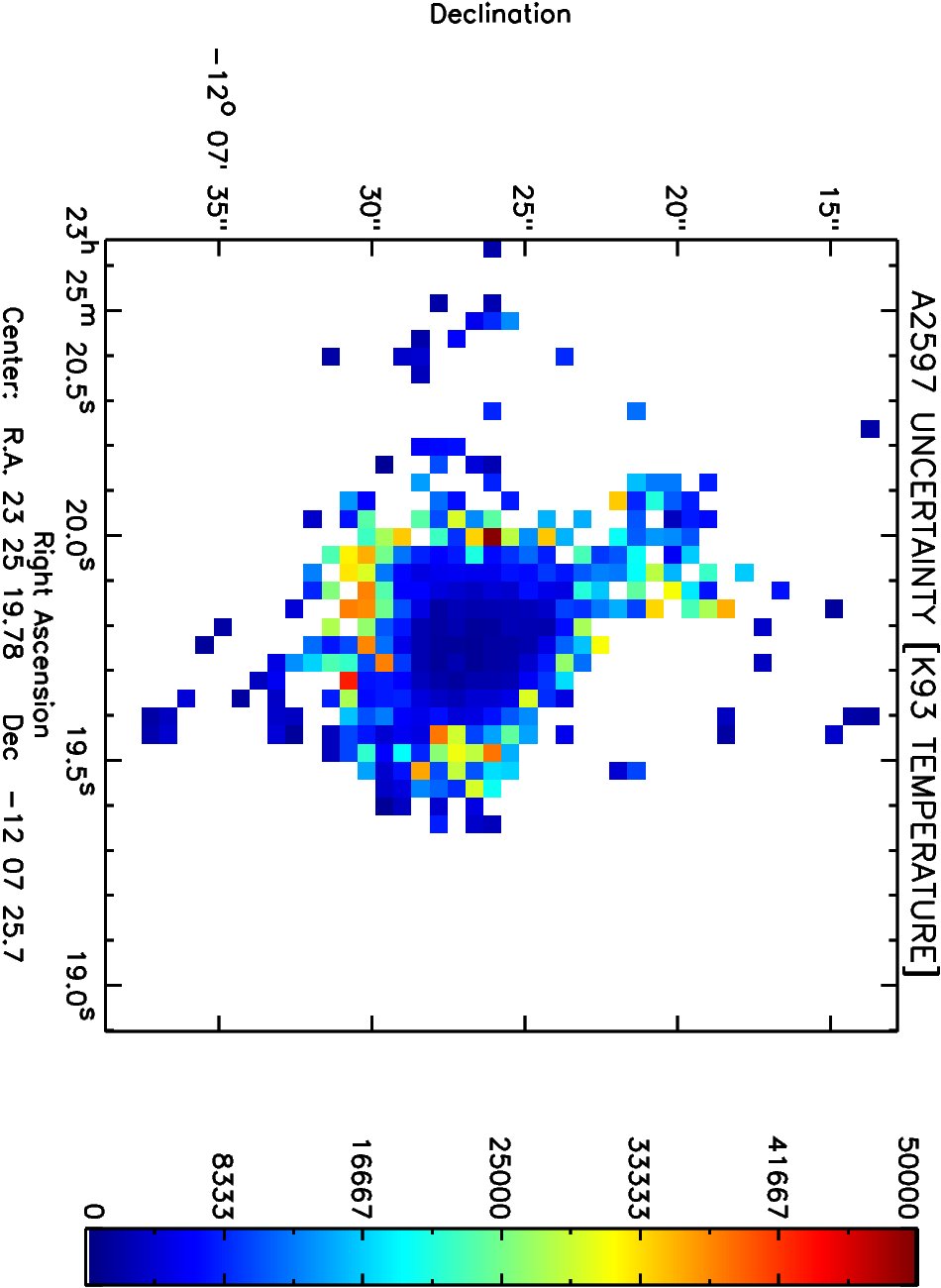}
  \vspace{0.5cm}
  \caption{FUV$_{\nu,exc}$/U$_{\nu,exc}$ ratio and K93 stellar temperature maps for A2597. (\textit{Top-Left}) FUV$_{\nu,exc}$/U$_{\nu,exc}$ ratio. (\textit{Top-Right}) Uncertainty in the FUV$_{\nu,exc}$/U$_{\nu,exc}$ ratio. (\textit{Bottom-Left}) K93 stellar temperature corresponding to the FUV$_{\nu,exc}$/U$_{\nu,exc}$ ratio. (\textit{Bottom-Right}) Uncertainty in K93 derived stellar temperature. Note that in calculating the K93 stellar temperature shown here we have only taken extinction due to the MW foreground into account, see Fig. \ref{f_bcg_bb_k93}. The data shown has been re-binned to 0.6$\times$0.6 arcsec$^{2}$ pixels.}\label{f_a2597_yngfu}
\end{figure*}

\begin{figure*}
    \includegraphics[width=0.36\textwidth, angle=90]{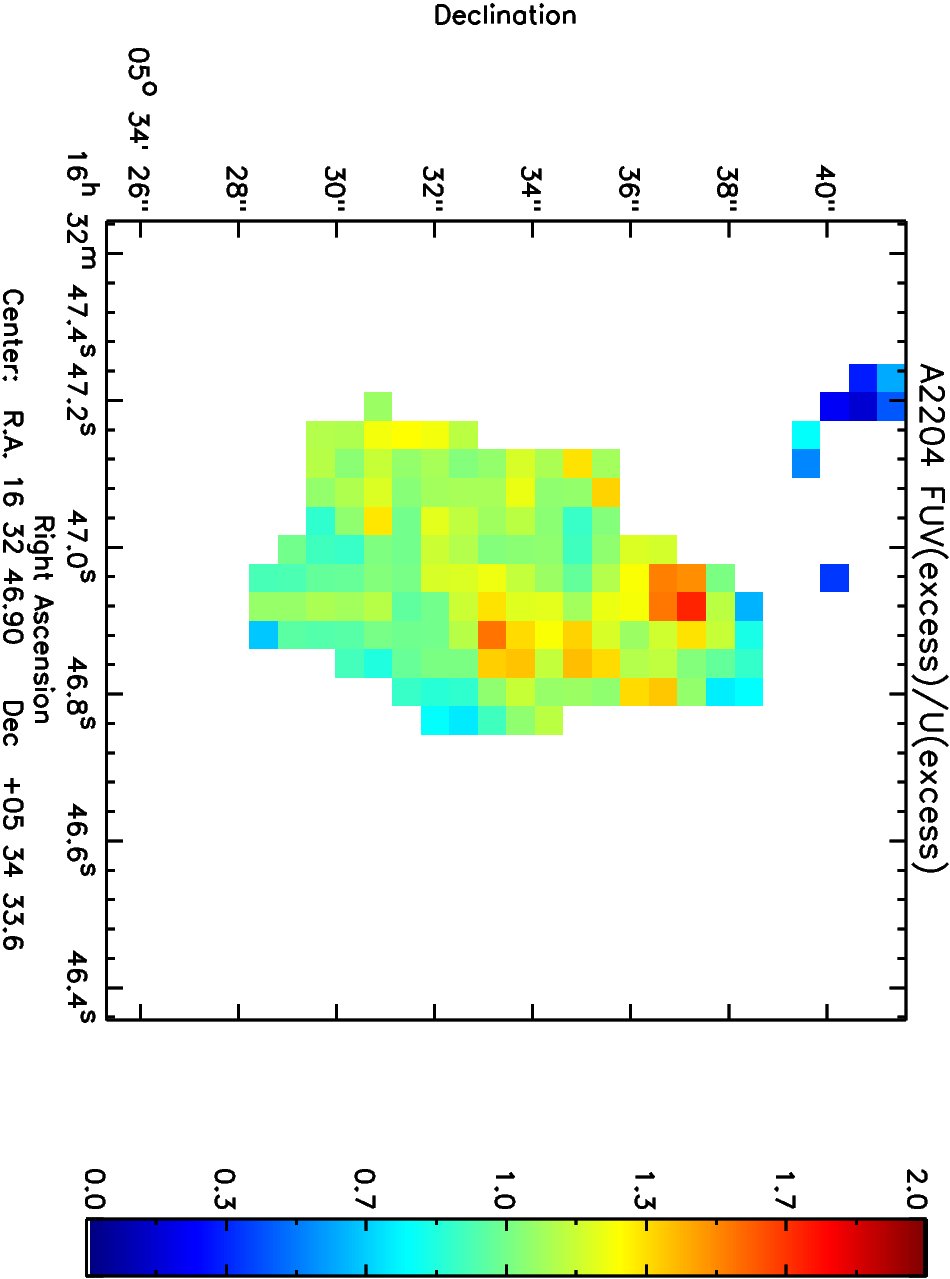}
    \includegraphics[width=0.36\textwidth, angle=90]{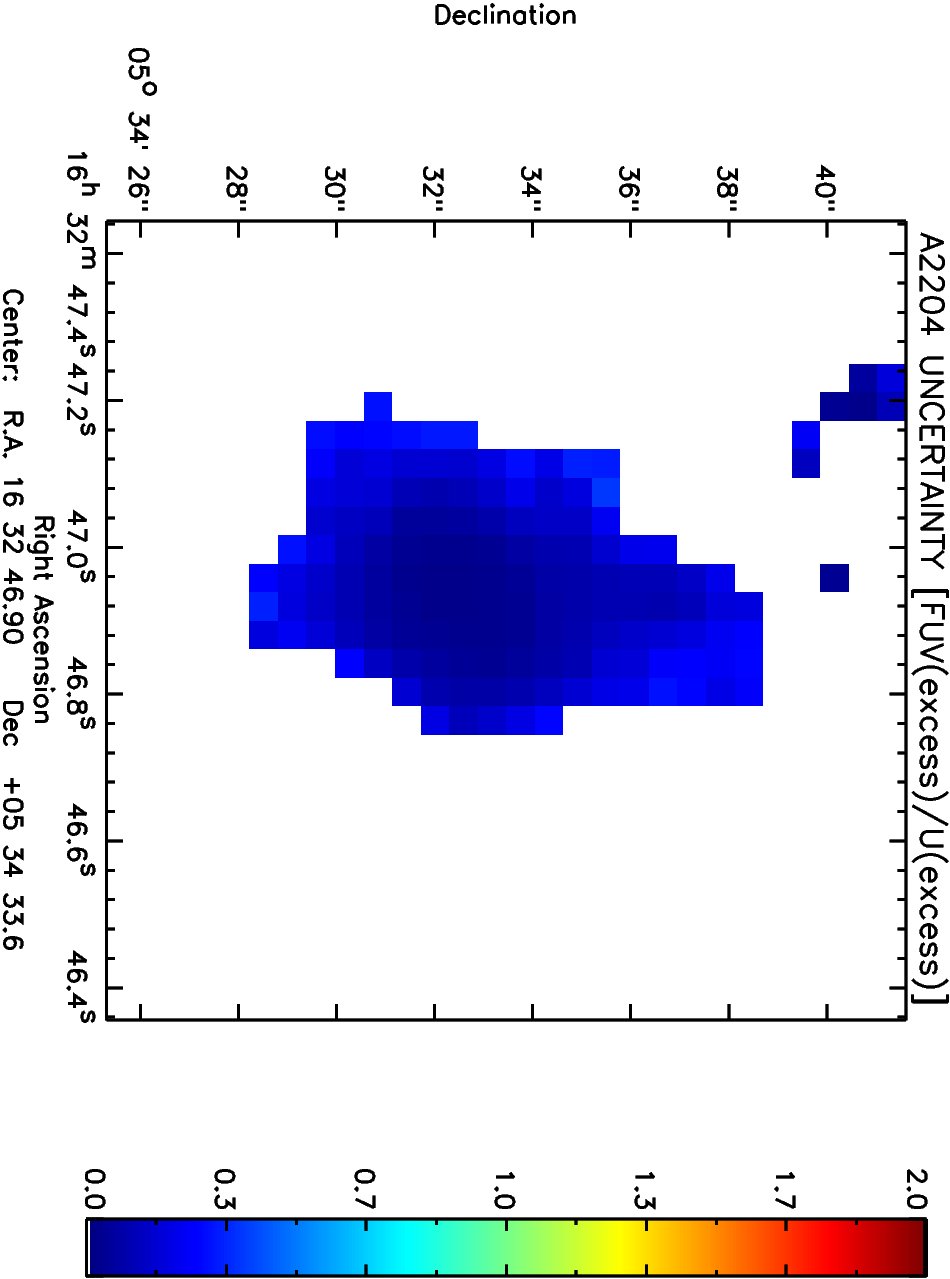}
    \includegraphics[width=0.36\textwidth, angle=90]{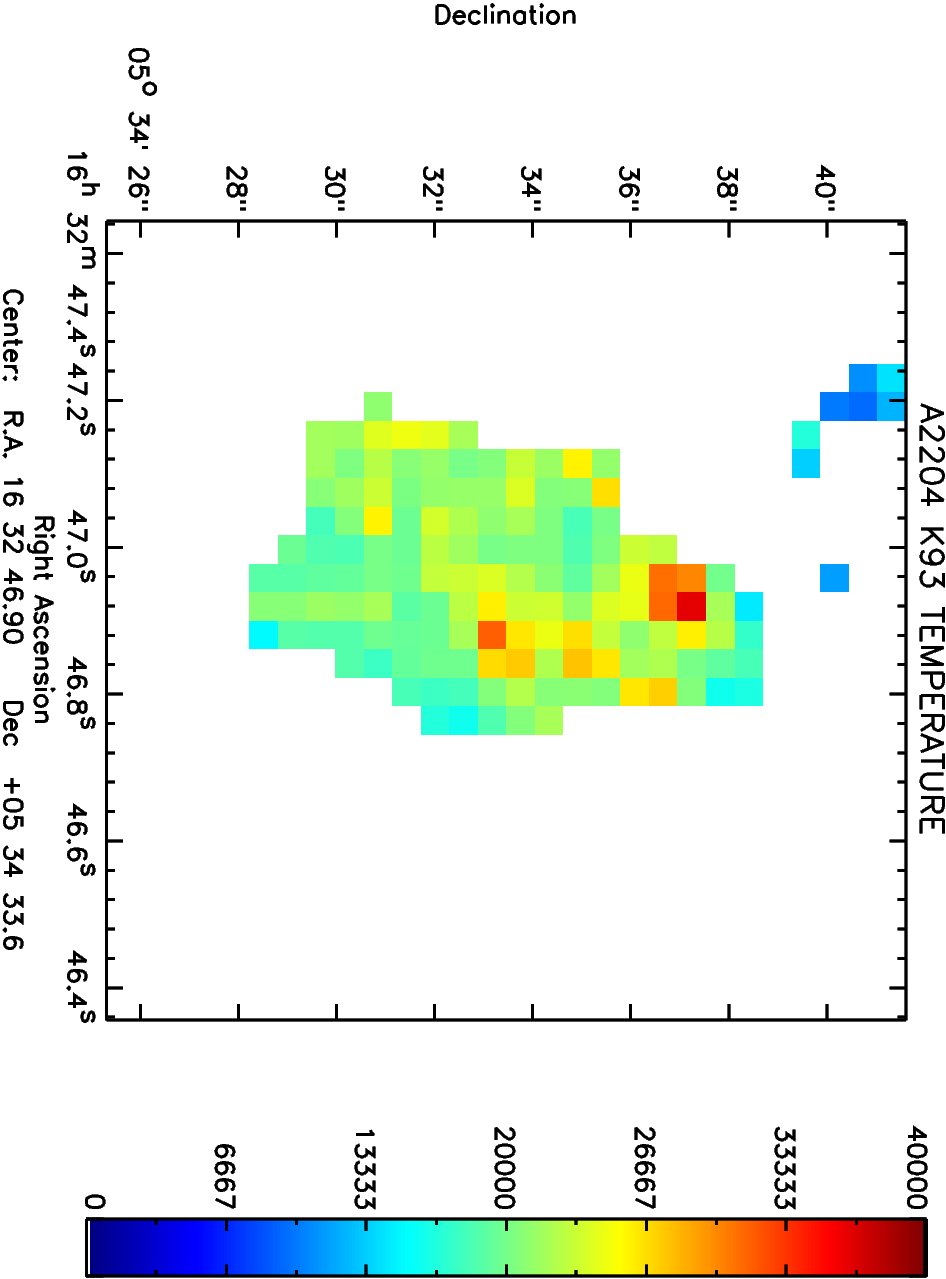}
    \includegraphics[width=0.36\textwidth, angle=90]{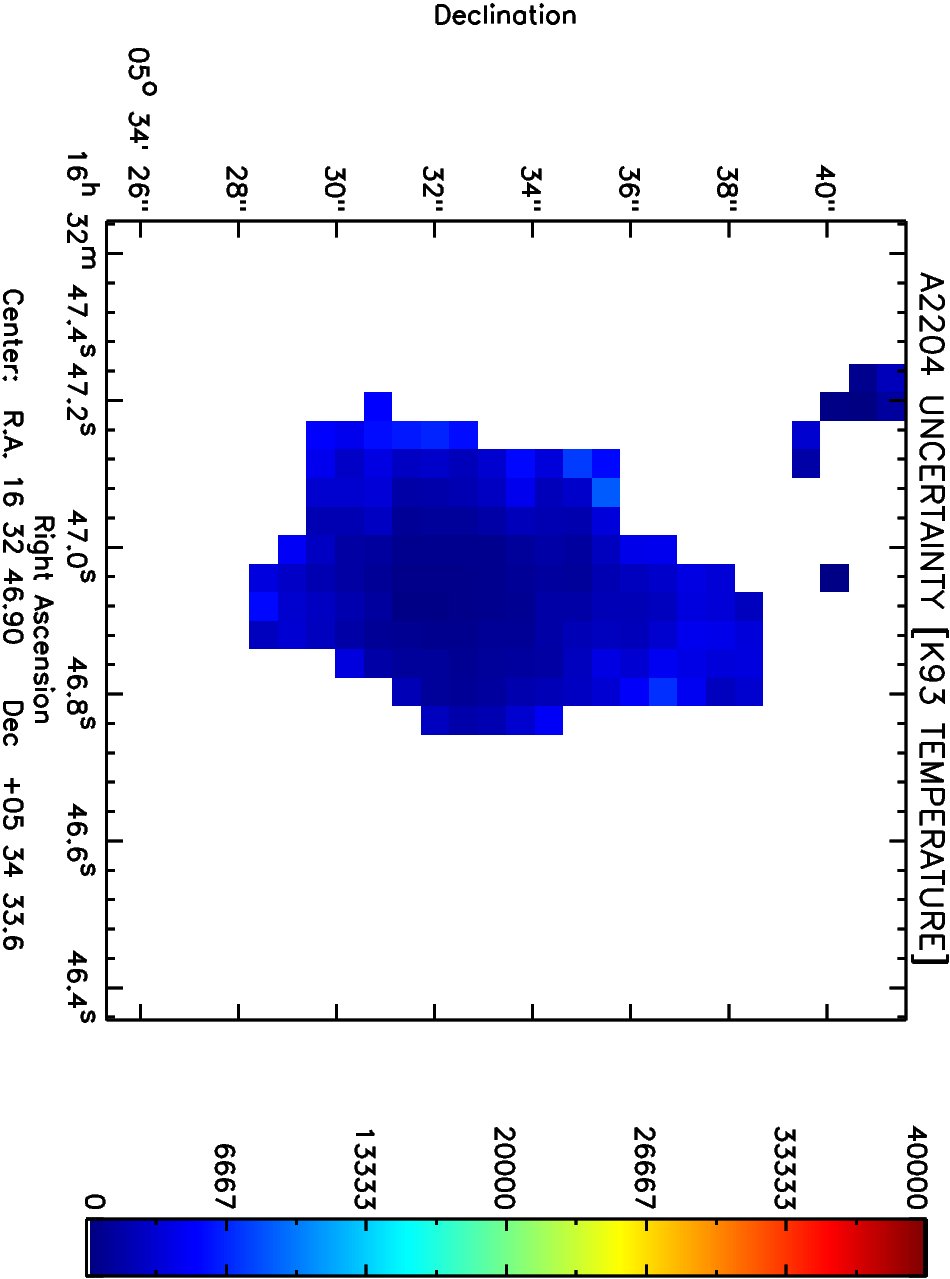}
  \vspace{0.5cm}
  \caption{FUV$_{\nu,exc}$/U$_{\nu,exc}$ ratio and K93 stellar temperature maps for A2204. (\textit{Top-Left}) FUV$_{\nu,exc}$/U$_{\nu,exc}$ ratio. (\textit{Top-Right}) Uncertainty in the FUV$_{\nu,exc}$/U$_{\nu,exc}$ ratio. (\textit{Bottom-Left}) K93 stellar temperature corresponding to the FUV$_{\nu,exc}$/U$_{\nu,exc}$ ratio. (\textit{Bottom-Right}) Uncertainty in K93 derived stellar temperature. Note that in calculating the K93 stellar temperature shown here we have only taken extinction due to the MW foreground into account, see Fig. \ref{f_bcg_bb_k93}. The data shown has been re-binned to 0.6$\times$0.6 arcsec$^{2}$ pixels.}\label{f_a2204_yngfu}
\end{figure*}

\begin{figure*}
\mbox{
    \includegraphics[width=0.6\textwidth, angle=90]{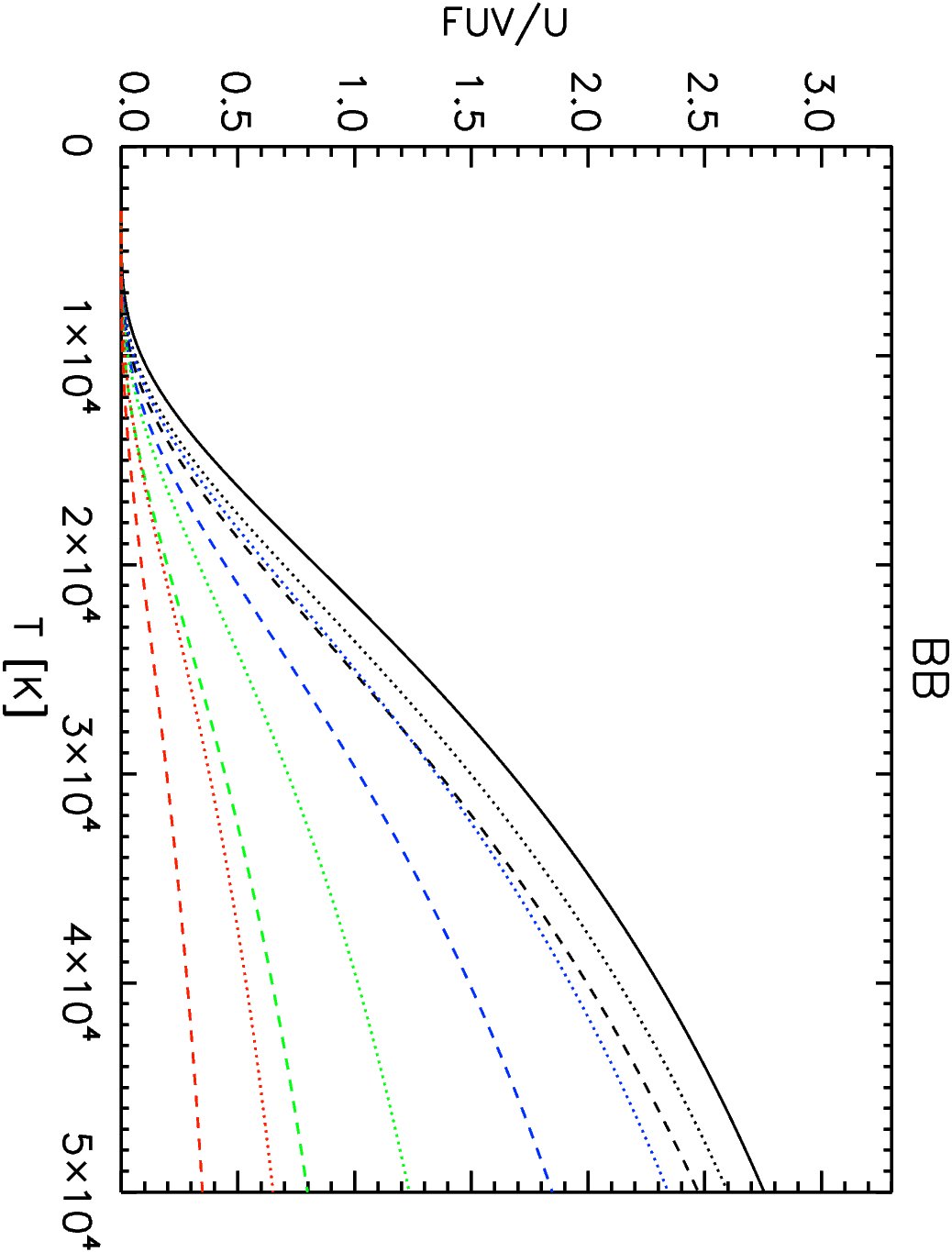}
}
\mbox{
    \includegraphics[width=0.6\textwidth, angle=90]{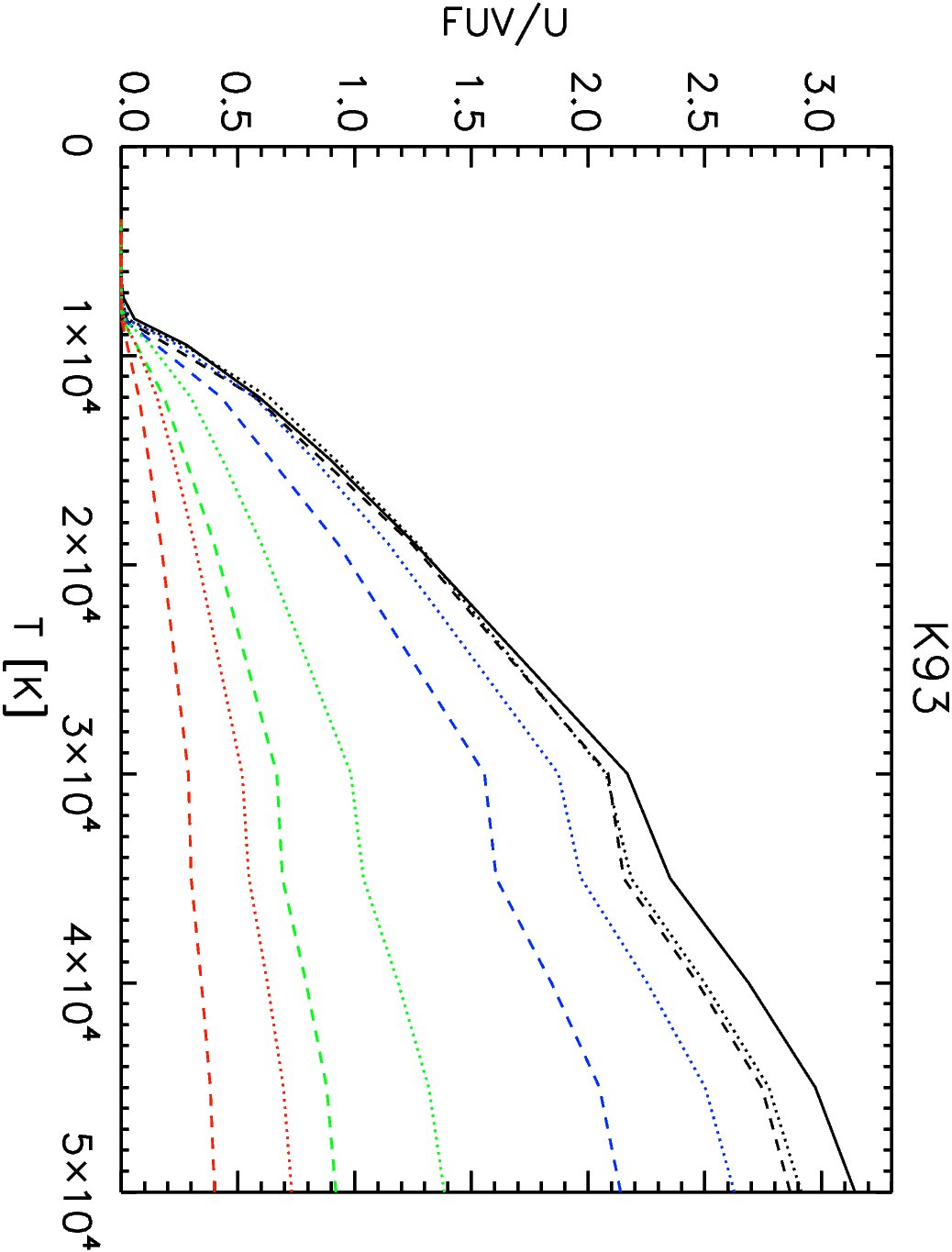}
}
  \vspace{0.5cm}
  \caption{FUV$_{\nu}$/U$_{\nu}$ ratio models for A2597 and A2204. Shown are BB models (\textit{top}) and K93 stellar models (\textit{bottom}). Curves are given for three redshifts z=0 (solid line), z=0.0821 (dotted line) and z=0.1517 (dashed line). For the latter two redshifts, corresponding to A2597 and A2204, we computed this flux ratio for varying amounts of extinction. The black curves refer to a dust-free environment. The blue curves correspond to extinction by the MW foreground only. The red curves assume that all of the extinction intrinsic to the BCG is in front of the FUV and U emitting regions. The green curves assume that only half of the extinction intrinsic to the BCG is in front of the FUV and U emitting regions. Dust extinction has been computed using the C89 extinction law for both the MW and the BCG component. The model ratios are computed for the HST-ACS/SBC F150LP and VLT-FORS U\_Bessel filters.}\label{f_bcg_bb_k93}
\end{figure*}

\begin{figure*}
    \includegraphics[width=0.60\textwidth, angle=90]{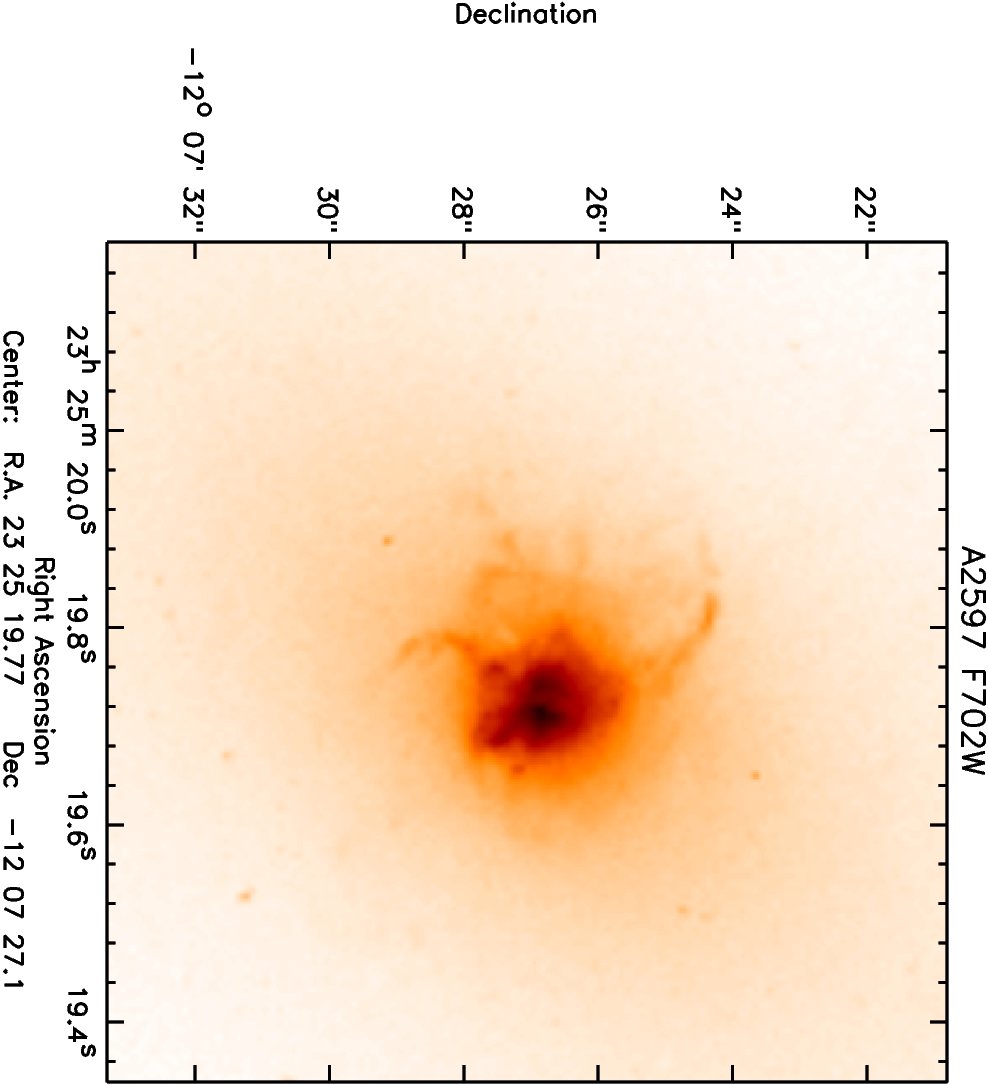}
    \includegraphics[width=0.60\textwidth, angle=90]{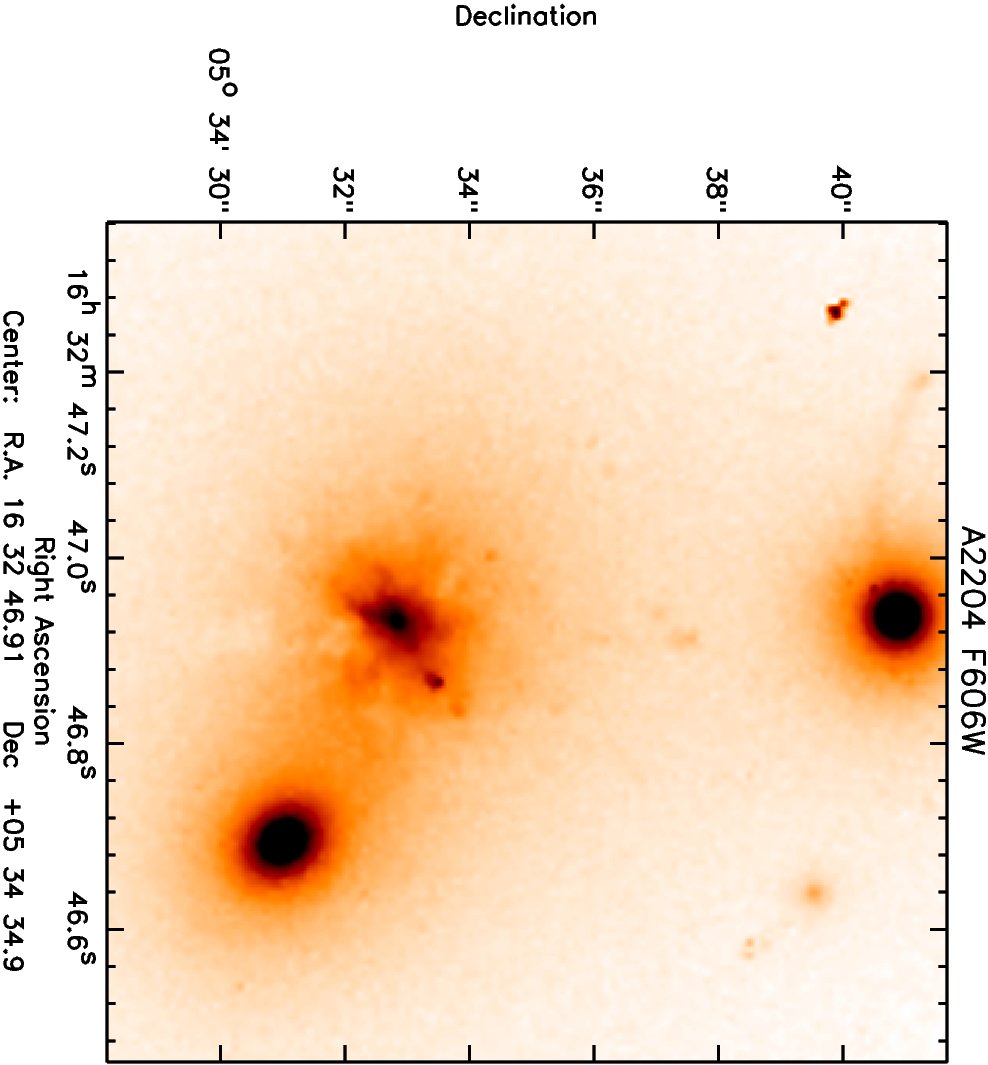}
  \vspace{0.5cm}
  \caption{HST optical imaging.  A2597 WFPC2 F702W image (\textit{Top}). A2204 WFPC2 F606W image (\textit{Bottom}). These images show that the BCGs have a disturbed morphology indicative of dust obscuration. Contamination by line emission also contributes to observed structures.}\label{f_hst_add}
\end{figure*}

Our optical spectra sample a small, but representative part of the FUV bright regions. Here we can rule out option (a) because we do not observe any regions with low extinction. Option (b) seems unlikely because the FUV to H$\alpha$ ratio would vary significantly if background HII regions exist where high Balmer decrements are observed but the FUV emission is highly extinguished. Some spread is observed in the FUV to H$\alpha$ emission, but generally they are found to be well correlated, see \citet{Od04} and Figs. \ref{f_a2597_ovl}, \ref{f_a2204_ovl} and \ref{f_pslit}.

While selection bias may have favored regions with low extinction in determining the FUV to U ratio, the FUV to H$\alpha$ ratio seems to indicate that this is not crucial in our analysis. Our optical spectra, ofcourse, only sample a subset of the regime where we have measured the FUV to U ratio. Two dimensional integral field observations of the relation between H$\alpha$/H$\beta$ versus FUV/H$\alpha$ could clarify this issue.

\section{Star formation}\label{s_starformation}
In the previous section we showed that very hot O-stars are able to explain to observed FUV and U emission. In this section we investigate whether these stars can power the ionised gas nebulae observed in H$\alpha$ line emission and what the FUV and H$\alpha$ implied star formation rates are.

\subsection{The H$\alpha$ nebula}\label{s_ha_nebula}
Optical emission line nebulae, usually observed in H$\alpha$ emission, are typical for cool-core clusters \citep[e.g.][]{He89,Cr99}. \citet{Ja05} show that both A2597 and A2204 contain large H$\alpha$ nebulae. Higher spatial resolution observations of the central few kpc in A2597 by \citet{Ko99} and \citet{Do00} show that the central ionised gas nebula is very clumpy and filamentary. If we assume that the FUV light has a stellar origin then we can ask ourselves whether the observed H$\alpha$ emission is quantitatively consistent with the same stellar origin.

The total FUV fluxes in 18 kpc radial apertures for A2597 and A2204 are given in Table \ref{t_laper}. Upon correcting for Milky Way foreground dust and distance we find luminosities L(FUV,A2597)~=~2.5~$\times$~10$^{28}$~erg~s$^{-1}$~Hz$^{-1}$ and L(FUV,A2204)~=~7.2~$\times$~10$^{28}$~erg~s$^{-1}$~Hz$^{-1}$. We then assume that the stars responsible for the FUV emission are on average hot O-stars and calculate their FUV flux. Here we will perform this calculation for an O3 star with a stellar radius 15~R$_{\odot}$. Such a star has a FUV flux F(F150LP,O3)~$\approx$~0.064~erg~s$^{-1}$~cm$^{-2}$~Hz$^{-1}$ in the F150LP band which then corresponds to a luminosity L(F150LP,O3)$\approx$8.74$\times$1-$^{23}$~erg~s$^{-1}$~Hz$^{-1}$. The observed FUV luminosities thus require about 30000 O3 stars for A2597 and about 80000 O3 stars for A2204.

Following \citet*{Be01} and \citet*{Va96} the total Lyman continuum (Ly$c$) flux of an O3 main sequence star is F(Ly$c$)=10$^{49.87}$~photons~s$^{-1}$ or equivalently a luminosity L(Ly$c$)=10$^{39.17}$~erg~s$^{-1}$. Assuming Case B downward conversion this yields L(H$\alpha$)=10$^{38}$~erg~s$^{-1}$. The O3 stars in A2597 and A2204 thus produce a H$\alpha$ luminosities L(O3-H$\alpha$,A2597)=3$\times$10$^{42}$~erg~s$^{-1}$ and L(O3-H$\alpha$,A2204)=8$\times$10$^{42}$~erg~s$^{-1}$. 

This is consistent with the H$\alpha$ luminosities measured by \citet{Ja05} for A2597 and A2204, i.e. L(H$\alpha$,A2597)=0.9$\times$10$^{42}$~erg~s$^{-1}$ and L(H$\alpha$,A2204)=2.8$\times$10$^{42}$~erg~s$^{-1}$. We conclude that if all the FUV light is due to hot O-stars, then the overall photon budget of these stars is sufficient to produce the observed H$\alpha$ emission. For A2597 this result is in agreement with earlier results by \citet{Od04}.

Only the Milky Way foreground extinction was taken into account in the above calculation. Dust extinction intrinsic to the BCGs is discussed in more detail in Sect. \ref{s_dust}. Increasing the amount of dust will increase the amount of extinguished FUV emission relative to H$\alpha$ emission and thus even more O-stars will be available to produce ionising photons. Not all of the total FUV flux can be attributed to young stars. Below we show that about 20 per cent of the FUV flux in F150LP is due to the old stellar population in A2597 and A2204. Such a small old star contribution to the FUV flux does not affect the above conclusion.

\subsection{FUV and H$\alpha$ star formation rates}\label{s_sfr}
The extinction uncertainty translates to a large uncertainty for the derived star formation rates (SFR) in these systems. In the case of A2597, using the C89 extinction law with a two component dust analysis, we find that the maximum fractional extinction at the wavelength of the FUV F150LP filter ($\lambda$=1612~$\rm \AA$) and the H$\alpha$ line ($\lambda$=6563~$\rm \AA$) is 28.7 and 2.9 in flux respectively. The minimum fractional extinction due to the Milky Way foreground is only 1.3 and 1.1 at the FUV and H$\alpha$ wavelengths respectively. 

Thus depending on how much of the dust is actually in front of the FUV and H$\alpha$ emitting structures means that the FUV derived star formation rate is SFR(FUV,A2597)=3.5-77.5~M$_{\odot}$~yr$^{-1}$ and the H$\alpha$ derived star formation rate is SFR(H$\alpha$,A2597)=7.0-18.6~M$_{\odot}$~yr$^{-1}$. Here we invoked the SFR relations discussed in \citet*{Bel01}. Similarly, for A2204 we find SFR(FUV,A2204)=10.2-476.9~M$_{\odot}$~yr$^{-1}$ and SFR(H$\alpha$,A2204)=22.1-75.4~M$_{\odot}$~yr$^{-1}$. 

The H$\alpha$ SFR range falls within the corresponding FUV SFR range, but both are poorly constrained due to the uncertainty in the dust distribution. Previous SFR estimates range from 4-12~M$_{\odot}$~yr$^{-1}$ for A2597 \citep*{Mc93,Od04,Do07} and 15~M$_{\odot}$~yr$^{-1}$ for A2204 \citep{Od08}. SFRs deduced by different methods are not expected to agree as they trace different physical processes, regions and timescales in the overall star formation scheme. A recent overview of the different star formation rates estimated for A2597 and a discussion of them can be found in \citet{Od08}.

In Section \ref{s_res_rx} we showed deep radio maps of our BCGs revealing interesting new structures. Here we will briefly discuss the northern 3~sigma feature in the 8~GHz observations of A2204. Although it is possible that this feature is simply related to a background radio source it is interesting that its location agrees with the bright northern FUV knot. Assuming for now that the radio emission in this knot is due to synchrotron emission from supernova explosions we can estimate a SFR by invoking the far-infrared radio correlation \citep*{Co91} and the SFR relations discussed in \citet*{Bel01}. 

The peak emission of the radio feature is 82~$\pm$~20~$\mu$Jy and it has an integrated flux of 110~$\pm$~42~$\mu$Jy at 8~GHz. Assigning a spectral index $\alpha$=-0.5 (F$\sim\nu^{\alpha}$) to this synchrotron emission we find a 1.49~GHz radio luminosity 1.4$\times$10$^{22}$~W~Hz$^{-1}$. Applying the above mentioned relations this translates to a SFR(radio,n-knot)=5.8~M$_{\odot}$~yr$^{-1}$. This is much higher than the FUV implied SFR i.e. SFR(FUV)=0.2~M$_{\odot}$~yr$^{-1}$. This difference can be accounted for by extinction if most of the dust lies in front of the emitting area. The radio emission can not be associated with free-free emission from young O-stars as this would require far too many of such stars. 

\subsection{Dust and gas mass estimates from A$\rm_{V}$}\label{s_gdmass}
Under the assumption of MW type dust, case B recombination and that the dust is optically thin everywhere we can derive rough gas and dust mass estimates from the amount of visual A$_{V}$ extinction. In the diffuse interstellar medium of the Milky Way \citep*{Bo78} find N(H)/A$\rm_{V}$=1.9$\times$10$^{21}$~atoms~cm$^{-2}$~mag$^{-1}$ where N(H)=N(HI+H$\rm_{2}$) is the number of nucleons.

We find A$\rm_{V,BCG}$=1.3 for A$\rm_{V,BCG}$=1.6 over the central 10$\times$10~kpc$^{2}$ regions in A2597 and A2204 respectively. Integrating N(H) over this region and multiplying by two to account for the dust on the far side of the galaxy we find M(H,A2597)=4.0$\times$10$^{9}$~M$_{\odot}$ and M(H,A2204)=4.8$\times$10$^{9}$~M$_{\odot}$. Our total gas mass estimates for the central region of A2597 are about an order of magnitude higher than the atomic gas mass estimated by \citet{Od94} for this region. \citet{Ed01} indeed find a much higher molecular gas mass (M(H$\rm_{2}$,A2597)$\sim$10$^{10}$~M$_{\odot}$), but the lower spatial resolution of this measurement means that it probes a much larger region and makes it difficult to compare it to the other measurements above. All these gas mass estimates suffer from systematic uncertainties and further observational constraints are necessary to refine them. There are no cold gas mass measurements published for A2204.

\citet{Ed10} show that the global gas to dust ratio for BCGs is similar to the MW average i.e. M(H)/M(dust)=140. This would imply that M(dust,A2597)=1.4$\times$10$^{7}$~M$_{\odot}$ and M(dust,A2204)=1.7$\times$10$^{7}$~M$_{\odot}$. For A2597 this dust mass estimate is in good agreement with the value deduced from far-infrared emission by \citet{Ed10}. If the dust is optically thick in some places in the BCG then the above calculated masses are lower limits for the true gas and dust masses. There are no dust mass measurements published for A2204.

If we assume constant star formation at a rate of 10$^{1-2}$~M$_{\odot}$~yr$^{-1}$ then the gas masses calculated by us above imply a gas depletion time of about 10$^{7-8}$~yr. The (residual) mass deposition rates allowed for by X-ray spectroscopy is similar to the range in star formation rates observed here and may thus be able to sustain star formation over longer time scales \citep[e.g.][]{Pe06,Pe98}.

\section{Excess FUV/U: Non-stellar Origin ?}\label{s_non_stellar}
In the previous sections we have showed that young, hot stars can explain the observations, but only if the dust content and nebular continuum emission is low enough. Especially in A2597 it is difficult, but not impossible, to reconcile the observed FUV$_{\nu}$/U$_{\nu}$ ratio with stars alone. Below we show that the common non-stellar processes that are known to take place in cool-core clusters can not provide an alternative explanation for the observations. Stars thus remain as the only evident option.

\subsection{Active Galactic Nuclei}\label{s_agn}
Using models by \citet{Na01} we have investigated the FUV$_{\nu}$/U$_{\nu}$ ratio for AGN emission. These models correspond to F$_{\nu}\sim \nu^{-0.5}$ in FUV-Optical wavelength range. This means that this model predicts a FUV$_{\nu}$/U$_{\nu}$ ratio which is always less than unity. Adding the additional blackbody, infrared and X-ray terms in equation (2) by \citet{Na01} does not change this. We thus conclude that the observed FUV$_{\nu,exc}$/U$_{\nu,exc}$ ratio in A2597 and A2204 can not be explained by an AGN.

\subsection{Non-thermal processes}\label{s_nontherm}
A significant contribution to the U band excess light by synchrotron emission and/or scattering of light by dust or hot electrons is ruled out by \citet{Mc99} using U-band polarisation studies. Since the FUV and U excess light trace each other well we conclude that neither synchrotron emission nor scattering contributes significantly to the excess FUV and U emission. 

Bremsstrahlung is also ruled out because under optically thin conditions it produces a spectrum that is flat or decreases slowly in flux with increasing frequency in the FUV to optical wavelength range. Only in optically thick conditions can bremsstrahlung produce a spectrum that increases with frequency. In this case the emergent intensity will obey the Rayleigh-Jeans limit of the Planck black body function i.e. F$_{\nu} \propto \nu^{2}$. However, this can also be ruled out as the integrated bremsstrahlung flux measured with ROSAT (0.1-2.4~keV) for A2597 and A2204 is an order of magnitude below the measured FUV fluxes here \citep{Br94,Eb98}.  

\section[]{Discussion}\label{s_discuss}
Below we will shortly discuss our results. In the first two sections we will compare our results to two previous investigations into the FUV to optical excess emission in cool-core BCGs. In the third section we discuss the extinction law and in the final section we compare the results for A2597 and A2204.

\subsection{Crawford \& Fabian 1993}\label{s_cf93}
First we will discuss the CF93 results. They use measurements from the International Ultraviolet Explorer (IUE) and combine it with optical spectroscopy to study the excess emission from 1300-5000~$\rm \AA$ in 4 BCGs. Their spectral coverage of the blue emission in BCGs is wider and better sampled than our investigation. However, they do not have spatial information, being limited by the 10$\times$20~arcsec$^{2}$ IUE aperture. Their analysis is similar to ours in that dust, the old stellar population and line emission is taken into account. For the dust correction they use the \citet*{Se79} average MW extinction curve, which is similar to the C89 curve used by us (see Fig. 4 in C89). Their correction for the old stellar population is based on a template model derived from optical spectroscopy of the Abell 4059 BCG and extending it into the UV.

Their main result is that they find that the excess blue emission in BCGs, even if they allow for reddening by dust, can be explained with star formation dominated by B5 stars i.e. $\sim$15000~K. Their results are based on the Kurucz 1979 stellar models which are identical to the Kurucz 1993 models used by us.

Considering an aperture with 7 arcsec radius centered on the A2597 BCG we compute the FUV$_{\nu,tot}$/U$_{\nu,tot}$ ratio from our data and correct it only for foreground MW dust. We then find FUV$_{\nu,tot}$/U$_{\nu,tot}$$\approx$0.7. This corresponds to a K93 stellar model temperature of 14000~K. This means that at low dust extinction we are in reasonable agreement with the CF93 result for A2597. However, our results are no longer consistent with CF93 after correcting for the dust intrinsic to the BCG and the emission contributed by old stars. There are a number of reasons for this difference. 

First CF93 assume an extinction of A$\rm_{V,BCG}$=0.3 based on Ly$\alpha$/H$\alpha$ results by \citet*{Hu92}. We use A$\rm_{V,BCG}$=1.3 based on optical spectroscopy of the Balmer decrements. An additional magnitude of intrinsic extinction at visual wavelength as shown above has dramatic effects on the FUV/U ratio. We will discuss the extinction law in more detail below. The second important difference concerns the ratio of excess to total emission in the U-band. CF93 find that this ratio is 0.55 at 3500~$\rm \AA$, whereas we find that this ratio is 0.17 in the U-band (3600~$\rm \AA$). This is a large difference and means that either; (i) the old stellar population in A2597 differs significantly from the old stellar population template used by CF93, or (ii) there is significant excess emission contributing to the total light in our V-band observations. 

To test the second option we have also computed the FUV$_{\nu,exc}$/U$_{\nu,exc}$ ratio by using the R and I band data, which should contain less excess emission as compared to the V-band. Contamination due to line emission was set to 12 per-cent and 7 per-cent for the R and I band respectively. In these numbers we have included contamination due to the SII 6716~$\rm \AA$ and 6731~$\rm \AA$ lines that were not included in our spectra, but were measured by \citet*{Vo97}. We find no significant difference in the FUV$_{\nu}$/U$_{\nu}$ ratio when either the V, R or I band is used to remove the old stellar population, see Fig. \ref{f_add_rat}. We thus conclude the old stellar population in A2597 is different from the template used by CF93.

A further uncertainty in the CF93 work concerns the large aperture mismatch between their optical and UV data which requires them to invoke large, uncertain, scaling factors that can affect their results. We also note that their 1600~$\rm \AA$ FUV value for A2597 is about 25 per-cent larger than predicted by their B5 model value. The CF93 FUV 1600~$\rm \AA$ value would thus allows for higher stellar temperatures than B5 as indicated by our result. Taking into account the differences due the dust, old stellar population model and their FUV 1600~$\rm \AA$ flux value then our results are in rough agreement with CF93. 

CF93 discuss that the blue excess emission can be equivalently well described by a power-law F$_{\nu}\sim\nu^{\alpha}$ with $\alpha$=[-0.5:0.5]. As noted by CF93 this is no surprise because hot stars can be approximated by a power law with this slope in the FUV-optical regime. \citet{Mc93} showed evidence for the blue excess emission to be correlated with the AGN radio lobes in two objects a.o. A2597. From our higher resolution data we find that although the FUV emission is roughly extended along the radio axis, that there is no detailed correlation between the radio and the FUV for A2597. In Sections \ref{s_agn} and \ref{s_nontherm} we already discussed that the AGN and non-thermal processes can be ruled out for our observations of blue excess emission in A2597. Stars thus remain as the most plausible option.

\subsection{Hicks et al. 2010}\label{s_hi10}
\citet{Hi10} (hereafter H10) recently presented a BCG sample observed with GALEX in the FUV ($\sim$1500~$\rm \AA$) and the NUV ($\sim$2500~$\rm \AA$). In their Table 2 they give FUV and NUV photometry for 16 BCGs within 7 arcsec radial apertures centered on the BCG nuclei. These numbers have been corrected for foreground MW extinction but not for the dust intrinsic to the BCG or for emission due to the old stellar population. These apertures correspond on average to the central $\sim$10 kpc of the BCG which is similar to the area investigated by us.

We have computed the temperature corresponding to tabulated FUV$_{\nu,tot}$/NUV$_{\nu,tot}$ ratio in H10 in the same way as done above for our FUV$_{\nu}$/U$_{\nu}$ ratios using the K93 stellar models. The results are given in Table \ref{t_hicks}. Contamination of the FUV band by Ly$\alpha$ emission for systems with z$>$0.1 has not been taken into account and if present would lead to an overestimate of the stellar temperature estimated by us.

The Ly$\alpha$ line is not a problem for A2597 and the H10 FUV$_{\nu,tot}$/NUV$_{\nu,tot}$ ratio implies a K93 stellar temperature 14000~K. This value, being uncorrected for intrinsic dust and the old stellar population, is in good agreement with our results and the results by CF93, see Section \ref{s_cf93}.

H10 also perform a correction for emission contributed by the old stellar population. This correction is based on the average UV-J band colors measured for a control sample of elliptical galaxies measured in the same bands with GALEX. The corrected FUV, NUV excess emission is tabulated in their table 7. Interestingly the excess FUV$_{\nu,exc}$/NUV$_{\nu,exc}$ ratios do not increase significantly with respect their corresponding total FUV$_{\nu,tot}$/NUV$_{\nu,tot}$ ratios for most BCGs in their sample, including A2597.

\begin{figure*}
    \includegraphics[width=0.36\textwidth, angle=90]{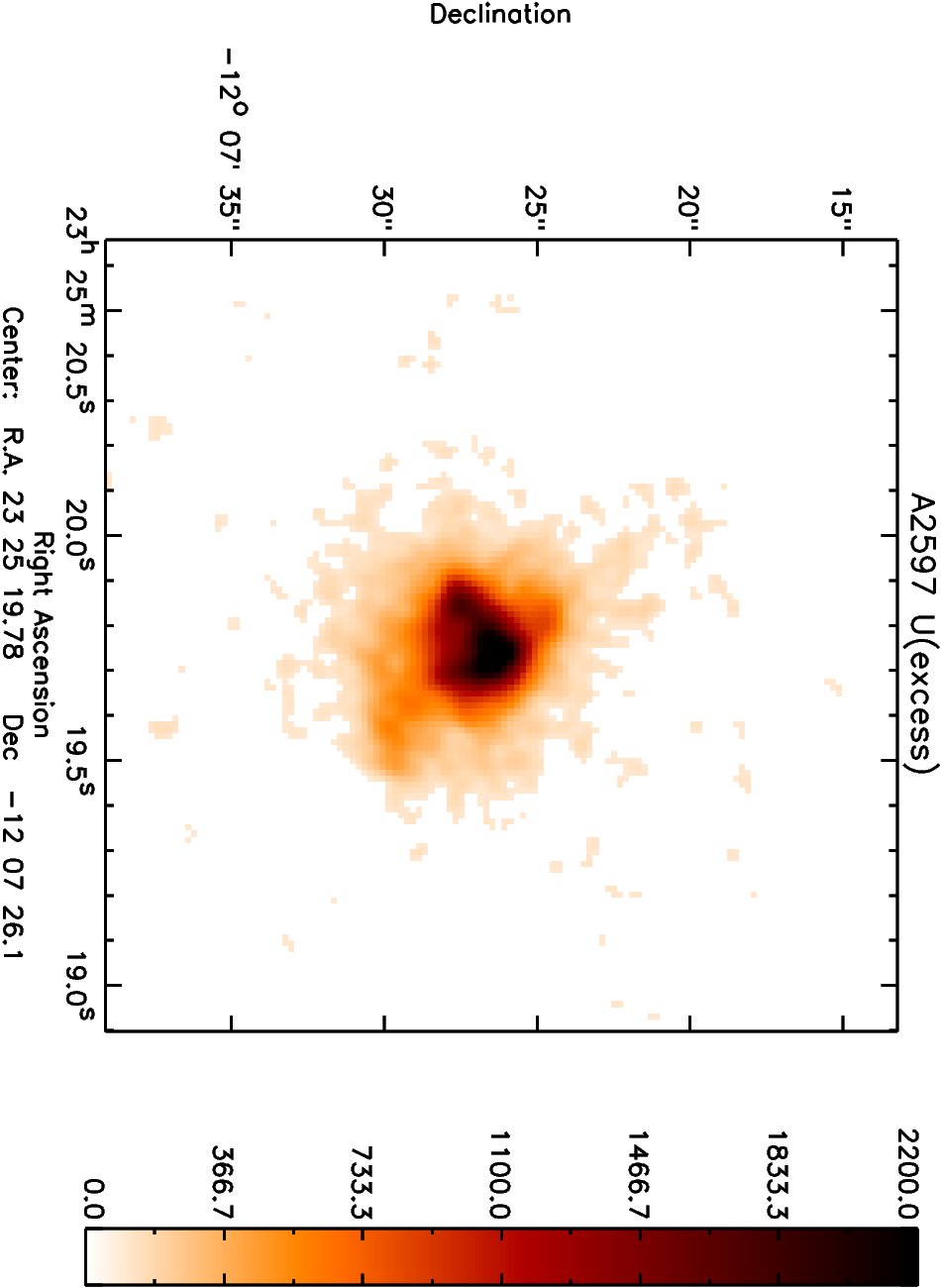}
    \includegraphics[width=0.36\textwidth, angle=90]{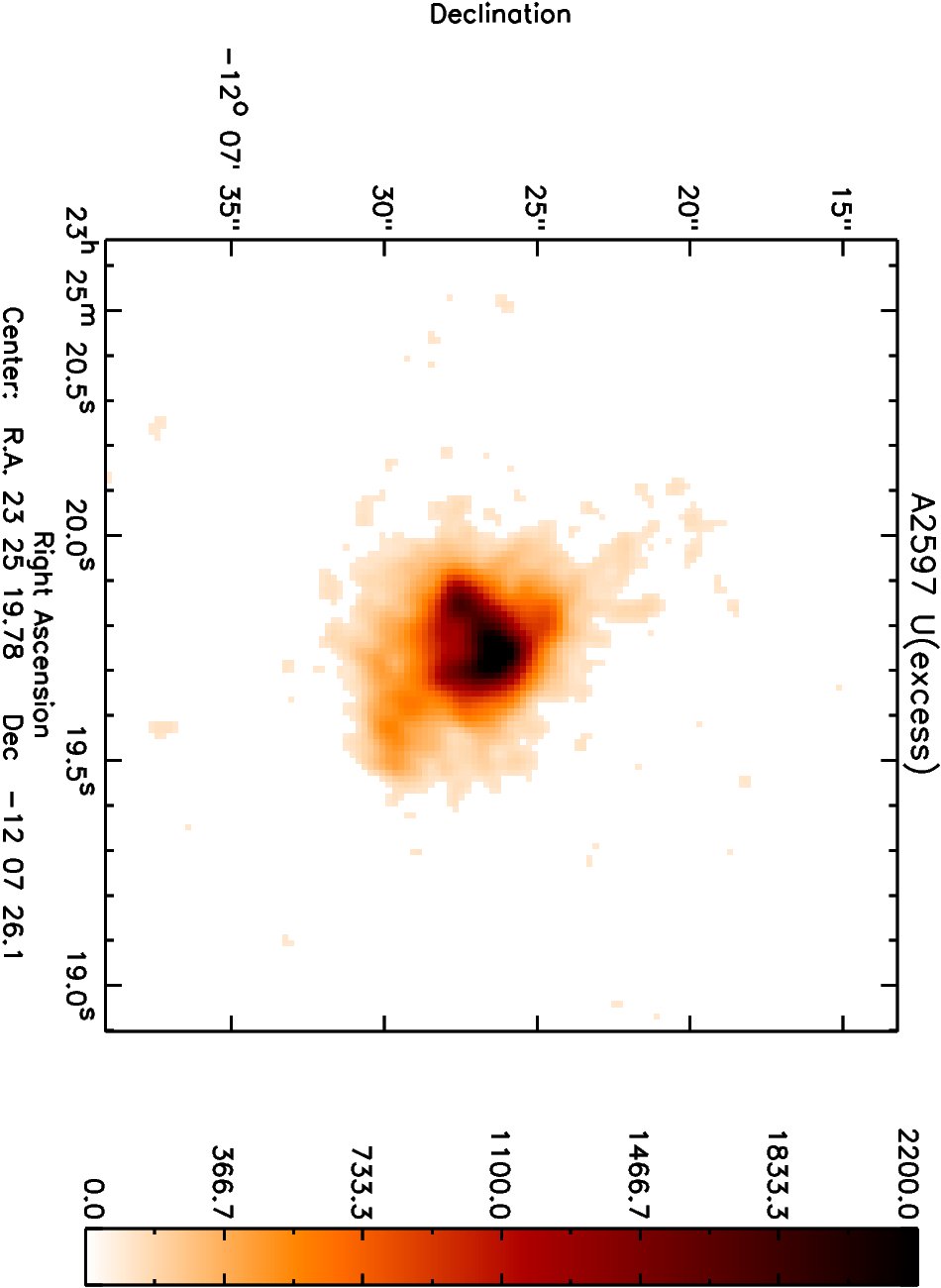}
    \includegraphics[width=0.36\textwidth, angle=90]{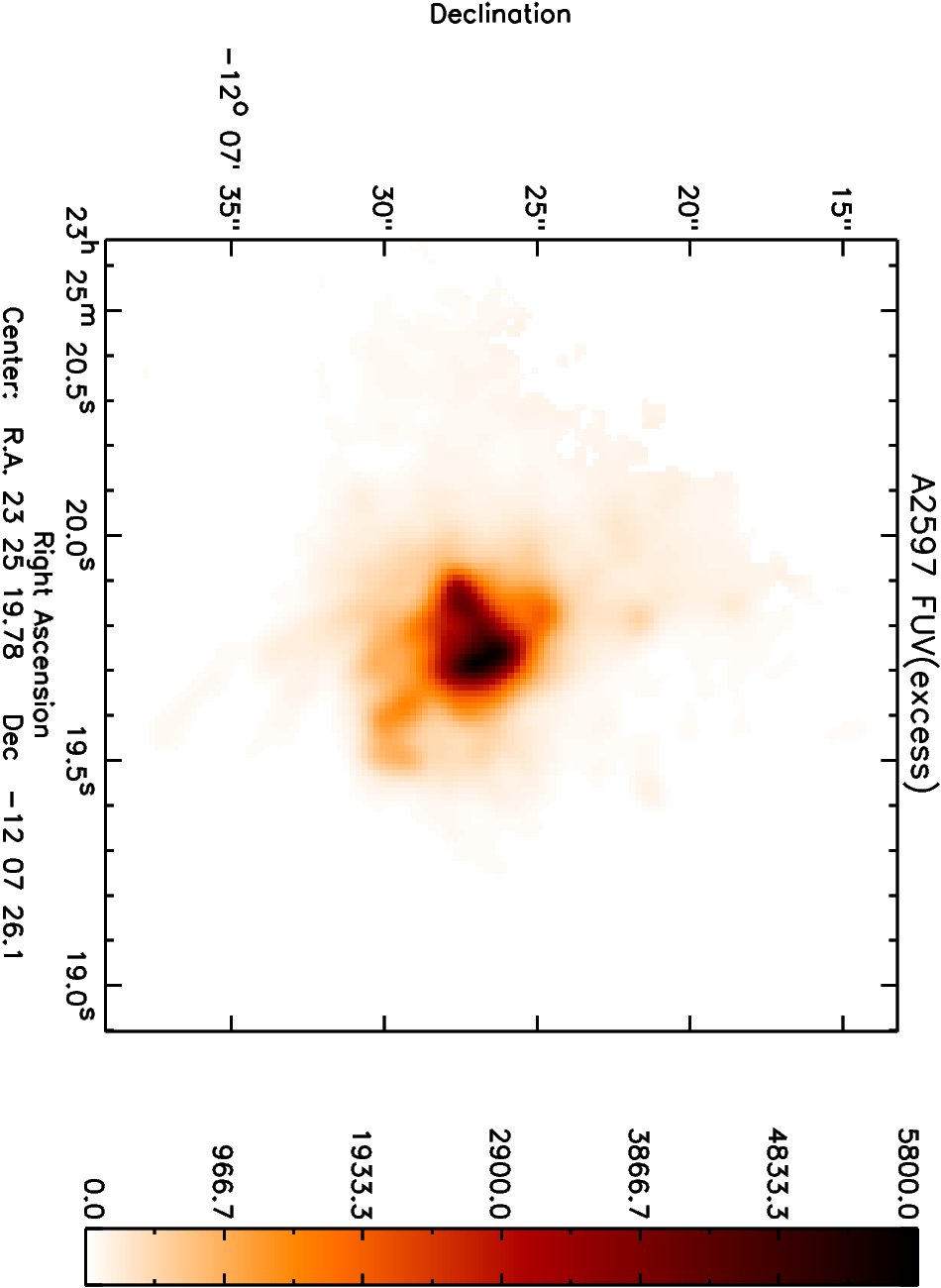}
    \includegraphics[width=0.36\textwidth, angle=90]{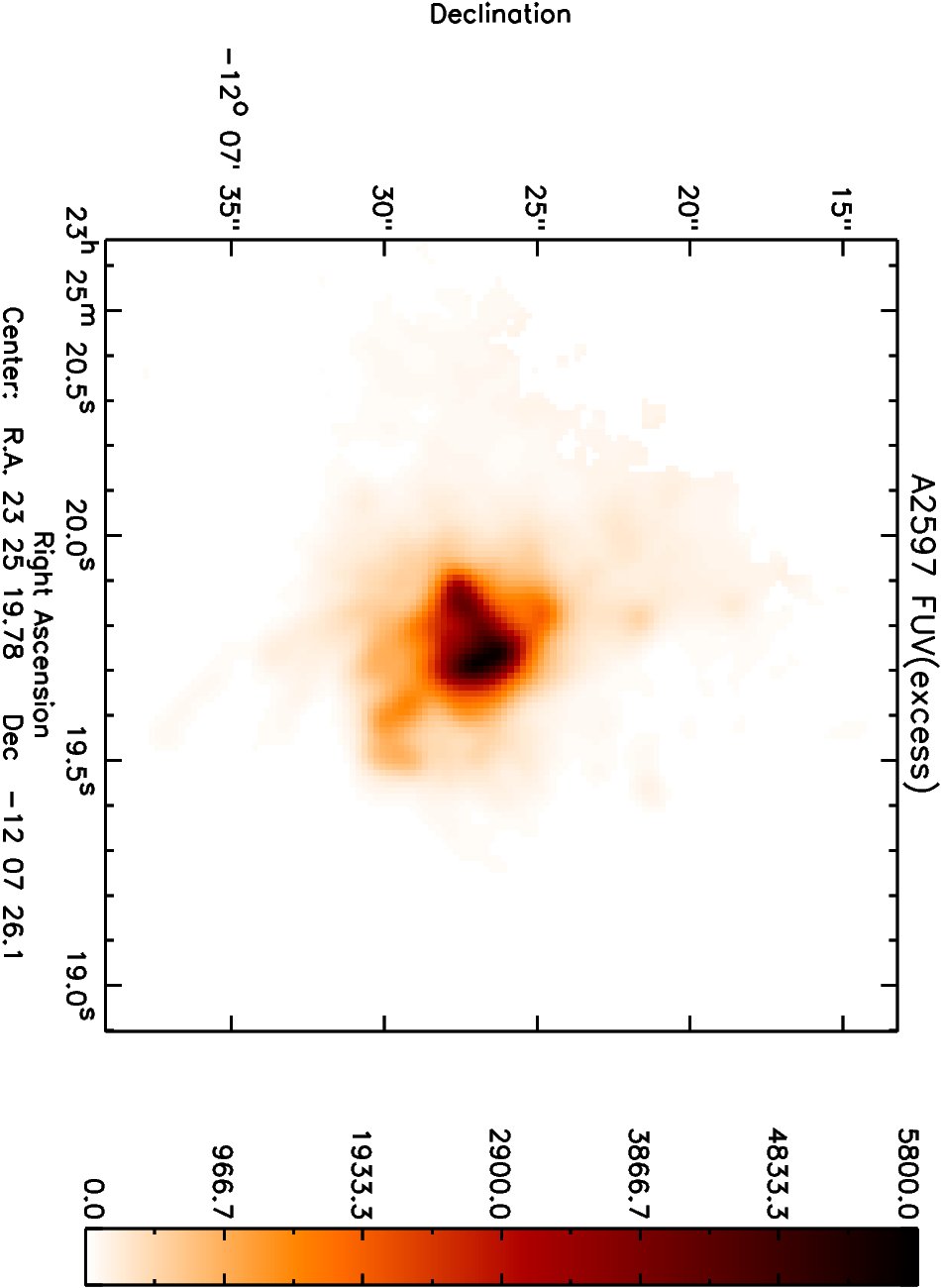}
    \includegraphics[width=0.36\textwidth, angle=90]{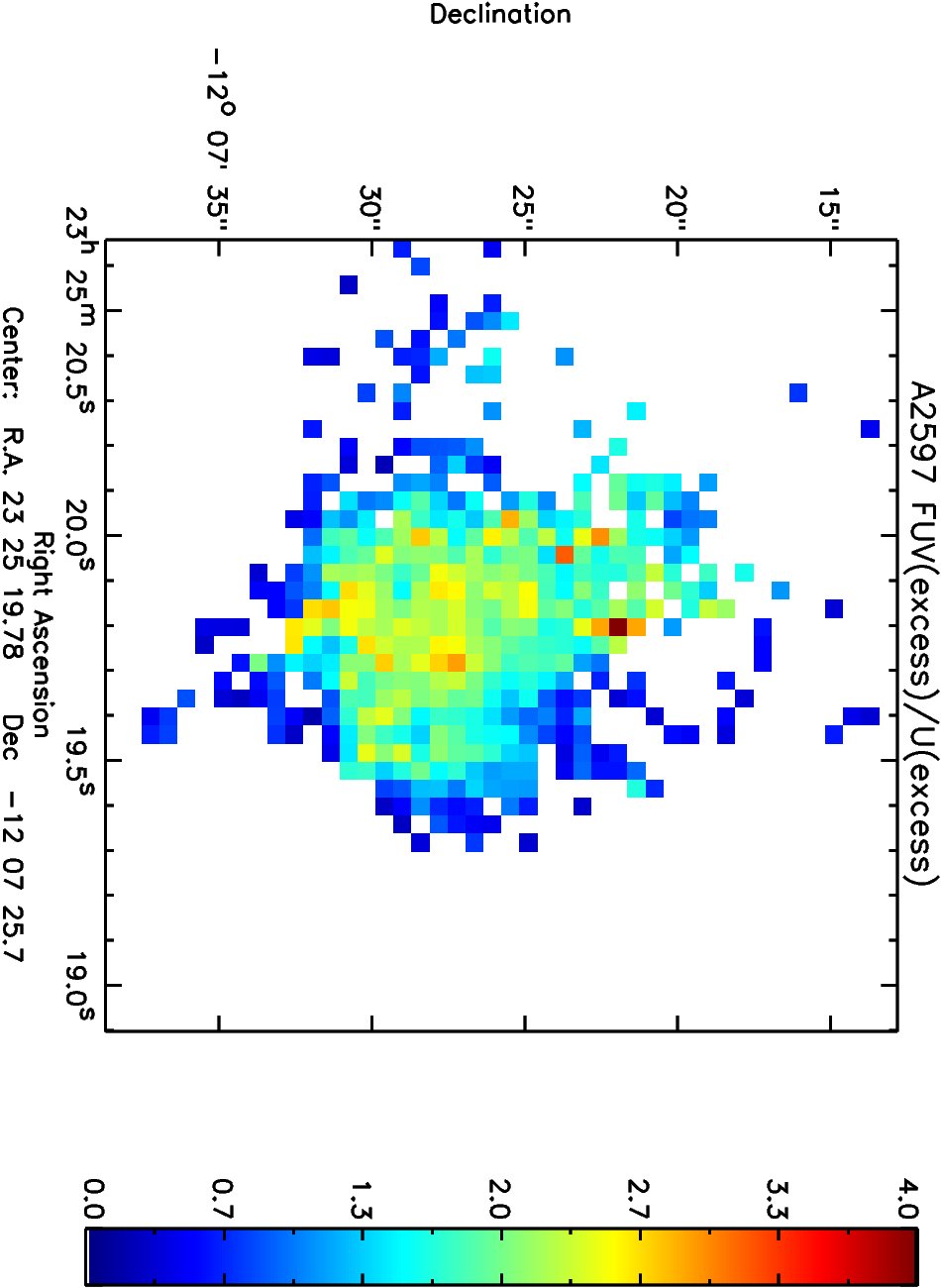}
    \includegraphics[width=0.36\textwidth, angle=90]{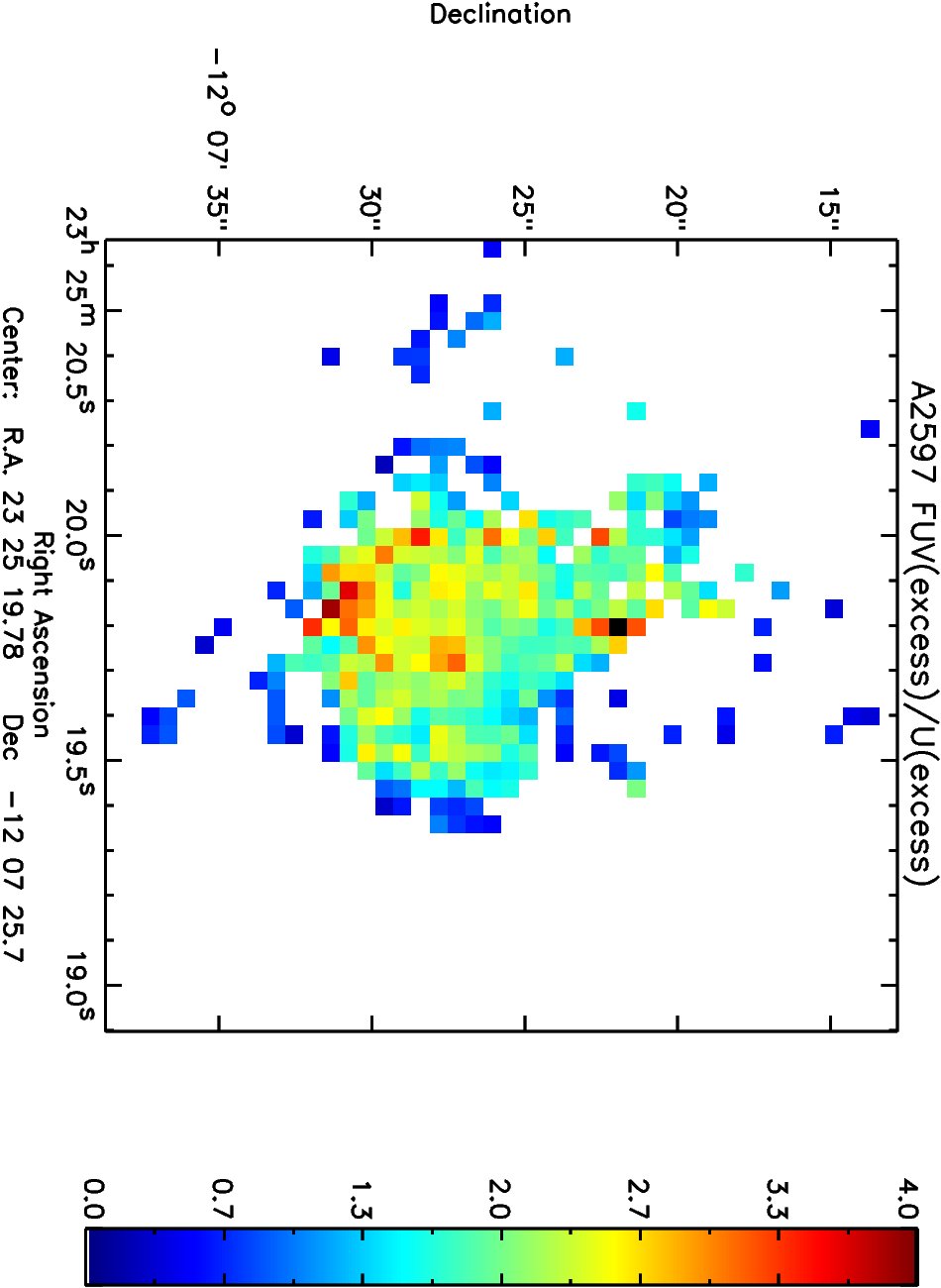}
  \vspace{0.5cm}
  \caption{Using the R (\textit{Left}) and I (\textit{Right}) band images to remove the old stellar population in A2597. (\textit{Top-Left}) Excess U$_{\nu,exc}$ emission when using the R band image to remove emission from old stars. (\textit{Top-Right}) Excess U$_{\nu,exc}$ emission when using the I band image to remove emission from old stars. (\textit{Middle-Left}) Excess FUV$_{\nu,exc}$ emission when using the R band image to remove emission from old stars. (\textit{Middle-Right}) Excess FUV$_{\nu,exc}$ emission when using the I band image to remove emission from old stars. (\textit{Bottom-Left}) FUV$_{\nu,exc}$/U$_{\nu,exc}$ ratio when the R band image is used to remove the old stellar population. (\textit{Bottom-Right}) FUV$_{\nu,exc}$/U$_{\nu,exc}$ ratio when the I band image is used to remove the old stellar population. The bottom images show that the final FUV to U excess ratio does not depend on whether the V, R or I band image is used to remove the old stellar population. This shows that neither of these bands contains a significant contribution by excess emission and that all three are an adequate representation of the old stellar population.}\label{f_add_rat}
\end{figure*}

We do find a significant increase in the FUV$_{\nu,exc}$/U$_{\nu,exc}$ ratio as compared to the FUV$_{\nu,tot}$/U$_{\nu,tot}$ ratio which appears to be in disagreement with H10 result for this particular system. We believe that our results are correct because (i) we obtain the correction for the old stellar population from the BCG itself and (ii) the correction does not depend on which band (V, R or I) is used, see Section \ref{s_cf93} and Fig. \ref{f_add_rat}. This would indicate that the old stellar population in A2597 differs from the template used in H10. The average old star correction performed by H10 is uncertain for individual systems because the UV-J colors of BCGs have a significant scatter, particularly in the FUV band (priv. comm. A. Hicks, M. Donahue).

\subsection{The extinction law in BCGs}
An average MW type extinction law certainly does not apply to all galaxies. Detailed observations of (i) different sightlines within the Milky Way, (ii) local group galaxies and (iii) luminous infrared galaxies show significant variations in the extinction law \citep[e.g. C89;][]{Ca00}. This is particularly true for emission in the UV regime. Currently we do not know what the detailed shape of the extinction law in BCGs is. This lack of information is the most severe limitation in interpreting the FUV-optical excess emission presented here. However, even without knowing the exact shape of the extinction law we have obtained interesting limits on the origin of this emission in the previous sections. In this section we will discuss the information that we have on the extinction law in BCGs in a bit more detail.

In this work we have used deep, optical $\sim$3500-7500~$\rm \AA$ long-slit spectroscopy to determine the amount of extinction and the shape of the extinction law from the Balmer decrements. The reason for using this data set is because of the large wavelength coverage and thus being able to measure the Balmer ratios under the same observing conditions and within the same slit observation. We have shown that these ratios can be described by an average MW type extinction law such as described by C89, with an A$_{V}\sim$1-2, for A2597 and A2204. 

The Balmer lines are optically thin and thus we do not know how much of the dust implied by these lines is in front of our line of sight to the FUV and U emitting regions. Star formation is usually accompanied by dust. Thus if young stars are responsible for the observed excess emission then a significant amount of the total extinction should be associated with these stars, see also the discussion in Section \ref{s_dust}.

Including MW type dust in our analysis of the excess emission in these BCGs has profound affects on the nature of this emission. If at least half of the dust is in front of the emitting regions, then stellar temperatures above 50000~K are required to explain the observed excess emission in the central 20~kpc regions of A2597 and A2204. Such high temperatures can not be accounted for by normal O-stars. It is unlikely that exotic, hotter stars such as Wolf-Rayet stars are present in sufficient amounts to dominate the observed excess emission over such large areas. White dwarfs are an option but these will be distributed in the same way as the old stellar population and as such should not contribute to the excess emission.   

Measurements of the Ly~$\alpha$ to H$\alpha$ ratio appear to contradict the dust properties inferred from the optical Balmer ratios. \citet*{Hu92} and \citet{Od04} find Ly~$\alpha$/H$\alpha\sim$6 for A2597. A simple calculation for A2597 based on eq. 7.4 and the standard R=3.1 MW extinction law tabulated in Table 7.1 in \citet*{Os06} shows that this can not be in agreement with our optical Balmer results. Invoking our measured value of H$\alpha$/H$\beta$=4.5 in A2597 and assuming that the un-extincted value of this ratio is 2.87 we find that the constant c in eq. 7.4 is 0.56. Applying this value of c and the R=3.1 MW extinction law to the dust-free Ly~$\alpha$/H$\alpha\approx$13 ratio \citet*{Fe86} we find that the measured value of this ratio in A2597 should be 0.4. This strongly disagrees with the measurements by \citet{Od04}, which by itself would imply c$\approx$0.125 for a R=3.1 MW extinction law.

The C89 extinction law used in the work above corresponds to the standard R=3.1 MW extinction law. It would seem that applying this extinction law in the manner described above overestimates the Ly~$\alpha$/H$\alpha$ ratio by a factor 15. However, there are a number of possible explanations for this discrepancy.

It is possible that there is a selection effect in that the observed Ly $\alpha$ emission is weighted towards regions of lower extinction. We have discussed a similar case of this selection effect, involving the FUV emission and the Balmer lines, in Section \ref{s_dust}. The same arguments apply here, (i) our optical spectroscopy does not show evidence for inhomogeneous extinction, and (ii) there is a good correlation between Ly~$\alpha$ and H$\alpha$ emission \citep{Od04}. This indicates to us that both come from the same regions and are subject to the same extinction.

Another option to explain the observed Ly~$\alpha$/H$\alpha$ ratio is that the shape of the extinction curve for the dust intrinsic to the BCG is different from the average MW curve. We experimented with a much flatter R=5.5 extinction curve, such as observed along the line of sight towards Orion nebula \citep[e.g.][]{Os06}. This curve, with c$\approx$0.56, is able to explain the observed Ly~$\alpha$/H$\alpha$ ratio and also gives a reasonable fit for all Balmer ratios except H$\alpha$/H$\beta$ which would have to be 3.9 instead of the measured 4.5. 

The standard R=3.1 MW type extinction law, with c=0.56 i.e. assuming all dust is in front of the emitting regions, implies that the extinction in the FUV(1600~$\rm \AA$) band is a factor 4 and 10 higher than in the U(3600~$\rm \AA$) and H$\alpha$(6563~$\rm \AA$) bands respectively. However, a R=5.5 extinction law with the same amount of extinction (i.e. c=0.56) would imply that the extinction for the FUV(1600~$\rm \AA$) and U(3600~$\rm \AA$) bands are practically the same and that the FUV(1600~$\rm \AA$) would be a factor 2 more extincted than the H$\alpha$ emission. Lowering the relative extinction between the FUV and U band also means that it is possible to reconcile the observed excess emission with normal stars and that the range in allowed star formation rates decreases.

Finally it is well known that stars alone are not able to explain all the details of the emission spectra observed from BCGs in cool-core clusters \citep[e.g.][]{Vo97,Ja05}. In particular the strong neutral atomic lines (e.g. OI~6300~$\rm \AA$) in combination with the weak high excitation lines (e.g. OIII~5007~$\rm \AA$) are a problem (Oonk et al. in prep.). A strong source of heating with modest ionisation is required, but the details of this process are currently not understood. Lyman and Balmer line ratios are known to be affected by non-stellar heating processes such as strong shocks and AGN \citep{Os06}. Our BCGs are not dominated by either of these processes. Alternative non-thermal models such low velocity shocks and/or heating by energetic particles have been proposed and may play a role\citep{Fa06, Fe09}. It is possible that these processes could also influences the line ratios, but this requires further investigation in the future.

A detailed investigation taking into account both the continuum and line emission over a broad wavelength regime is needed to investigate the shape of the extinction curve in BCGs and to clarify which of the above explanations for the Lyman and Balmer lines is appropriate. Such an investigation is beyond the scope of this paper.

\subsection{A2597 versus A2204}
The BCGs in A2597 and A2204 are in many ways typical for the class of nearby (z$\sim$0.1) BCGs in massive cool-core clusters. They have blue FUV-optical colors, strong optical line emission and balmer decrements implying significant dust obscuration. The peak fluxes of the emission processes discussed here, i.e. X-ray, FUV, optical and radio emission, all agree and line up with the BCG nucleus. In both A2597 and A2204 the FUV morphology consists of a clumpy, filamentary part and a more diffuse part. Each of these accounts for about half of the total FUV flux. The FUV morphology for A2597 differs from A2204 in that A2597 has a more spiral-like filamentary structure as opposed to the more radially extending filaments in A2204.

The total FUV to U emission ratio, i.e. prior to any corrections for emission contributed by old stars, peaks in FUV bright clumps and is indicative of very young star formation. Correcting for emission contributed to the U-band by old stars shows that the old stellar component in A2597 is responsible for a larger part of the total U-band emission than is the case in A2204. In neither object does the old stellar population contribute significantly to the total FUV emission. Excess FUV and U-band emission is defined by correcting the total observed emission for the emission due to the underlying older stellar population. In both cases the location of the maximum amount of excess FUV and U-band emission corresponds with the BCG nucleus.

The excess FUV to U emission ratio is found to be significantly higher in A2597 than in A2204. It also has less spatial sub-structure in A2597 as compared to A2204. The maximum ratio of the excess FUV to U emission is found to be located in off-nuclear regions in both objects. In A2204 this ratio peaks in the bright knots north and north-west of the nucleus. In A2597 this ratio is high across most of the central region with small increases to the south and south-east of the nucleus.

Interpreteting the excess FUV to U ratio in terms of main-sequence stars, taking into account only Milky Way foreground extinction, we find that we require stellar temperatures up to 40000~K for the bright knots in A2204 and up to the 50000~K for the central region in A2597. Correcting for emission contributed by the nebular continuum and/or additional extinction by dust intrinsic to the BCG means that the regions with highest ratio of excess FUV to U emission can no longer be explained by normal stars. Contributions from non-thermal processes that have not been treated in this work should be investigated.

\section[]{Conclusion}
Using new, deep FUV imaging with HST and optical imaging and spectroscopy with FORS on the VLT we present the first two dimensional analysis of the FUV-optical colors at 1 arcsec spatial resolution in the A2597 and A2204 BCGs. 

\begin{itemize}
\item  We have mapped the FUV continuum emission with HST ACS/SBC in the F150LP (1612~$\rm \AA$) filter in 30$\times$30~arcsec$^{2}$ regions centered on the BCGs in A2597 and A2204. The FUV continuum emission is distributed in clumpy, filamentary structures and is observed out to 20 kpc from the BCG nuclei.
\item The total FUV to U-band emission ratio shows that the A2597 and A2204 BCG have blue cores typical of BCGs in cool-core clusters.
\item The excess FUV and U emission, obtained by removing emission due to the old stellar population, is mapped at 1~arcsec resolution throughout the central $\sim$20~kpc regions of the BCGs. The FUV and U excesses are shown to trace each other well on kpc-scales.
\item Taking only Milky Way foreground extinction into account the excess FUV to U band emission ratio ranges between 0.6-3.0 in A2597 and 0.7-1.8 A2204 respectively. Interpreting this in terms of K93 main-sequence stellar models we find that stellar temperatures between 10000~K and 50000~K in A2597 and 15000~K and 40000~K in A2204 are required to explain the observations.
\item Most of the central 20~kpc regions in A2597 and A2204 have an excess FUV to U band ratio greater than 1. This implies that the AGN and non-thermal emission processes, such as synchrotron emission or bremsstrahlung can not account for the observations.
\item The morphology of the excess FUV and U band emission, as well the polarisation limits deduced for A2597 by \citet{Mc99}, rule out scattering of light coming from the AGN. Stars thus remain as the most plausible explanation for the observed excess emission.
\item Spectroscopy shows that the FUV and U-band emission observed by us for A2597 is not contaminated by line emission. Recent claims of significant CIV 1549~$\rm \AA$ line emission in M87 are investigated and found not to be necessary in order to explain the observations. By extension of these results we believe that our FUV and U-band observations of A2204 are also free from line emission.
\item Nebular continuum emission can account for up to 10 per-cent of the excess FUV flux and up to 40 per-cent of the U flux in A2597 and A2204. Correcting for the maximum amount of allowed nebular continuum means that most of the central 20~kpc excess emission in A2597 and the brightest excess knots in A2204 can no longer be explained by stars.  
\item Optical Balmer line ratios indicate an average Milky Way extinction law with A$_{V}$=1.3 for A2597 and A$_{V}$=1.6 for A2204. Extending the extinction law to the FUV regime means that the observed FUV$_{\nu,exc}$/U$_{\nu,exc}$ ratio in the central 20~kpc regions of A2597 and A2204 can no longer be explained by K93 stellar models, if half or more of this dust is in front of the excess regions. 
\item The FUV continuum emission matches the morphology of the extended H$\alpha$ nebula in A2597 and A2204 on large scales. If the FUV emission is attributed to hot O-stars then these stars are able to account for the observed H$\alpha$ emission in both systems. For A2597 this is in agreement with earlier results by \citep{Od04}.
\item The minimum star formation rate deduced from the FUV continuum is 3.5~M$_{\odot}$~yr$^{-1}$ for A2597 and 10.2~M$_{\odot}$~yr$^{-1}$ for A2204. The maximum star formation rate depends strongly on the unknown dust distribution and extinction law.
\item Assuming optically thin dust with Milky Way properties we estimate the gas and dust masses of the BCGs based on their measured A$_{V}$. We find M$_{gas}\sim$10$^{9}$~M$_{\odot}$ and M$_{dust}\sim$10$^{7}$~M$_{\odot}$ for the central 10$\times$10~kpc$^{2}$ regions in both A2597 and A2204.
\end{itemize}

This investigation shows that under special conditions on the nebular continuum emission and the dust obscuration it is possible to reconcile the FUV to optical colors with a stellar origin. The conditions are especially stringent for A2597 and in this regard contributions from non-thermal processes not treated here should be investigated in more detail.

A more detailed investigation of the possible stellar origin for the individual knots observed at the intrinsic spatial resolution of HST requires further imaging with HST in line-free bands. Also careful investigation into the dust properties and the shape of the extinction law will need to be carried out in order to resolve the uncertainties in determining the exact nature of the FUV to optical excess emission in BCGs.

We note with interest that detailed observations of the morphology of the FUV continuum, the excess FUV to U band ratio as well as the properties of the cool T$<$10$^{4}$~K gas in BCGs, do not imply a strong interaction with the central AGN or its outflows on scales larger than the central few kpc \citep[e.g.][]{Pi09,Od10,Wi09,Oo10}. A minor merger or ICM cooling scenario may be more appropriate to explain the extended 30~kpc-scale star formation, cool gas and dust properties in these objects. 

If the origin of the cool gas at the centres of BCGs is the (residual) cooling of the hot ICM then the interaction between this hot gas and the AGN outflows, thought to regulate the cooling process of this hot phase, may still be indirectly responsible for the observations at lower temperatures.

\section*{Acknowledgements}
JBRO wishes to thank J. Brinchmann, B. A. Groves, G. R. Tremblay and C. P. O'Dea for useful discussions. NAH acknowledges support from STFC, and the University of Nottingham through the Anne McLaren Fellowship. 

This research has made use of NASA's Astrophysics Data System. The FORS observations were taken at the Very Large Telescope (VLT) facility of the European Southern Observatory (ESO) as part of projects 69.A-0444, 67.A-0597 and 63.N-0485. The ACS/SBC observations were taken as part of project 11131 with the NASA/ESA Hubble Space Telescope which is operated by AURA, Inc., under NASA contract NAS~5-26555.

The National Radio Astronomy Observatory is a facility of the National Science Foundation operated under cooperative agreement by Associated Universities, Inc. The Chandra X-ray Observatory Center, which is operated by the Smithsonian Astrophysical Observatory on behalf of the National Aeronautics and Space Administration under contract NAS~8-03060.

\pagebreak

%%%%%%% TABL %%%%%%%

%table: a2597 date, band and exptime.
\begin{table*}
 \centering
  \begin{tabular}{|l|l|l|l|} \hline
  Date & Filter & Exptime & Zeropoint \\ \hline
  2008-07-21 & F150LP & 8141 & 22.4484 \\
  2008-06-23 & F150LP & 5419$^{a}$ & 22.4484 \\
  2002-06-16 & U\_Bessel & 1320 & 23.4824$\pm$0.0530 \\
  2002-06-16 & B\_Bessel & 1080 & 26.4815$\pm$0.0503 \\
  2001-07-26 & V\_Bessel & 330 & 27.2367$\pm$0.0129 \\
  2001-07-26 & R\_Bessel & 330 & 27.2484$\pm$0.0185 \\
  2001-07-26 & I\_Bessel & 330 & 26.4829$\pm$0.0133 \\ \hline
  \end{tabular}
 \caption[]{A2597 observations log. Column 1 lists the date of the observation in YYYY-MM-DD. Column 2 lists the filter name. Column 3 lists the exposure time in units of seconds. Column 4 lists the FORS Bessel system zeropoint in units of magnitudes. The zeropoint is based upon the average of at least 3 Landolt standard stars within each observation. The uncertainty in the zeropoint is taken as the maximum deviation of these standard stars have with respect to the quoted average zeropoint. The SBC F150LP zeropoint is taken from the ACS handbook and given in AB magnitudes. The photometric uncertainty according to the most recent investigation is 2\% for the F150LP filter \citep{Ma05}. All observations listed here were taken in photometric observing conditions. \\ $^{a}$ this data set is affected by low level background glow and has been dropped from further analysis.)}\label{t_obslog_a2597}
\end{table*}

%table: a2204 date, band and exptime.
\begin{table*}
 \centering
  \begin{tabular}{|l|l|l|l|} \hline
  Date & Filter & Exptime & Zeropoint \\ \hline
  2008-04-30 & F150LP & 8126 &  22.4484 \\
  2008-04-22 & F150LP & 5409 &  22.4484 \\
  2002-06-10 & U\_Bessel & 1440 & 23.5860$\pm$0.0197 \\
  2002-06-10 & B\_Bessel & 780 & 26.5420$\pm$0.0627 \\
  2002-06-10 & R\_Bessel & 400 & 27.1567$\pm$0.0329 \\
  2001-04-19 & V\_Bessel & 330 & 27.2315$\pm$0.0158 \\ 
  2001-04-19 & R\_Bessel & 330 & 27.2503$\pm$0.0218 \\
  2001-04-19 & I\_Bessel & 330 & 26.5046$\pm$0.0087 \\ \hline
  \end{tabular}
 \caption[]{A2204 observations log. The column headers are the same as in Table \ref{t_obslog_a2597}.}\label{t_obslog_a2204}
\end{table*}

%table: Bessel to AB conversion (from HYPERZ, H. Hildebrandt)
\begin{table*}
 \centering
  \begin{tabular}{|l|l|} \hline
  Filter & (FORS\_AB)$\rm_{conv}$ \\ \hline
  U\_Bessel & 0.931 \\
  B\_Bessel & -0.042 \\
  V\_Bessel & 0.060 \\
  R\_Bessel & 0.258 \\
  I\_Bessel & 0.458 \\ \hline
  \end{tabular}
 \caption[]{FORS Bessel to AB magnitude conversion factors. Column 1 lists the filter name. Column 2 lists the FORS Bessel to AB conversion factor in units of magnitude. It converts the FORS Bessel magnitudes to AB magnitudes. This conversion factor is calculated for FORS Bessel filter curves with the HYPERZ program \citep[][; priv. comm. H. Hildebrandt]{Bo00}}\label{t_ab_conv}
\end{table*}

%table: line contamination
\begin{table*}
 \centering
  \begin{tabular}{|l|l|l|l|l|l|} \hline
  Name  & U\_Bessel & B\_Bessel & V\_Bessel & R\_Bessel & I\_Bessel \\ \hline
  A2597 & - & 0.185 & 0.050 & $>$0.101 & $>$0.034 \\
  A2204 & - & 0.253 & 0.067 & $>$0.115 & $>$0.239 \\ \hline
  \end{tabular}
 \caption[]{Line contamination of the FORS Bessel images. Column 1 list the target name. Columns 2-6 list the fraction of the total emission in the filter which is due to line emission.  This line contamination fraction is calculated using the FORS spectra described in Section \ref{s_fors_spec}. None of the spectral lines probed by our spectra fall in the U filter, see also Fig. \ref{f_fors_spec_filt}. The spectra do not probe the SII lines at 6716~$\rm \AA$ and 6731~$\rm \AA$ and thus we can only provide a lower limit to the line contamination fraction of the R and I Bessel filters.}\label{t_contam}
\end{table*}

%table: dust
\begin{table*}
 \centering
  \begin{tabular}{|l|l|l|l|l|l|} \hline
  Name  & A$\rm_{V,MW}$ & A$\rm_{V,BCG}$ & H$\delta$/H$\beta$ & H$\gamma$/H$\beta$ & H$\alpha$/H$\beta$ \\ \hline
  A2597 & 0.1 & 1.3 & 0.18 (0.19) & 0.37 (0.40) & 4.52 (4.52) \\
  A2204 & 0.3 & 1.6 & 0.16 (0.15) & 0.34 (0.38) & 5.27 (5.29) \\ \hline
  \end{tabular}
 \caption[]{Visual extinction A$_{V}$ and Balmer decrements. Column 1 lists the target name. Column 2 lists the extinction in the V-band for the MW foreground to \citep{Sc98}. Column 3 lists the extinction in the V-band intrinsic to the BCG. This value is calculated by decomposing the Balmer decrements observed in the FORS spectra (Section \ref{s_fors_spec}) into two dust components, one related to the BCG and one related to the MW foreground. Describing both components by the C89 extinction law allows us to derive A$\rm_{V,BCG}$. The intrinsic dust-free Balmer decrements were set to their stellar case B recombination values i.e. H$\delta$/H$\beta$=0.26, H$\gamma$/H$\beta$=0.47 and H$\alpha$/H$\beta$=2.87 \citep{Os06}. Columns 4, 5 and 6 list the reddened Balmer decrement values obtained from this two component dust analysis for A$\rm_{V,MW}$ and A$\rm_{V,BCG}$ given in columns 2 and 3. The Balmer decrements calculated in this manner agree well with the measured decrements from our FORS spectra. The latter are given in the parentheses in columns 4, 5 and 6.}\label{t_dust}
\end{table*}

%table: old star fuv ratios
\begin{table*}
 \centering
  \begin{tabular}{|l|l|l|l|l|} \hline
  Name  & FUV$_{\nu,tot}$/U$_{\nu,tot}$ & FUV$_{\nu,tot}$/V$_{\nu,tot}$ & FUV$_{\nu,tot}$/R$_{\nu,tot}$ & FUV$_{\nu,tot}$/I$_{\nu,tot}$ \\ \hline
  A2597-OFF & 0.1564 & 0.0153 & 0.0102 & 0.0068 \\
  A2204-SW & 0.1526 & 0.0105 & 0.0066 & 0.0043 \\ \hline
  \end{tabular}
 \caption[]{FUV flux ratios for old star regions. Column 1 gives the name of the region, see Section \ref{old_star_corr}. Columns 2-5 lists the flux ratios. The subscript \textit{tot} refers to the total flux observed in this region.}\label{t_fuv_ratio}
\end{table*}

%table: old star u ratios
\begin{table*}
 \centering
  \begin{tabular}{|l|l|l|l|} \hline
  Name  & U$_{\nu,tot}$/V$_{\nu,tot}$ & U$_{\nu,tot}$/R$_{\nu,tot}$ & U$_{\nu,tot}$/I$_{\nu,tot}$ \\ \hline
  A2597-OFF & 0.0988 & 0.0654 & 0.0439 \\
  A2597-COMP & 0.0969 & 0.0664 & 0.0470 \\
  A2204 SW & 0.0706 & 0.0437 & 0.0281 \\
  A2204-COMP & 0.0710 & 0.0452 & 0.0297 \\
  \end{tabular}
 \caption[]{U flux ratios for old star regions. The column headers are the same as in Table \ref{t_fuv_ratio}.}\label{t_u_ratio}
\end{table*}

%table: aperture fluxes
\begin{table*}
 \centering
  \begin{tabular}{|l|r|r|r|r|r|r|r|} \hline
  Name  & F(FUV)$_{\nu,tot}$ & F(FUV)$_{\nu,exc}$ & F(FUV)$_{\nu,NC}$ & F(U)$_{\nu,tot}$ & F(U)$_{\nu,exc}$ & F(U)$_{\nu,NC}$ & F(H$\alpha$)$\rm_{\nu}$ \\ \hline
  A2597 & 12.04 & 9.60 & 0.57 & 21.20 & 3.60 & 1.36 & 41.5 \\
  A2204 & 6.16 & 5.20 & 0.67 & 10.50 & 4.00 & 1.57 & 48.5 \\ \hline
  \end{tabular}
 \caption[]{FUV, U and H$\alpha$ fluxes. These fluxes are integrated over 18~kpc radial apertures centered on the BCG and given in units 10$^{-28}$~erg~s$^{-1}$~cm$^{-2}$~Hz$^{-1}$. Columns 1-3 give the FUV F150LP fluxes. Columns 4-6 give the U\_Bessel fluxes.  The subscripts \textit{tot}, \textit{exc} and \textit{NC} correspond respectively to the total flux, the excess flux and the flux expected from the nebular continuum. The latter is calculated using the NEBCONT program \citep{Ho04}, see also Section \ref{s_nebcont}. Column 7 gives the H$\alpha$ flux in  units 10$^{-15}$~erg~s$^{-1}$~cm$^{-2}$ \citep{Ja05}.}\label{t_laper}
\end{table*}

%table: Hicks et al. 2010 FUV/NUV
\begin{table*}
 \centering
  \begin{tabular}{|l|r|r|r||} \hline
  name & z & FUV$_{\nu,tot}$/NUV$_{\nu,tot}$ & T$_{K93}$ \\ \hline
A85    & 0.0557 &   0.60   &    10656 \\
A1204  & 0.1706 &   1.12   &    23226 \\
A2029  & 0.0779 &   0.61   &    10973 \\
A2052  & 0.0345 &   0.54   &    10000 \\
A2142  & 0.0904 &   0.63   &    11273 \\
A2597  & 0.0830 &   0.85   &    13771 \\
A3112  & 0.0761 &   0.74   &    11858 \\
HercA  & 0.1540 &   0.73   &    14069 \\
HydrA  & 0.0549 &   0.77   &    11876 \\
MKW3s  & 0.0453 &   0.67   &    11083 \\
MKW4   & 0.0196 &   0.67   &    10870 \\
MS0839 & 0.1980 &   0.90   &    17951 \\
MS1358 & 0.3272 &   0.91   &    20747 \\
MS1455 & 0.2578 &   1.13   &    26313 \\
RX1347 & 0.4500 &   0.76   &    21622 \\
Zw3146 & 0.2906 &   1.33   &    37883 \\ \hline
  \end{tabular}
 \caption[]{\citet{Hi10} GALEX BCG sample. Column 1 lists the cluster name for the BCG observed. Column 2 lists the redshift as given in \citet{Hi10}. Column 3 lists the FUV$_{\nu,tot}$/NUV$_{\nu,tot}$ ratio corresponding to the observed 7~arcsec radial apertures and have been corrected for foreground MW extinction only. Column 4 lists the K93 stellar temperatures corresponding to the flux ratio given in column 3. The z$>$0.1 BCGs are potentially affected by Ly$\alpha$ emission, see also Section \ref{s_hi10}.}\label{t_hicks}
\end{table*}

\clearpage

%%%%%%% FIGS %%%%%%%

%%fig: a2597 HST

\clearpage

%%fig: a2204 HST

\clearpage

%%fig: fors uvr

\clearpage

%%fig: vla 8ghz

\clearpage

%%fig: pseudo slits

\clearpage

%%fig: total FUV/U vs. BC03 SSP.

\clearpage

%%fig: FUV/U BC03 SBRST

%\clearpage

%%fig: BB-MODELS.

\clearpage

%%fig: a2597,a2204 FORS NUCL SPECTRA AND FILTERS (normalised).

\clearpage

%%fig: a2597 V-ratio and excess maps. 

\clearpage

%%fig: a2204 V-ratio and excess maps.

\clearpage

%%fig: a2597 FUV/U-EXCESS-RATIO AND TEMPERATURE MAPS. 

\clearpage

%%fig: a2204 FUV/U-EXCESS-RATIO AND TEMPERATURE MAPS. 

\clearpage

%%fig: FUV/U BB and K93 models

\clearpage

%%fig: hst imaging

\clearpage

%%fig: a2597 FUV/U based R, I band old star subtraction.

%--- APPENDICES ---
\clearpage
\appendix

\section[]{A2597, A2204 FUV continuum knots.}\label{fuv_knots}
The FUV emission in A2597 and A2204 is observed to consists of two components, a filamentary component and a more diffuse component. To estimate the contribution to the total FUV flux from each component we assign pixels belonging to the filamentary component as those pixels having values at least twice the mean flux calculated within a 5 arcsec aperture centered on the BCG. We find that the filamentary component contributes about half of the total FUV flux in our objects. This agrees with the FUV observations by \citet{Od10} for a sample of BCGs.

Below we have identified knots within the filamentary component. These knots have been selected by eye and are required to have a mean flux per pixel within the specified aperture of at least twice the root-mean-square (RMS) background level.

%%table: (A2597)
%\begin{table*}
\begin{table}
 \centering
  \begin{tabular}{|l|r|r|r|r|} \hline
  No. & $\alpha$ (J2000) & $\delta$ (J2000) & Area & F$_{F150LP,tot}$ \\ \hline
       0 & 23:25:19.890 & -12:07:18.26 & 21$\times$21 &     0.0341735 \\
       1 & 23:25:19.864 & -12:07:21.11 & 21$\times$21 &     0.0499983 \\
       2 & 23:25:19.781 & -12:07:25.81 & 41$\times$41 &       1.58047 \\
       3 & 23:25:19.745 & -12:07:30.23 & 21$\times$21 &     0.0772838 \\
       4 & 23:25:19.637 & -12:07:29.88 & 21$\times$21 &      0.130110 \\
       5 & 23:25:19.668 & -12:07:29.11 & 21$\times$21 &      0.147277 \\
       6 & 23:25:19.547 & -12:07:28.81 & 21$\times$21 &     0.0955810 \\
       7 & 23:25:19.550 & -12:07:29.83 & 21$\times$21 &     0.0730317 \\
       8 & 23:25:19.762 & -12:07:26.73 & 41$\times$41 &       1.72703 \\
       9 & 23:25:19.743 & -12:07:27.56 & 41$\times$41 &       1.20056 \\
      10 & 23:25:19.941 & -12:07:27.36 & 41$\times$41 &      0.903357 \\
      11 & 23:25:19.885 & -12:07:26.93 & 41$\times$41 &       1.21367 \\
      12 & 23:25:19.833 & -12:07:26.36 & 41$\times$41 &       1.39439 \\
      13 & 23:25:19.873 & -12:07:24.48 & 41$\times$41 &      0.551081 \\
      14 & 23:25:20.021 & -12:07:24.93 & 21$\times$21 &     0.0503924 \\
      15 & 23:25:19.815 & -12:07:27.58 & 41$\times$41 &      0.998714 \\
      16 & 23:25:19.656 & -12:07:27.06 & 21$\times$21 &      0.136095 \\
      17 & 23:25:19.777 & -12:07:33.46 & 15$\times$15 &     0.0155612 \\
      18 & 23:25:19.842 & -12:07:32.58 & 15$\times$15 &     0.0138937 \\
      19 & 23:25:19.885 & -12:07:30.53 & 21$\times$21 &     0.0607091 \\
      20 & 23:25:19.486 & -12:07:20.98 & 15$\times$15 &     0.0148215 \\
      21 & 23:25:20.050 & -12:07:30.41 & 15$\times$15 &     0.0139930 \\
      22 & 23:25:19.856 & -12:07:25.61 & 21$\times$21 &      0.166120 \\
      23 & 23:25:19.552 & -12:07:26.68 & 21$\times$21 &     0.0585360 \\
      24 & 23:25:19.549 & -12:07:25.18 & 21$\times$21 &     0.0363437 \\
      25 & 23:25:19.840 & -12:07:28.86 & 21$\times$21 &      0.146558 \\
      26 & 23:25:19.854 & -12:07:21.58 & 21$\times$21 &     0.0433445 \\ \hline
  \end{tabular}
 \caption[]{A2597 FUV continuum knots as measured in the ACS/SBC F150LP filter. Column 1 lists the knot number, see Fig. \ref{f_fuv_knots}. Column 2 lists the right ascension coordinate. Column 3 lists the declination coordinate. Column 4 lists the aperture used in units of pixels (one pixel has a length of 0.025~arcsec). Column 5 lists the flux integrated over the aperture in units elec~$s^{-1}$.}\label{t_a2597_knots}
\end{table}
%\end{table*}

%%table: (A2204)
%\begin{table*}
\begin{table}
 \centering
  \begin{tabular}{|l|r|r|r|r|} \hline
  No. & $\alpha$ (J2000) & $\delta$ (J2000) & Area & F$_{F150LP,tot}$ \\ \hline
       0$^{a}$ & 16:32:47.076 & +05:34:45.97 & 21$\times$21 &     0.0488258 \\
       1$^{b}$ & 16:32:46.973 & +05:34:40.87 & 21$\times$21 &      0.1734531 \\
       2 & 16:32:46.801 & +05:34:36.72 & 21$\times$21 &     0.0399950 \\
       3 & 16:32:46.948 & +05:34:37.42 & 21$\times$21 &      0.209281 \\
       4 & 16:32:46.977 & +05:34:37.05 & 21$\times$21 &     0.0986732 \\
       5 & 16:32:46.873 & +05:34:35.35 & 21$\times$21 &      0.116265 \\
       6 & 16:32:46.945 & +05:34:36.07 & 21$\times$21 &     0.0687493 \\
       7$^{c}$ & 16:32:46.731 & +05:34:30.95 & 21$\times$21 &     0.0598593 \\
       8 & 16:32:47.049 & +05:34:31.10 & 21$\times$21 &     0.0863461 \\
       9 & 16:32:47.044 & +05:34:32.05 & 21$\times$21 &      0.172684 \\
      10 & 16:32:47.020 & +05:34:32.55 & 21$\times$21 &      0.337899 \\
      11 & 16:32:46.977 & +05:34:33.37 & 21$\times$21 &      0.455177 \\
      12 & 16:32:46.943 & +05:34:33.05 & 21$\times$21 &      0.664247 \\
      13 & 16:32:46.903 & +05:34:33.37 & 21$\times$21 &      0.849183 \\
      14 & 16:32:46.868 & +05:34:33.80 & 21$\times$21 &      0.358830 \\
      15 & 16:32:46.970 & +05:34:32.40 & 41$\times$41 &       2.00912 \\
      16 & 16:32:46.908 & +05:34:32.07 & 21$\times$21 &      0.164597 \\
      17 & 16:32:46.967 & +05:34:31.32 & 21$\times$21 &      0.107333 \\
      18 & 16:32:46.997 & +05:34:34.87 & 21$\times$21 &     0.0610536 \\
      19 & 16:32:46.937 & +05:34:29.75 & 21$\times$21 &     0.0322799 \\
      20 & 16:32:46.798 & +05:34:37.30 & 15$\times$15 &     0.0166241 \\
      21 & 16:32:47.082 & +05:34:34.67 & 21$\times$21 &     0.0338256 \\
      22 & 16:32:47.054 & +05:34:32.45 & 21$\times$21 &      0.145208 \\
      23 & 16:32:46.851 & +05:34:34.75 & 21$\times$21 &     0.0631433 \\
      24 & 16:32:46.923 & +05:34:34.75 & 21$\times$21 &      0.103866 \\
      25 & 16:32:47.009 & +05:34:31.30 & 21$\times$21 &     0.0626998 \\
      26 & 16:32:46.862 & +05:34:36.80 & 15$\times$15 &     0.0146095 \\ \hline
  \end{tabular}
 \caption[]{A2204 FUV continuum knots as measured in the ACS/SBC F150LP filter. The columns headers are the same as in Table \ref{t_a2597_knots}. \\ $^{a}$ this is a star. \\ $^{b}$ this is an elliptical galaxy with a bright FUV point source. \\ $^{c}$ this is an elliptical galaxy with diffuse FUV emission.}\label{t_a2204_knots}
\end{table}
%\end{table*}

%%fig: a2597,a2204 knots
\begin{figure*}
    \includegraphics[width=0.62\textwidth, angle=90]{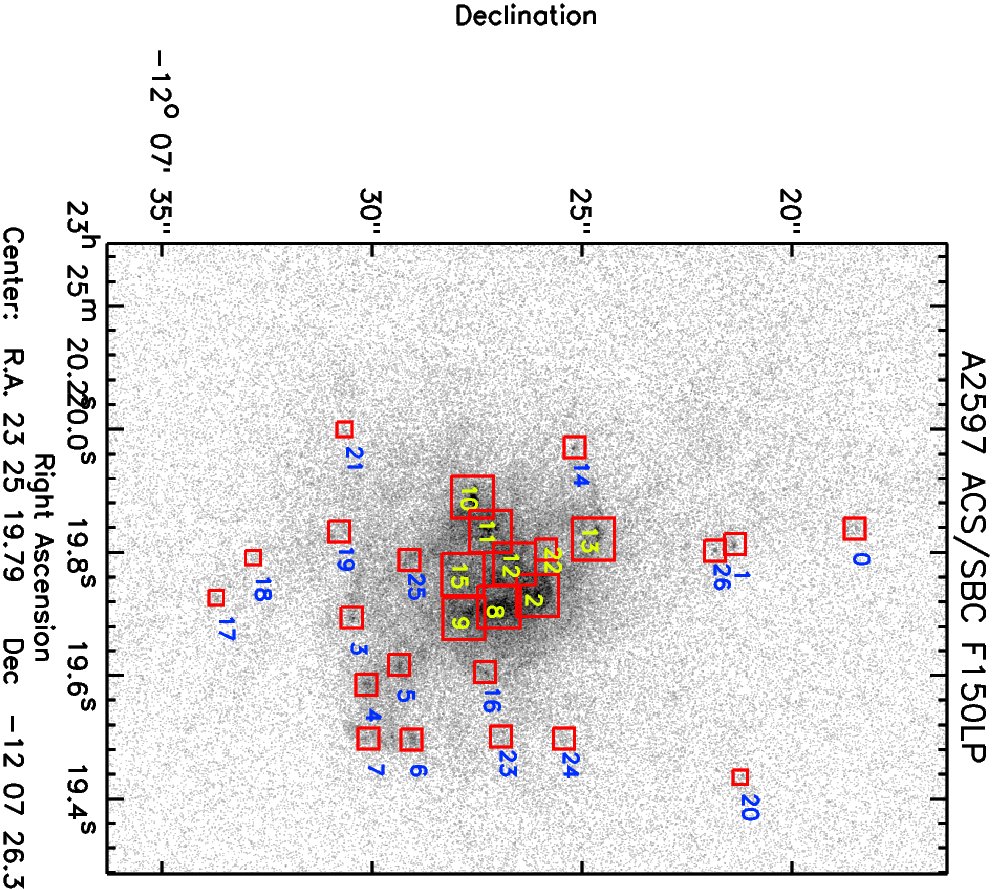}
    \includegraphics[width=0.62\textwidth, angle=90]{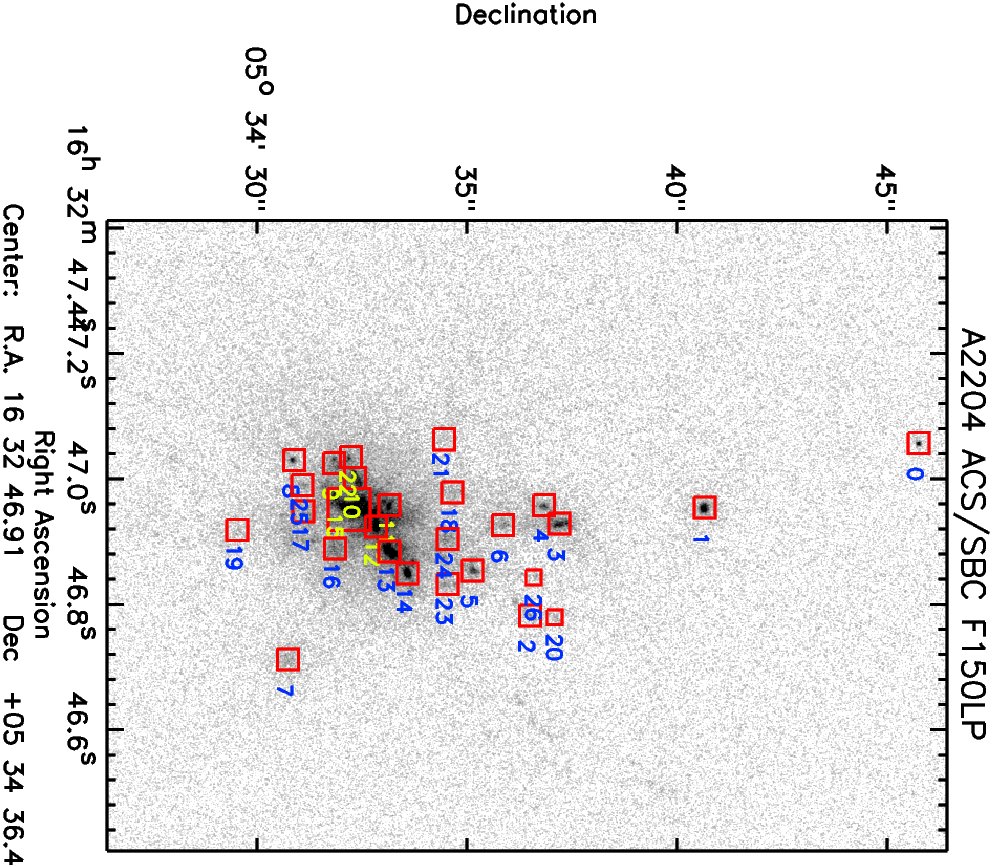}
  \vspace{0.5cm}
  \caption{FUV continuum knots for A2597 (\textit{Top}) and A2204 (\textit{Bottom}) as measured in the ACS/SBC F150LP filter. The red squares show apertures used to extract the knot fluxes. Knots are labelled according to the numbers written next to these apertures and the extracted fluxes are given in Tables \ref{t_a2597_knots} and \ref{t_a2204_knots}.}\label{f_fuv_knots}
\end{figure*}

\clearpage

\section[]{NGC 604 a Test Case}\label{app_n604}
\begin{figure}
    \includegraphics[width=0.35\textwidth, angle=90]{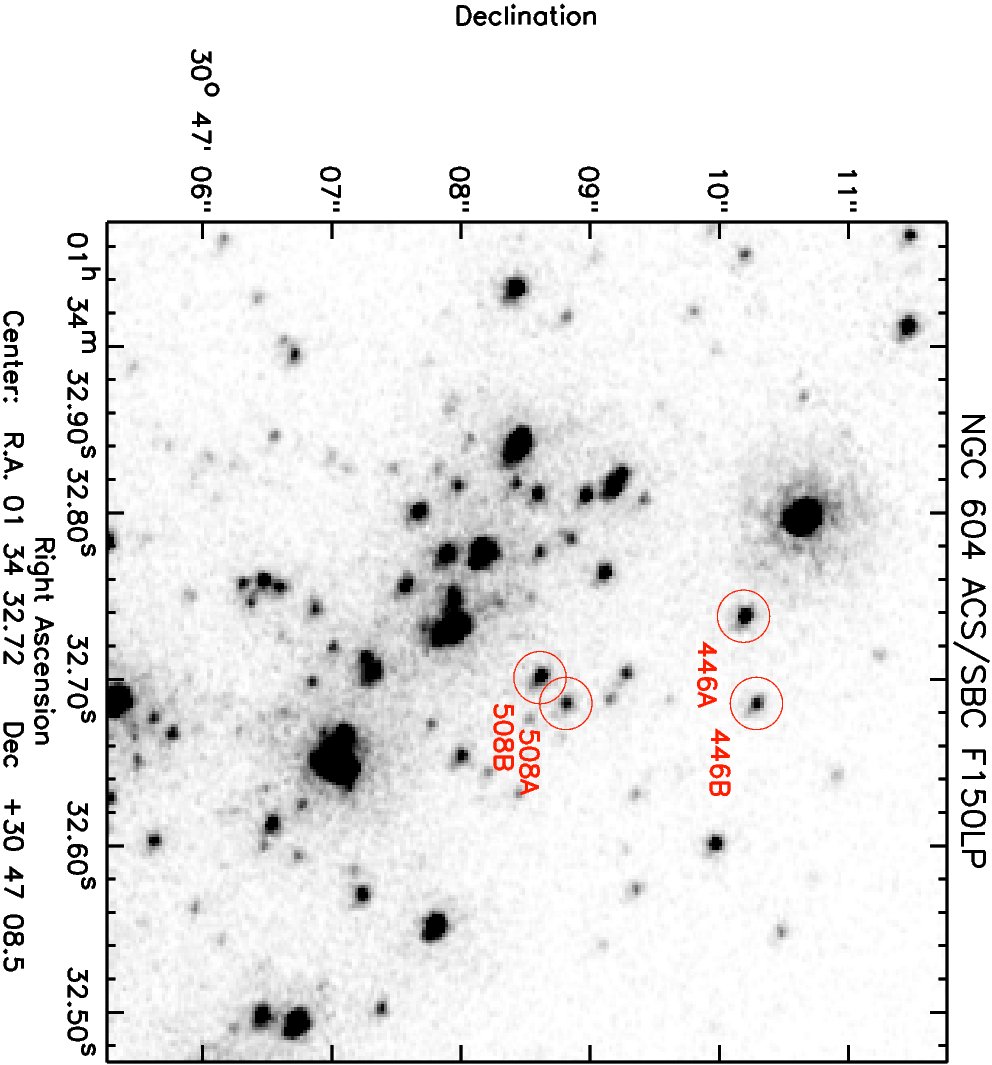}
    \includegraphics[width=0.35\textwidth, angle=90]{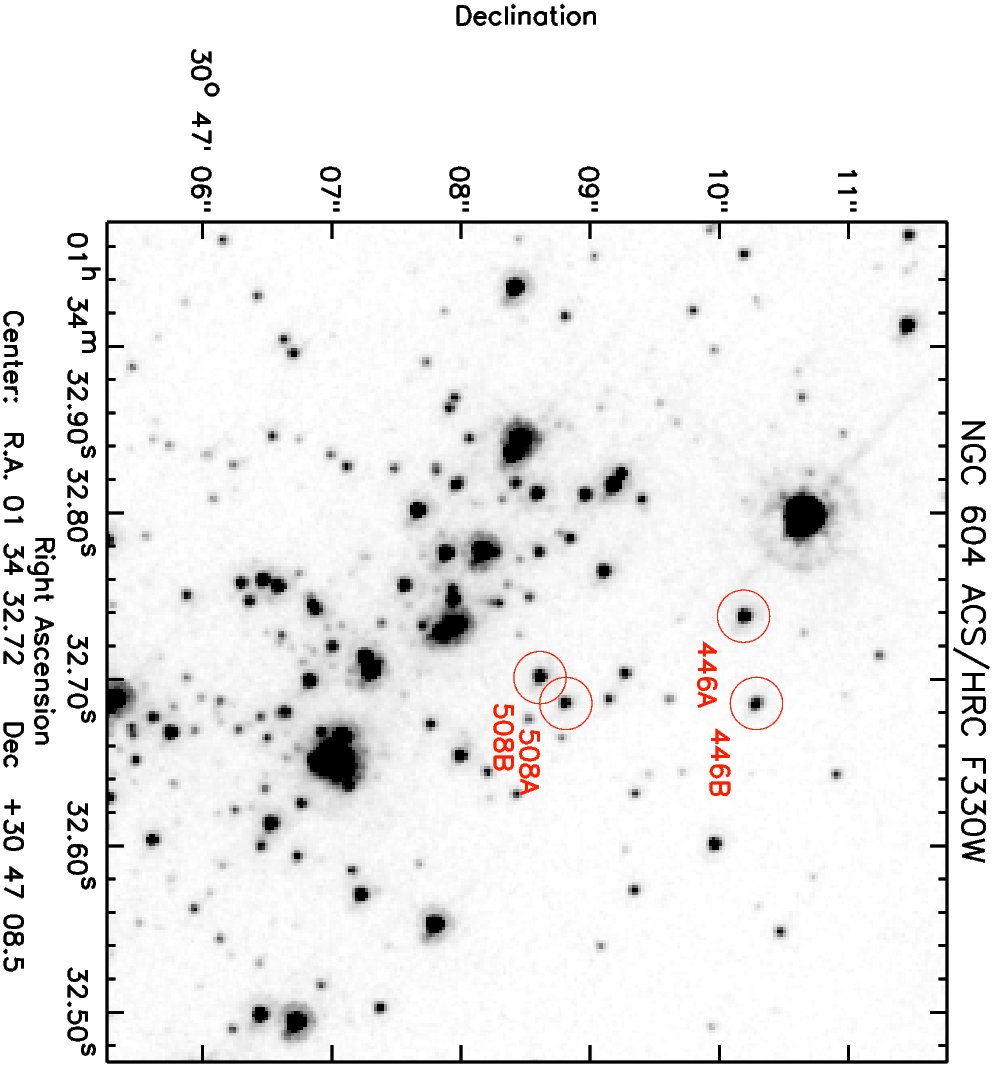}
  \vspace{0.5cm}
  \caption{HST ACS/SBC imaging of NGC~604. (Top) F150LP. (Bottom) F330W. Selected main-sequence stars are marked by the red circles.}\label{f_n604}
\end{figure}

\begin{figure}
    \includegraphics[width=0.32\textwidth, angle=90]{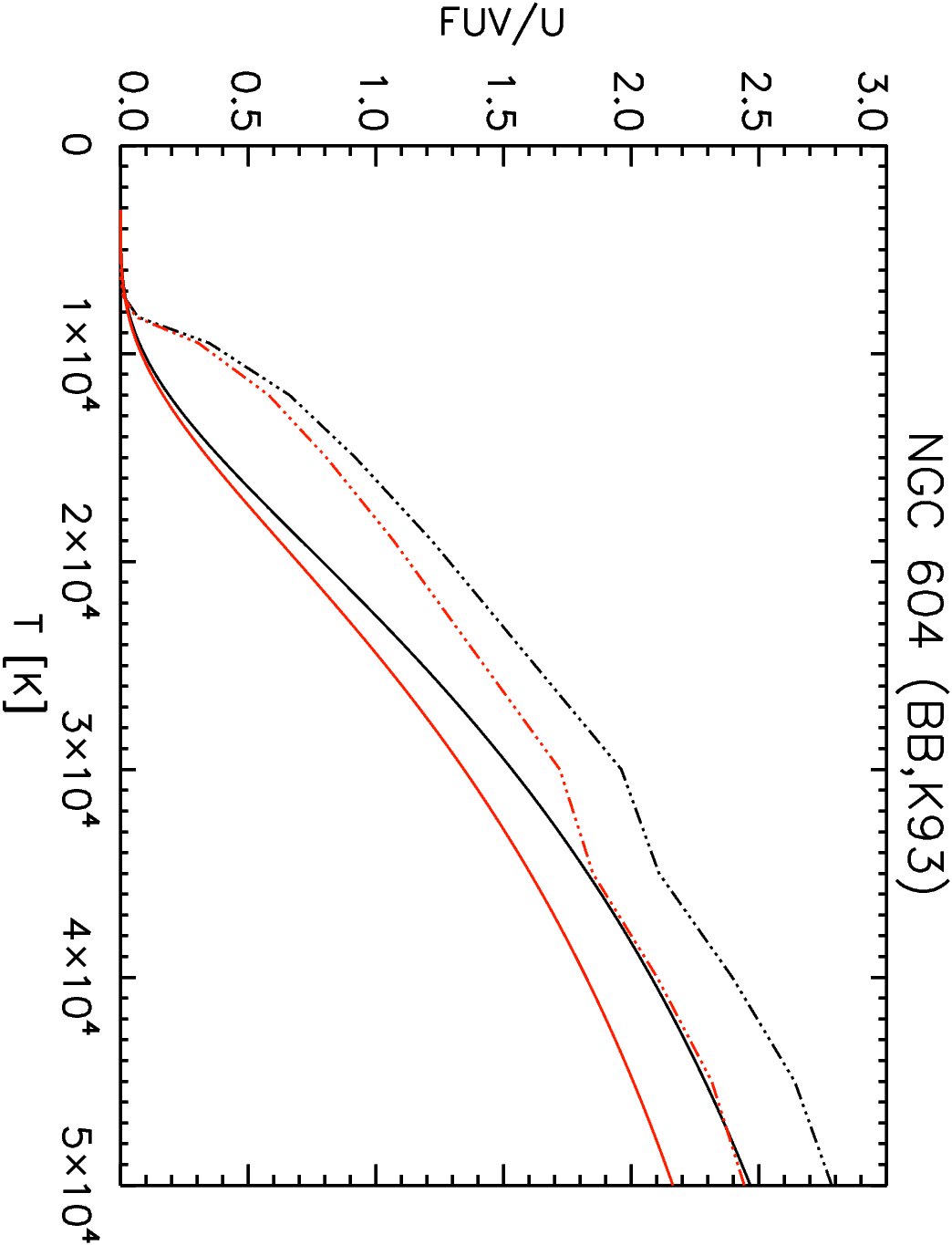}
  \vspace{0.5cm}
  \caption{NGC~604 FUV$_{\nu}$/U$_{\nu}$ ratio models. Black Body models (solid line) and the K93 stellar models (dash-dot line) are shown. Black lines refer to a dust-free environment. Red lines refer to models reddened by MW foreground dust only.}\label{f_n604_bb_k93}
\end{figure}

In Section \ref{s_stellar} we have shown that the FUV/U broad band ratio is a good discriminator of stellar temperature. In order to verify that the temperatures derived from the broad band ratio are consistent with spectroscopically derived temperatures we have tested our method on a few main sequence stars in the nearby giant HII region NGC~604 in M33. We retrieved archival HST/ACS SBC-F150LP and HRC-F330W observations of this object. We note that the F330W filter shape is similar to that of our FORS U\_Bessel filter and that it does not contain any line emission.
\\ \\
We selected four, isolated, main-sequence stars for which FUV spectroscopic temperatures have been determined by \citet{Bruh03}, see Fig. \ref{f_n604}. The FUV$_{\nu}$/U$_{\nu}$ ratio derived for these stars, takes into account corrections for the background, the difference in point spread function between the two filters. The stellar temperature is then derived by comparing the measured FUV$_{\nu}$/U$_{\nu}$ ratio to the ratio determined for BB and K93 stellar models. The extinction towards each star is taken into account when computing the models. The results for NGC~604 are shown in Fig. \ref{f_n604_bb_k93} and Table \ref{t_n604}. 
\\ \\
We find that the K93 temperatures derived by our broad band method are consistent with the spectroscopic temperatures determined by \citet{Bruh03}. This is expected as these authors also use the K93 stellar models to derive their temperatures. This thus shows the equivalence of the two methods and that there are no spectral features unaccounted for when using the broad band approach. However, this investigation by no means verifies the K93 models as an adequate model for real stars. 
\\ \\
Interpreting the measured FUV$_{\nu}$/U$_{\nu}$ ratio in terms of BB models leads to much higher temperatures for the stars investigated. The difference of the slope in FUV to U-band regime means that above 10$^{4}$~K the BB inferred temperatures are systematically below the K93 inferred temperatures whereas below 10$^{4}$ K they are systematically above the K93 inferred temperatures.

%%fig: FUV/U BB-MODEL.
%\begin{figure*}
%\end{figure*}

%%fig: FUV/U MODELS N604.
%\begin{figure*}
%\end{figure*}

%\clearpage

%\begin{table*}
\begin{table}
 \centering
  \begin{tabular}{|l|l|l|l|l|l|l|} \hline
  Name & Type & FUV$_{\nu}$/U$_{\nu}$ & A$_{V,tot}$ & T$_{BB}$ & T$_{K93}$ & T$_{B03}$ \\ \hline
  446A & O7V & 1.70 & 0.461 & 52000 & 42700 & 40100 \\
  446B & O8V & 1.79 & 0.384 & 51000 & 41900 & 38000 \\
  508A & O7V & 1.82 & 0.306 & 47400 & 40200 & 38000 \\
  508B & O6V & 1.94 & 0.229 & 46900 & 39900 & 40100 \\ \hline
  \end{tabular}
 \caption[]{Stellar Temperatures for main-sequence stars in NGC~604. Column 1 and 2 give the name and type of the star \citep{Bruh03}. Column 3 lists the FUV$_{\nu}$/U$_{\nu}$ ratio as measured for each star. Column 4 lists the total visual extinction in magnitudes towards each star. Columns 5 and 6 list the stellar temperatures derived from BB and K93 stellar models. Column 7 lists the spectroscopic temperatures derived by \citet{Bruh03}.}\label{t_n604}
\end{table}
%\end{table*}

\clearpage

\section[]{M87 filaments.}\label{app_m87_film}
\begin{figure}
    \includegraphics[width=0.34\textwidth, angle=90]{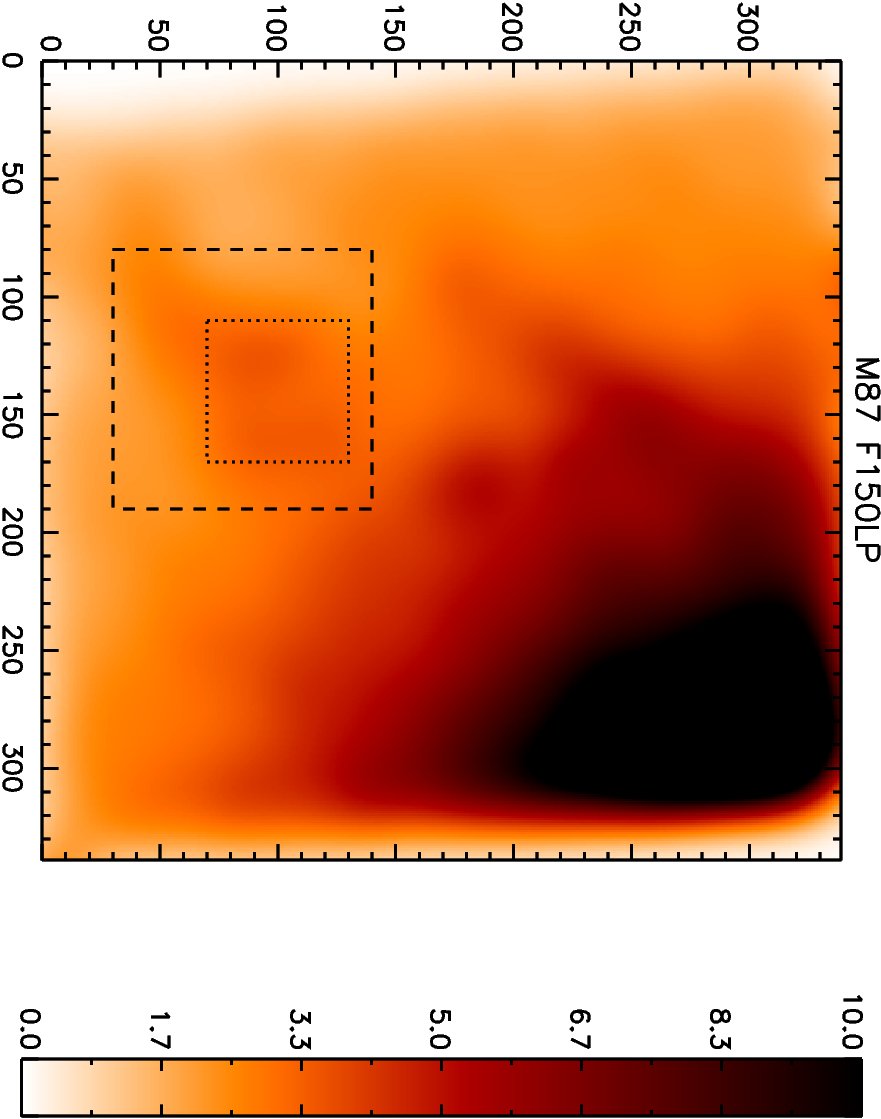}
  \vspace{0.5cm}
  \caption{M87 south-east. Re-binned and smoothed F150LP image. The dashed rectangle shows the location of the filament and the dotted rectangle shows the region used by us to calculate the stellar temperature. The units on the x and y axes are re-binned pixels with each pixel having a projected size on the sky of 0.1 arcsec.}\label{f_m87_f150}
\end{figure}

\begin{figure}
    \includegraphics[width=0.34\textwidth, angle=90]{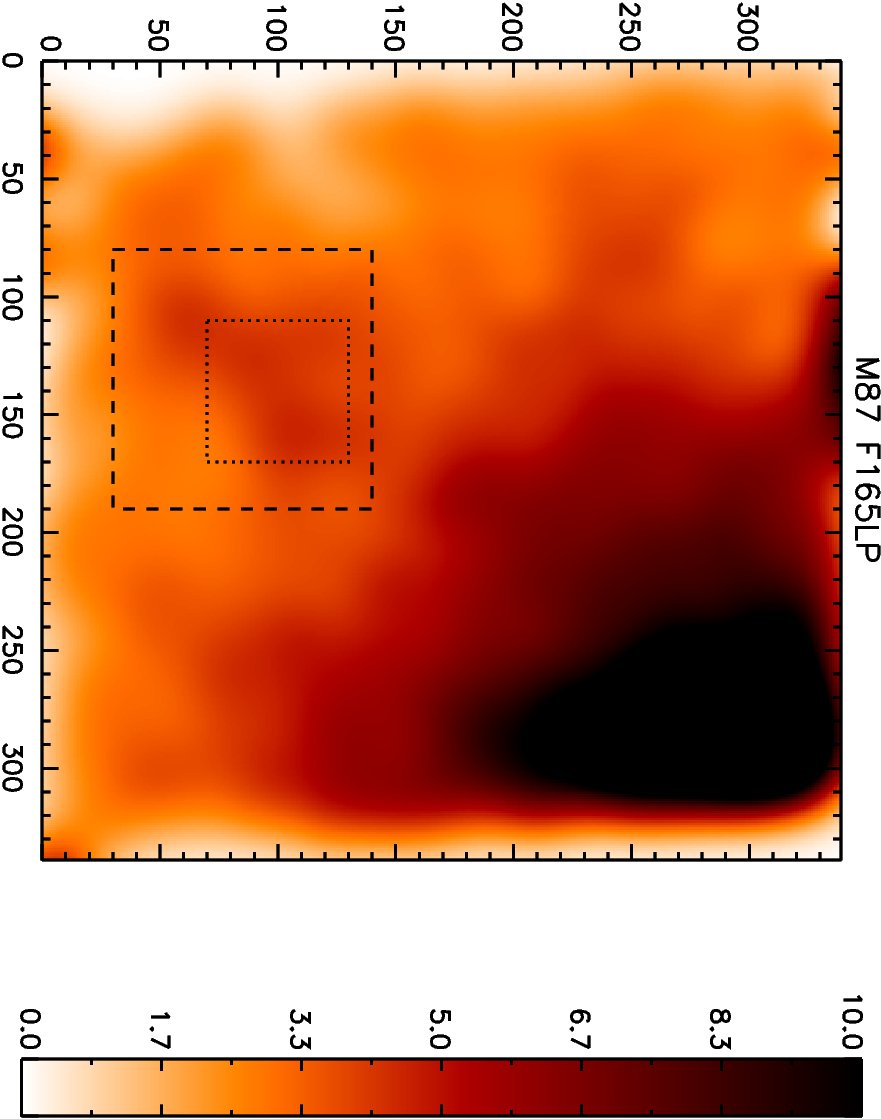}
  \vspace{0.5cm}
  \caption{M87 south-east. Re-binned and smoothed F165LP image. The rectangles and axes drawn are the same as in Fig. \ref{f_m87_f150}.}\label{f_m87_f165}
\end{figure}

\begin{figure}
    \includegraphics[width=0.34\textwidth, angle=90]{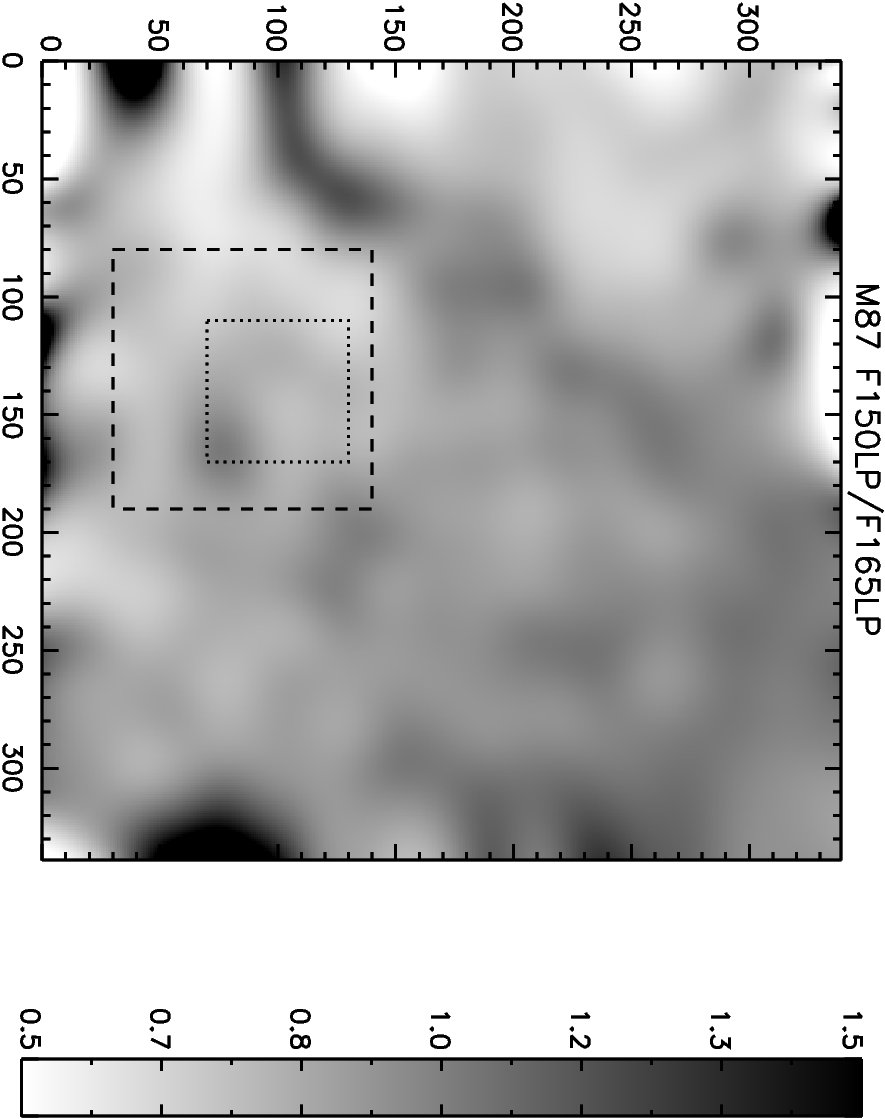}
  \vspace{0.5cm}
  \caption{M87 south-east. F150LP$_{\nu}$/F165LP$_{\nu}$ ratio image. The rectangles and axes drawn are the same as in Fig. \ref{f_m87_f150}.}\label{f_m87_ratio}
\end{figure}

\begin{figure}
    \includegraphics[width=0.30\textwidth, angle=90]{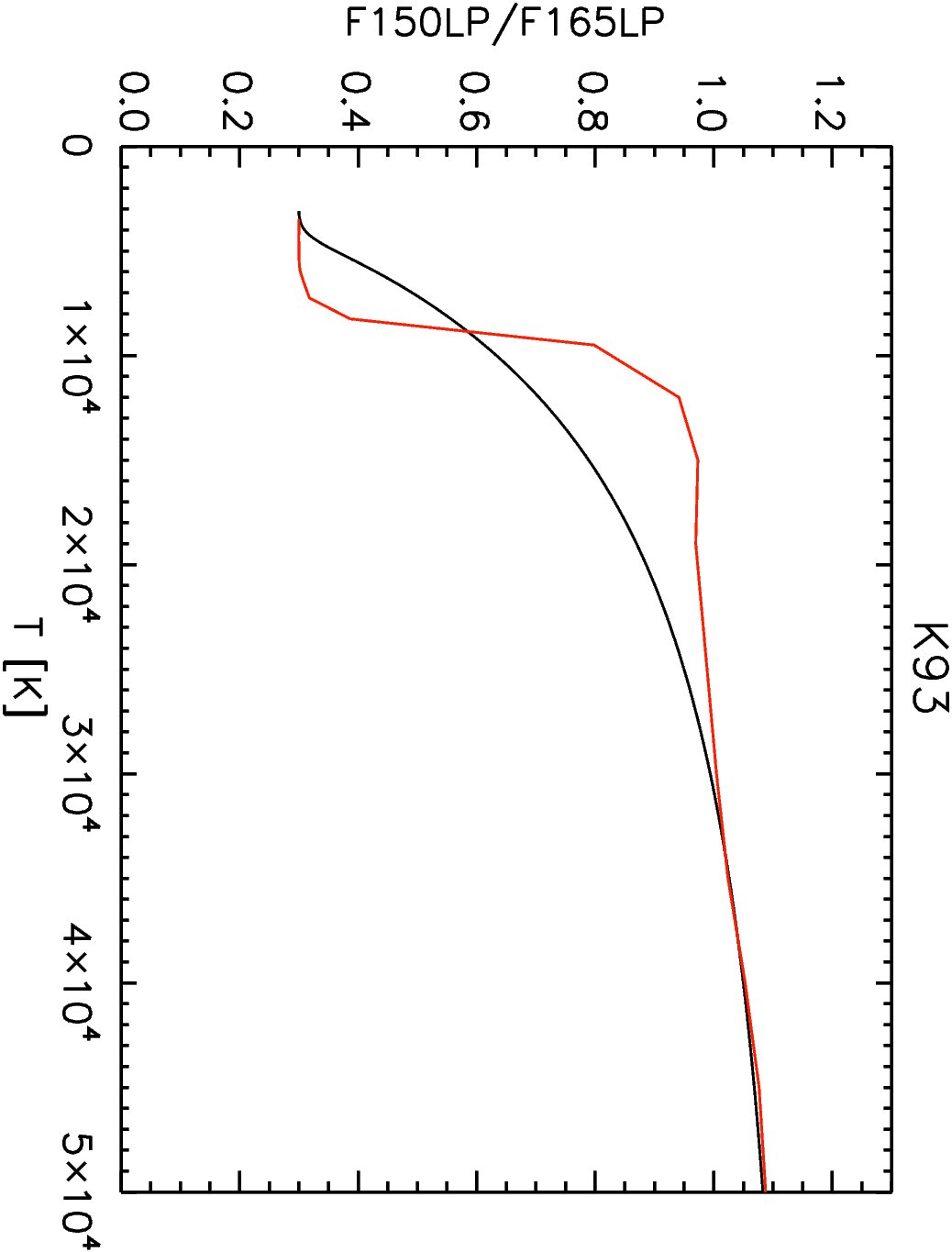}
  \vspace{0.5cm}
  \caption{M87 south-east. BB models (black line) and K93 stellar models (red line) showing the F150LP$_{\nu}$/F165LP$_{\nu}$ ratio versus temperature.}\label{f_m87_bb_k93}
\end{figure}

Line emission from [CIV] 1549~$\rm \AA$ is claimed by \citet{Sp09} to represent the bulk of the FUV emission observed from the south-eastern filaments in M87. They present HST FUV imaging using the F150LP and F165LP filters. Based on the apparent absence of these filaments in the F165LP image they suggest that the bulk of the filamentary FUV emission is due to the CIV line at 1549~$\rm \AA$. 
\\ \\
We have re-processed their images (project code 11681) and come to a different conclusion. Both the throughput and the exposure time of the F165LP image is lower than for the F150LP image. This means that the sensitivity of the F165LP image is significantly lower than the F150LP image. After re-binning the pixels by a factor four on each side and smoothing the data with a 30 pixel FWHM Gaussian kernel we do detect the south-eastern filament in both images. The filament is indicated by the dashed square region in Figs. \ref{f_m87_f150} and \ref{f_m87_f165}. 
\\ \\
We find F150LP$_{\nu}$/F165LP$_{\nu}$=0.80$\pm$0.04 for the integrated flux in the dotted, 6 arcsec by 6 arcsec region shown in Fig. \ref{f_m87_ratio}. This region contains the brightest part of the south-east filament. We computed the F150LP$_{\nu}$/F165LP$_{\nu}$ ratio for BB and K93 stellar models, see Fig. \ref{f_m87_bb_k93}. We find that the observed ratio is consistent with T$\approx$10000~K stars in the context of the K93 stellar models. We thus conclude that [CIV] emission is possible, but the observed FUV emission can also be explained by normal stars.

%%fig: filaments f150lp/f165lp
%\begin{figure*}
%\end{figure*}

%%fig: filaments f150lp/f165lp
%\begin{figure*}
%\end{figure*}

%%fig: filaments f150lp/f165lp
%\begin{figure*}
%\end{figure*}

%%fig: filaments f150lp/f165lp
%\begin{figure*}
%\end{figure*}

\bsp

\label{lastpage}

\end{document}